\documentclass[11pt]{article}

\usepackage{JOEconfig}

\usepackage[T1]{fontenc}
\usepackage{graphicx}

\usepackage[svgnames]{xcolor}
\usepackage{hyperref}
\hypersetup{
    linktocpage,  
    colorlinks,
    citecolor=DarkBlue,
    filecolor=blue,
    linkcolor=blue,
    urlcolor=DarkBlue
}

\usepackage{accents} 
\usepackage{bm} 

\usepackage{amsmath, amsthm} 
\numberwithin{equation}{section} 
\usepackage{array}
\usepackage{longtable,tabularx} 
\usepackage{cleveref}
\crefformat{equation}{Eq.(#2#1#3)} 
\Crefformat{equation}{Eq.(#2#1#3)} 
\crefformat{section}{\S#2#1#3}
\Crefformat{section}{\S#2#1#3}
\crefformat{remrm}{Remark #2#1#3}
\Crefformat{remrm}{Remark #2#1#3}

\usepackage{enumitem} 
\setlist[itemize,1]{
left=2ex,
labelsep=0.75ex,
itemsep=3pt,
topsep=1pt,
label={\scriptsize{\textbullet}},
after={\vspace{0.5ex}},
listparindent=0.8em, 
parsep=0pt, 
}

\setlist[itemize,2]{left=1.5ex,nosep}

\setlist[enumerate,1]{
left=1.5ex,
labelsep=0.75ex,
itemsep=3pt,
topsep=1pt,
after={\vspace{0.5ex}},
listparindent=0.8em, 
parsep=0pt, 
}

\setlist[enumerate,2]{left=1ex,nosep}

\usepackage{stackengine}
\usepackage{xfrac}
\usepackage{tensor}
\usepackage{bigints}
\usepackage{xcolor} 
\usepackage{cancel} 
\allowdisplaybreaks 
\usepackage{mathtools} 
\usepackage{extarrows} 

\usepackage[slantedGreek,trueslanted,libertine,vvarbb]{newtx}
    
\DeclareMathAlphabet{\mathsfb}{OT1}{\sfdefault}{b}{sl}

\usepackage[scale=1.05]{sourcesanspro} 

\DeclareRobustCommand{\textsb}[1]{{\fontseries{sb}\selectfont #1}} 
\newcommand{\sbseries}{\fontseries{sb}\selectfont} 

\newsavebox\foobox

\newsavebox\fooboxx

\usepackage[frak=euler,scr=cm]{mathalpha}

\usepackage{upgreek} 
\usepackage[artemisia]{textgreek} 

\usepackage{mtcoff} 

\title{\vspace{-1em}\textsb{Hamiltonian dynamics with nonconservative forces\\ on extended phase space}\\
{\Large{\textmd{An analytical mechanics derivation}}}}
\hypersetup{
pdftitle={
    Hamiltonian dynamics with nonconservative forces on extended phase space: An analytical mechanics derivation},
pdfauthor={Joseph T.A. Peterson}
}

\author{\normalsize Joseph T.A. Peterson\thanks{Work conducted while affiliated with the Department of Aerospace Engineering, Texas A\&M University, College Station, TX.} ,\;\; 
Manoranjan Majji\thanks{Professor of Aerospace Engineering, Texas A\&M University, College Station, TX.} ,\;\;
and John L.~Junkins\footnotemark[2]
}

\begin{document}
\date{}
\maketitle
\dominitoc[n]  
\doparttoc 



\newcommand{\pgrf}[1]{\noindent\textbf{#1}}
\newcommand\blfootnote[1]{%
  \begingroup
  \renewcommand\thefootnote{}\footnote{#1}%
  \addtocounter{footnote}{-1}%
  \endgroup
}

\newcommand{\cf}[0]{l}%
\newcommand{\cff}[0]{{\sl{f}}}

\newcommand{\nind}[0]{\noindent}
\newcommand{\verx}[1][1.0]{ \vspace{#1ex}}
\newcommand{\Verx}[1][2]{ \vspace{#1ex}\noindent}

\newcommand{\tx}[1]{{\text{#1}}}
\newcommand{\txi}[1]{{\textnormal{\textit{#1}}}}
\newcommand{\tbf}[1]{{\textbf{\textup{\textrm{#1}}}}}
\newcommand{\trm}[1]{{\textup{\textrm{#1}}}}
\newcommand{\txnonb}[1]{{\textmd{#1}}} 
\newcommand{\txup}[1]{{\textup{#1}}} 
\newcommand{\bfi}[1]{{\textbf{\textit{#1}}}}
\newcommand{\Bfi}[1]{{ \scalebox{1.18}{\textbf{\textit{#1}}}  }}

\newcommand{\rmsb}[1]{{\textup{\textsb{#1}}}}%
\newcommand{\itsb}[1]{{\textit{\textsb{#1}}}}

\renewcommand{\sl}[1]{{\textnormal{\textrm{\textsl{#1}}}}}
\newcommand{\slb}[1]{{\textbf{\textsl{#1}}}}

\newcommand{\txsfb}[1]{\textbf{\textit{\textsf{#1}}}}
\newcommand{\sfi}[1]{{\textit{\textsf{#1}}}}%
\newcommand{\sfup}[1]{{\textup{\textsf{#1}}}}
\newcommand{\sfbup}[1]{ {\textbf{\textup{\textsf{#1}}}} }%

\renewcommand{\sc}[1]{{\textsc{#1}}}
\newcommand*{\rmsc}[1]{{\textrm{\textup{\textsc{#1}}}}}
\newcommand*{\bsc}[1]{{\textbf{\textsc{#1}}}}
\newcommand*{\rmbsc}[1]{{\textbf{\textrm{\textsc{#1}}}}}
\newcommand*{\itsc}[1]{{\textit{\textrm{\textsc{#1}}}}}
\newcommand*{\bisc}[1]{{\textit{\textbf{\textsc{#1}}}}}
\newcommand*{\sfsc}[1]{\textsf{\textit{\textsc{#1}}}}
\newcommand*{\sfbsc}[1]{\textbf{\textsf{\textit{\textsc{#1}}}}}
\newcommand{\scg}[0]{{\trm{\textsc{g}}}}
\newcommand{\sck}[0]{{\trm{\textsc{k}}}}
\newcommand{\scm}[0]{{\trm{\textsc{m}}}}
\newcommand{\scb}[0]{{\trm{\textsc{b}}}}

\newcommand{\isc}[1]{{\scriptstyle{\textit{\textrm{\textsc{#1}}}}}} 
\newcommand{\ssc}[1]{{\scalebox{0.6}{$#1$}}} 

\newcommand{\eq}[1]{\text{$#1$}}

\newcommand{\bemph}[1]{\itsb{#1}} 
 \newcommand{\sbemph}[1]{\rmsb{#1}} 



\newcommand{\switchToTenPoint}{%
  \renewcommand{\normalsize}{\fontsize{10}{12}\selectfont}%
  \renewcommand{\small}{\fontsize{9}{11}\selectfont}%
  \renewcommand{\footnotesize}{\fontsize{8}{9}\selectfont}%
  \renewcommand{\scriptsize}{\fontsize{7}{8}\selectfont}%
  \renewcommand{\tiny}{\fontsize{5}{6}\selectfont}%
  \normalsize 
}

\newcommand{\smsize}[1]{%
{\text{\small{#1}}}}

\newcommand{\fnsize}[1]{%
{\text{\footnotesize{#1}}} }%

\newcommand{\scrsize}[1]{%
{\text{\scriptsize{#1}}}}

\newcommand{\nssize}[1]{%
{\text{\normalsize{#1}}}}

\newcommand*{\nmsz}[1]{{\displaystyle{#1}}}
\newcommand*{\scrsz}[1]{{\scriptstyle{#1}}}
\newcommand*{\ssz}[1]{{\scriptstyle{#1}}}
\newcommand*{\ii}[1]{{\scriptscriptstyle{#1}}} 
\renewcommand*{\ss}[1]{ {\scalebox{0.7}{$#1$}} }

\newcommand*{\smsz}[1]{\text{\small{$#1$}} }
\newcommand*{\fnsz}[1]{\text{\footnotesize{$#1$}} }
\newcommand*{\ns}[1]{{\scalebox{1.15}{$#1$}} }
\renewcommand*{\lg}[1]{{\scalebox{1.25}{$#1$}} }

\newcommand{\mscale}[2][.7]{ {\text{\scalebox{#1}{$#2$}}} } 

\newcommand{\ttfrac}[2]{{\scriptstyle{\frac{\bs{#1}}{\bs{#2}}}}}


\newcommand{\til}[1]{\tilde{#1}}
\newcommand{\wt}[1]{ \widetilde{#1}  }
\newcommand{\wh}[1]{\widehat{#1}}
\newcommand{\ol}[1]{\overline{#1}}

\newcommand{\uln}[1]{{\underline{\smash{#1}\mkern-1mu}{\mkern1mu}}}
\newcommand{\ubar}[1]{ \underbar{$#1\hspace{-0.15mm}$}\hspace{0.15mm} }
\newcommand{\ub}[1]{{\underaccent{\text{—}}{\smash{#1}}}} 

\newcommand{\dt}[1]{\accentset{\mbox{\footnotesize\textbf{.}}}{#1}}
\newcommand*{\ddt}[1]{ \accentset{\mbox{\footnotesize\bfseries..}}{#1} }
\newcommand*{\Dt}[1]{\accentset{\mbox{\small\textbf{.}}}{#1}}
\newcommand*{\DDt}[1]{ \accentset{\mbox{\bfseries .\hspace{-0.03ex}.}}{#1} }

\newcommand{\rng}[1]{ \mathring{#1} }
\newcommand{\rrng}[1]{{\rlap{$\,\mathring{\vphantom{#1}}\mkern4.5mu\mathring{\vphantom{#1}}$}}#1}
\newcommand{\rring}[1]{ \accentset{\circ\circ}{#1} }
\newcommand{\ringp}[1]{\accentset{\circ'}{#1} }
\newcommand{\rringp}[1]{\accentset{\circ\circ'}{#1}}

\newcommand{\pdt}[1]{ \acute{#1} }
\newcommand{\pddt}[1]{ \rlap{$\;\,\acute{\vphantom{#1}}\!\acute{\vphantom{#1}}$}#1 }
\newcommand{\pdot}[1]{{\'{#1}}}
\newcommand{\pddot}[1]{{\H{#1}}}

\newcommand{\tridot}[1]{ \accentset{\triangle}{#1} }
\newcommand{\triddot}[1]{ \accentset{\triangle\triangle}{#1} }
\newcommand{\tridt}[1]{ \accentset{\blacktriangle}{#1} }
\newcommand{\triddt}[1]{ \accentset{\blacktriangle\!\blacktriangle}{#1} }
\newcommand{\deldot}[1]{ \accentset{\triangledown}{#1} }
\newcommand{\delddot}[1]{ \accentset{\triangledown\triangledown}{#1} }

\newcommand{\bxdot}[1]{ \accentset{\square}{#1} }
\newcommand{\bxddot}[1]{ \accentset{\square\square}{#1} }
\newcommand{\sqdt}[1]{ \accentset{\blacksquare}{#1} }
\newcommand{\sqddt}[1]{ \accentset{\blacksquare\blacksquare}{#1} }

\newcommand{\hdot}[1]{ \accentset{\mbox{\hspace{0.1mm}.}}{#1} }
\newcommand{\hddot}[2][.2ex]{\ddot{\raisebox{0pt}[\dimexpr\height+#1][\depth]{$#2$}}}
\newcommand{\ggvec}[2][-2pt]{\vec{\raisebox{0pt}[\dimexpr\height+#1][\depth]{$#2$}}}

\newcommand{\harp}[1]{ {\accentset{\rightharpoonup}{#1}} }
\newcommand{\varvec}[1]{ {\accentset{\to}{#1}} }


\let\oldcdot\cdot
\renewcommand*{\cdot}{{\mkern1.2mu\oldcdot\mkern1.2mu}}
\newcommand{\slot}[0]{{\oldcdot}} 
\newcommand{\cdt}[0]{ \pmb{\cdot} }
\newcommand{\bdt}[0]{ \bs{\cdot} }

\newcommand{\txblt}[0]{{\text{\small{\textbullet\hspace{.7ex}}}}}
\newcommand*{\sblt}[0]{ \raisebox{0.25ex}{$\scriptscriptstyle{\bullet}$}  }
\newcommand*{\bltt}[0]{ \text{\large{${\,\,\bullet\,\,}$}} }
\newcommand*{\nmblt}[0]{ \raisebox{0.15ex}{$\scriptstyle{\bullet}$}  }
\newcommand{\CDT}[0]{ {\mkern2mu \ss{\bs{\therefore}} \mkern2mu} }

\newcommand{\cddt}[0]{{\mkern2mu\pmb{\cdot}\mkern2mu}}
\newcommand{\cddot}[0]{ {\mkern2mu\uln{\pmb{\cdot}}\mkern2mu} }

\newcommand*{\otms}[0]{ {\raisebox{0.15ex}{$\scriptstyle{\,\otimes\,}$}}  }
\newcommand*{\opls}[0]{ \raisebox{0.15ex}{$\scriptstyle{\,\oplus\,}$}  }
\newcommand*{\wdg}[0]{ \raisebox{0.15ex}{\footnotesize{$\,\wedge\,$}}  }
\newcommand*{\tms}[0]{ \raisebox{0.1ex}{$\scriptstyle{\,\times\,}$}  }
\newcommand{\botimes}[0]{ \scalebox{1.3}{$\otimes$} }
\newcommand{\boplus}[0]{ \scalebox{1.3}{$\oplus$} }
\newcommand{\bwedge}[1]{ {\scalebox{1}[1.3]{$\wedge$}}^{\!#1} }

\newcommand*{\ssqr}[0]{ \raisebox{0.15ex}{$\scriptstyle{\blacksquare}$}  }
\newcommand*{\iisqr}[0]{ \raisebox{0.25ex}{$\,\scriptscriptstyle{\blacksquare}\,$}  }
\newcommand*{\sbx}[0]{ \raisebox{0.15ex}{$\scriptstyle{\square}$}  }
\newcommand*{\iibx}[0]{ \raisebox{0.25ex}{$\,\scriptscriptstyle{\square}\,$}  }

\newcommand{\dimd}[0]{\diamond}%
\newcommand*{\tri}[0]{\triangle}%
\newcommand*{\trid}[0]{\triangledown} %
\newcommand*{\rtri}[0]{\triangleright} %
\newcommand*{\ltri}[0]{\triangleleft} %
\newcommand*{\stri}[0]{ \raisebox{0.25ex}{$\scriptstyle{\,\triangle\,}$} }
\newcommand*{\strid}[0]{\raisebox{0.25ex}{$\scriptstyle{\,\triangledown\,}$}}%
\newcommand*{\srtri}[0]{ \raisebox{0.25ex}{$\scriptstyle{\,\triangleright\,}$} }
\newcommand*{\slri}[0]{\raisebox{0.25ex}{$\scriptstyle{\,\triangleleft\,}$}}%

\newcommand*{\rnlarrow}[0]{ \overset{\nLeftarrow\;}{\Rightarrow}}
\newcommand*{\Rnlarrow}[0]{ \overset{\Leftarrow \!/ \!=\,}{\Longrightarrow}}
\newcommand*{\lnrarrow}[0]{ \overset{\,\nRightarrow}{\Leftarrow}}
\newcommand*{\Lnrarrow}[0]{ \overset{=\! /\!\Rightarrow\,}{\Longleftarrow}}
\newcommand{\nDownarrow}[0]{\rotatebox[origin=c]{-90}{$\nRightarrow$}}
\newcommand{\nUparrow}[0]{\rotatebox[origin=c]{90}{$\nRightarrow$}}


\newcommand*{\shrp}[0]{{\scalebox{0.55}{$\sharp$}}}
\newcommand*{\flt}[0]{{\scalebox{0.6}{$\flat$}}}

\newcommand{\iio}[0]{{\scalebox{0.55}{$\mrm{o}$}} }
\newcommand{\iix}[0]{{\scriptscriptstyle{\times}} }
\newcommand{\iistr}[0]{{\scriptscriptstyle{\star}} }
\newcommand{\str}[0]{{\scriptscriptstyle{\star}} }
\newcommand{\drg}[0]{{\ss{\dagger}}} 

\newcommand{\hodge}[0]{\star}
\newcommand{\hdge}[1]{ #1^\ii{\star}  }
\newcommand{\hhdge}[1]{ #1^\ii{\star\star}  }
\newcommand{\hdg}[0]{{\scriptscriptstyle{\star}}}
\newcommand{\varhodge}[0]{\clubsuit} 
\newcommand{\varhdg}[0]{{\scriptscriptstyle{\varhodge}}}
\newcommand{\ax}[1]{ {#1^{\scriptscriptstyle{\times}}} }
\newcommand{\axx}[1]{ {#1^{\scriptstyle{\times}}} }

\newcommand{\inv}[1]{ #1^{\scriptscriptstyle{-\!1}} }
\newcommand{\negone}[0]{ {\text{-} 1} }
\newcommand{\negg}[0]{{\text{-}}}

 \newcommand{\trn}[1]{ #1^{\ii{\textsf{\textup{T}}}}  }
\newcommand{\invtrn}[1]{ #1^{\ii{\textsf{\textup{--T}}}} }
 
\newcommand*{\zer}[0]{ {\ss{\mathit{0}}} } %
\newcommand*{\zr}[0]{ {\ss{0}} }%
\newcommand{\nozer}[0]{ {\scriptscriptstyle{\emptyset}} }

\newcommand*{\sodot}[0]{ \raisebox{0.12ex}{$\scriptstyle{\odot}$}  }

\newcommand*{\nc}[0]{{\ii{\mrm{nc}}}}


\newcommand*{\mrm}[1]{ {\mathrm{#1}} } 
\newcommand*{\iimrm}[1]{ {\ii{\mathrm{#1}}} }
\newcommand*{\iirm}[1]{ {\ii{\textrm{#1}}} }

\newcommand*{\mbf}[1]{{\mathbf{#1}}} 

\newcommand{\rmb}[1]{\tbf{#1}}


\newcommand{\bs}[1]{{\boldsymbol{#1}}}
\newcommand{\bsi}[1]{\boldsymbol{\mit{#1}}} 
\newcommand{\accalign}[2]{ \mkern1mu #1{\mkern-2mu #2} \mkern-1mu }
\newcommand{\tbs}[1]{ {\accalign{\tilde}{\bs{#1}}}\vphantom{#1} }%
\newcommand{\wtbs}[1]{ {\accalign{\widetilde}{\bs{#1}}} }%
\newcommand{\hbs}[1]{ {\accalign{\hat}{\bs{#1}}}\vphantom{#1} }%
\newcommand{\whbs}[1]{ {\accalign{\widehat}{\bs{#1}}} }%
\newcommand{\barbs}[1]{ {\accalign{\bar}{\bs{#1}}}\vphantom{#1} }%
\newcommand{\vecbs}[1]{ {\accalign{\vec}{\bs{#1}}} }%
\newcommand*{\dtbs}[1]{ {\accalign{\dt}{\bs{#1}}} }%
\newcommand*{\ddtbs}[1]{ {\accalign{\ddt}{\bs{#1}}} }%

\newcommand{\tbm}[1]{\tilde{\bm{#1}}}
\newcommand{\wtbm}[1]{\widetilde{\bm{#1}}}
\newcommand{\hbm}[1]{\hat{\bm{#1}}}
\newcommand{\whbm}[1]{\widehat{\bm{#1}}}
\newcommand{\barbm}[1]{\bar{\bm{#1}}}

\newcommand*{\nbs}[1]{ \scalebox{1.15}{$\boldsymbol{#1}$} }
\newcommand*{\lbs}[1]{ \scalebox{1.4}{$\boldsymbol{#1}$} }
\newcommand*{\nbm}[1]{ \scalebox{1.2}{$\bm{#1}$} }
\newcommand*{\lbm}[1]{ \scalebox{1.4}{$\bm{#1}$} }


\newcommand{\msfb}[1]{{\mathsfb{#1}}} 

\newcommand{\sfb}[1]{{\msfb{#1}}}

\renewcommand{\accalign}[2]{ \mkern1mu #1{\mkern-1.5mu #2} \mkern-.75mu }
\newcommand{\tsfb}[1]{ {\accalign{\tilde}{\sfb{#1}}} }%
\newcommand{\wtsfb}[1]{ {\accalign{\widetilde}{\sfb{#1}}} }%
\newcommand{\hsfb}[1]{ {\accalign{\hat}{\sfb{#1}}} }%
\newcommand{\whsfb}[1]{ {\accalign{\widehat}{\sfb{#1}}} }%
\newcommand{\barsfb}[1]{ {\accalign{\bar}{\sfb{#1}}} }%
\newcommand{\bsfb}[1]{ {\accalign{\bar}{\sfb{#1}}} }%
\newcommand{\vecsfb}[1]{ {\accalign{\vec}{\sfb{#1}}} }%
\newcommand*{\dtsfb}[1]{ {\accalign{\dt}{\sfb{#1}}} }%
\newcommand*{\ddtsfb}[1]{ {\accalign{\ddt}{\sfb{#1}}} }%

\newcommand*{\sfg}[0]{{\bs{g}}}


\newcommand*{\mbb}[1]{ {\mathbb{#1}} } 
\newcommand*{\tmbb}[1]{ \tilde{\mbb{#1}} }
\newcommand*{\wtmbb}[1]{ \widetilde{\mbb{#1}} }
\newcommand*{\hmbb}[1]{ \hat{\mbb{#1}} }
\newcommand*{\whmbb}[1]{ \widehat{\mbb{#1}} }
\newcommand*{\bbk}[0]{\Bbbk} 

\newcommand*{\mfrak}[1]{ {\mathfrak{#1}} }
\newcommand*{\bfrak}[1]{ {\bs{\mathfrak{#1}}} }
\newcommand*{\tmfrak}[1]{ \tilde{\mfrak{#1}} }
\newcommand*{\wtmfrak}[1]{ \widetilde{\mfrak{#1}} }
\newcommand*{\hmfrak}[1]{ \hat{\mfrak{#1}} }
\newcommand*{\whmfrak}[1]{ \widehat{\mfrak{#1}} }


\newcommand*{\mscr}[1]{{\mathscr{#1}}}%
\newcommand*{\bscr}[1]{ \bs{\mathscr{#1}} }
\newcommand*{\tscr}[1]{\tilde{\mathscr{#1}}}
\newcommand*{\wtscr}[1]{\widetilde{\mathscr{#1}}}
\newcommand*{\hscr}[1]{ \hat{\mathscr{#1}} }
\newcommand*{\whscr}[1]{ \widehat{\mathscr{#1}} }
\newcommand*{\barscr}[1]{ \bar{\mathscr{#1}} }
\newcommand*{\olscr}[1]{ \overline{\mathscr{#1}} }

\newcommand{\noscr}[1]{#1}
\newcommand{\whnoscr}[1]{\widehat{#1}}

\newcommand*{\mcal}[1]{ {\mathcal{#1}} }
\newcommand*{\bcal}[1]{ \bs{\mathcal{#1}} }
\newcommand*{\tcal}[1]{\tilde{\mathcal{#1}}}
\newcommand*{\wtcal}[1]{\widetilde{\mathcal{#1}}}
\newcommand*{\hcal}[1]{ \hat{\mathcal{#1}} }
\newcommand*{\whcal}[1]{ \widehat{\mathcal{#1}} }
\newcommand*{\barcal}[1]{ \bar{\mathcal{#1}} }
\newcommand*{\olcal}[1]{ \overline{\mathcal{#1}} }

\newcommand{\iical}[1]{{\scriptscriptstyle{\mcal{#1}}}}
\newcommand{\sscal}[1]{ {\scalebox{0.7}{$\mcal{#1}$}} } 
\newcommand{\iiscr}[1]{ {\scriptscriptstyle{\mscr{#1}}} }
\newcommand*{\sscr}[1]{ {\scalebox{0.7}{$\mscr{#1}$}} } 



\newcommand{\en}[0]{{\itsc{n}}}
\newcommand{\emm}[0]{{\itsc{m}}}
\newcommand{\envec}[0]{ \hbe_\en }
\newcommand{\enform}[0]{\,\hat{\!\bs{\epsilon}}^\en}
\newcommand{\tilp}[0]{{\mkern2mu \til{\mkern-2mu\smash{p}}}}

\newcommand*{\tp}[1]{{\vphantom{#1}\uln{#1}}} 
\newcommand*{\btp}[1]{ {\bar{\tp{#1}}} }
\newcommand*{\bartp}[1]{ {\bar{\tp{#1}}} }
\newcommand*{\tiltp}[1]{ {\til{\tp{#1}}} }
\newcommand*{\ttp}[1]{ {\til{\tp{#1}}} }
\newcommand*{\wttp}[1]{ {\wt{\tp{#1}}} }
\newcommand*{\htp}[1]{ {\hat{\tp{#1}}} }
\newcommand*{\whtp}[1]{ {\wh{\tp{#1}}} }
\newcommand{\dottp}[1]{{\dot{\tp{#1}}} }

\newcommand*{\tup}[1]{{\bs{\mathrm{#1}}}}
 \newcommand*{\bartup}[1]{ {\bar{\tup{#1}}} } 
\newcommand*{\tiltup}[1]{ {\til{\tup{#1}}} }
\newcommand*{\wttup}[1]{ {\wt{\tup{#1}}} }
\newcommand*{\htup}[1]{ {\hat{\tup{#1}}} }
\newcommand*{\whtup}[1]{ {\wh{\tup{#1}}} }
\newcommand{\dottup}[1]{{\dot{\tup{#1}}} }
\newcommand{\nrmtup}[1]{ {#1} } %

\newcommand*{\nbtup}[1]{\mathrm{#1}}
\newcommand*{\nbtiltup}[1]{\til{\nbtup{#1}}}
\newcommand*{\nbwhtup}[1]{\wh{\nbtup{#1}}}


\newcommand*{\pt}[1]{{#1}}
\newcommand{\tilpt}[1]{ {\til{\pt{#1}}} }
\newcommand*{\barpt}[1]{ {\bar{\pt{#1}}} }
\newcommand{\hatpt}[1]{ {\hat{\pt{#1}}} }
\newcommand{\hpt}[1]{ {\hat{\pt{#1}}} }
\newcommand*{\ptvec}[1]{ {\vec{\pt{#1}}} }
\newcommand*{\dtpt}[1]{\dt{\pt{#1}}}


\renewcommand{\d}[0]{\mrm{d}}
\newcommand*{\rmd}[0]{{\text{\rmsb{d}}}}
\newcommand*{\dif}[0]{ {\sfb{d}} }
\newcommand*{\Dif}[0]{{ \scriptstyle{\textbf{\textsf{\textsl{D}}}} }}

\newcommand{\vard}[0]{\deltaup}

\newcommand{\difbar}[0]{{ \txsfb{\kern0.6ex\ooalign{\hidewidth\raisebox{0.63ex}{--}\hidewidth\cr{\kern-0.6ex\textsf{d}}\cr}} \vphantom{d} }}
\newcommand{\itdifbar}{%
   {\mkern4.5mu \txsfb{\ooalign{\hidewidth\raisebox{0.65ex}{--}\hidewidth\cr$\mkern-4.5mu \bs{d}$\cr}} \vphantom{d} }%
}

\newcommand*{\hdif}[0]{  \hspace{1mm}\hat{\hspace{-1mm}\dif} }
\newcommand*{\tdif}[0]{  \hspace{1mm}\tilde{\hspace{-1mm}\dif} }
\newcommand*{\deldif}[0]{\bs{\delta}}

\newcommand{\tinywedge}[0]{ \scalebox{0.45}{$\pmb{\wedge}$}}
\newcommand{\exd}[0]{\dif_{\hspace{-0.25mm}{\tinywedge}}}
\newcommand{\exdbar}[0]{\difbar_{\hspace{-0.25mm}{\tinywedge}}}

\newcommand{\codif}[0]{ {\dif_{\ii{\!\star}}} }

\newcommand*{\fibd}[0]{{\sfb{F}}}%

\newcommand{\difscale}[1]{\scalebox{0.72}{$#1$}}

\newcommand{\lderiv}[1]{ \ns{\textrm{\pounds}}_{\difscale{#1}} } 

\newcommand{\nab}[1][\mkern1mu]{ \nabla_{\mkern-4mu \difscale{#1}\mkern1mu} }
\newcommand{\del}[1]{ \nabla_{\mkern-4mu {#1}\mkern1mu}  }

\newcommand{\pderiv}[2]{ \tfrac{\uppartial #1}{\uppartial #2} }
\newcommand{\ppderiv}[3]{\tfrac{\uppartial^2 #1}{\uppartial #2 \uppartial #3}}

\newcommand{\pdiv}[2]{ \pd_{#2} #1 }

\newcommand{\myfrac}[2]{\tfrac{\displaystyle{#1}}{\displaystyle{#2}}}

\newcommand{\pder}[2]{\myfrac{\uppartial #1}{\uppartial #2}} 
\newcommand{\ppder}[3]{\myfrac{\uppartial^2 #1}{\uppartial #2 \uppartial #3}}

\newcommand{\Pderiv}[2]{\myfrac{\uppartial #1}{\uppartial #2}} 
\newcommand{\PPderiv}[3]{\myfrac{\uppartial^2 #1}{\uppartial #2 \uppartial #3}}

 \newcommand*{\diff}[2]{ \tfrac{\mathrm{d} #1}{\mathrm{d} #2} }
\newcommand{\ddiff}[2]{ \tfrac{\mathrm{d}^2{#1}}{\mathrm{d} {#2}^2} }
\newcommand{\Diff}[2]{ \frac{\mathrm{d}#1}{\mathrm{d}\,#2} }
\newcommand{\DDiff}[2]{ \frac{\mathrm{d}^2{#1}}{\mathrm{d}\,{#2}^2} }

\newcommand{\fdiff}[1]{ \,\tfrac{ {\!\!\!\!}^{ \text{\footnotesize{$#1$}} }\d }{\;\d\, t} }
\newcommand{\fDiff}[3]{ \,\frac{ \!\!\!\!^{ \text{\footnotesize{$#3$}} }\d #1 }{\;\d\, #2} }
\newcommand{\fddiff}[1]{ \,\tfrac{ \!\!\!\!^{ \text{\scriptsize{$#1$}} }\d^2 }{\;\d\, t^2} }
\newcommand{\fDDiff}[3]{ \,\frac{ \!\!\!\!^{ \text{\footnotesize{$#3$}} }\d^2 #1 }{\;\d\, #2^2} }

\newcommand{\mdiff}[2]{ \tfrac{ \mathrm{D} #1}{\mathrm{d} #2}   }
\newcommand{\nabdiff}[2]{ \tfrac{\nabla #1}{\mathrm{d} #2}   }
\newcommand{\nabpderiv}[2]{ \tfrac{\nabla #1}{\uppartial #2}   }


\newcommand{\pmat}[1]{ \begin{pmatrix} #1 \end{pmatrix} }
\newcommand{\bmat}[1]{ \begin{bmatrix} #1 \end{bmatrix} }
\newcommand{\fnpmat}[1]{ \fnsz{\begin{pmatrix} #1 \end{pmatrix}} }
\newcommand{\fnbmat}[1]{ \fnsz{\begin{bmatrix} #1 \end{bmatrix}} }

\newcommand*{\kd}[0]{{\mathit{1}\mkern-0.5mu}}
\newcommand*{\lc}[0]{{\itsc{e}}}
\newcommand{\ibase}[0]{\hat{\rmsb{\oldstylenums{1}}}}
\newcommand*{\imat}[0]{{\textrm{1}}\hspace{-.2mm}}
\newcommand*{\jmat}[0]{ J}

\newcommand{\Id}[0]{{\mathrm{Id}}}%
\newcommand{\iden}[0]{\sfb{1}}
\newcommand*{\gvol}[0]{\varepsilon}
\newcommand*{\spvol}[0]{\sigma} 
\newcommand*{\lagvol}[0]{\varsigma}
\newcommand*{\spform}[0]{\theta}  
\newcommand*{\spformup}[0]{\thetaup}
\newcommand*{\lagform}[0]{\vartheta} 
\newcommand*{\lagformup}[0]{\varthetaup}

\newcommand{\vf}[0]{v} 
\newcommand{\pf}[0]{p} 
\newcommand{\vlin}[0]{{\mathsc{\mathcal{V}}}}
\newcommand{\plin}[0]{{\pi}} 
\newcommand{\uplin}[0]{{\piup}} 


\newcommand{\rmat}[1]{ \mathrm{M}^{#1}_\ii{\mbb{R}} }
\newcommand{\mats}[1]{ \rmat{#1} }
\newcommand{\mat}[2]{ \mathrm{M}^{#1}_\ii{#2} }
\newcommand{\lgrp}[3]{\mathrm{#1}^{#2}_\ii{\mbb{#3}}}
\newcommand{\lalg}[3]{\mathfrak{#1}^{#2}_\ii{\mbb{#3}}}
\newcommand{\Afmat}[1]{ \mathrm{Af}^{#1}_\ii{\mbb{R}} }
\newcommand{\afmat}[1]{ \mathfrak{af}^{#1}_\ii{\mbb{R}} }
\newcommand{\Glmat}[1]{ \mathrm{Gl}^{#1}_\ii{\mbb{R}} }
\newcommand{\Slmat}[1]{ \mathrm{Sl}^{#1}_\ii{\mbb{R}} }
\newcommand{\glmat}[1]{ \mathfrak{gl}^{#1}_\ii{\mbb{R}} }
\newcommand{\slmat}[1]{ \mathfrak{sl}^{#1}_\ii{\mbb{R}} }
\newcommand{\Spmat}[1]{ \mathrm{Sp}^{#1}_\ii{\mbb{R}} }
\newcommand{\spmat}[1]{ \mathfrak{sp}^{#1}_\ii{\mbb{R}} }
\newcommand{\Omat}[1]{ \mathrm{O}^{#1}_\ii{\mbb{R}} }
\newcommand{\Somat}[1]{ \mathrm{SO}^{#1}_\ii{\mbb{R}} }
\newcommand{\omat}[1]{ \mathfrak{o}^{#1}_\ii{\mbb{R}} }
\newcommand{\somat}[1]{ \mathfrak{so}^{#1}_\ii{\mbb{R}} }
\newcommand{\Umat}[1]{ \mathrm{U}^{#1}_\ii{\mbb{C}} }
\newcommand{\Sumat}[1]{ \mathrm{SU}^{#1}_\ii{\mbb{C}} }
\newcommand{\umat}[1]{ \mathfrak{u}^{#1}_\ii{\mbb{C}} }
\newcommand{\sumat}[1]{ \mathfrak{su}^{#1}_\ii{\mbb{C}} }
\newcommand{\Emat}[1]{ \mathrm{E}^{#1}_\ii{\mbb{R}} }
\newcommand{\Semat}[1]{ \mathrm{SE}^{#1}_\ii{\mbb{R}} }
\newcommand{\emat}[1]{ \mathfrak{e}^{#1}_\ii{\mbb{R}} }
\newcommand{\semat}[1]{ \mathfrak{se}^{#1}_\ii{\mbb{R}} }

\newcommand{\Aften}[0]{ \mathrm{Af} }
\newcommand{\aften}[0]{ \mathfrak{af} }
\newcommand{\Glten}[0]{ \mathrm{Gl} }
\newcommand{\Slten}[0]{ \mathrm{Sl} }
\newcommand{\glten}[0]{ \mathfrak{gl} }
\newcommand{\slten}[0]{ \mathfrak{sl} }
\newcommand{\Spten}[0]{ \mathrm{Sp}}
\newcommand{\spten}[0]{ \mathfrak{sp} }
\newcommand{\Oten}[0]{ \mathrm{O} }
\newcommand{\Soten}[0]{ \mathrm{SO} }
\newcommand{\oten}[0]{ \mathfrak{o} }
\newcommand{\soten}[0]{ \mathfrak{so} }
\newcommand{\Eten}[0]{ \mathrm{E} }
\newcommand{\Seten}[0]{ \mathrm{SE} }
\newcommand{\eten}[0]{ \mathfrak{e} }
\newcommand{\seten}[0]{ \mathfrak{se} }

\newcommand{\Spism}[0]{\mathscr{S}\mkern-1mu p}
\newcommand{\Dfism}[0]{\mathscr{D}\mkern-2mu f}
\newcommand{\Hmism}[0]{\mathscr{H}m }
\newcommand{\Ctism}[0]{\mathscr{C}t}
\newcommand{\Isom}[0]{\mathcal{I}\mkern-3mu s}


\newcommand{\man}[1]{ \mathcal{#1} }  
\newcommand{\tman}[1]{ \til{\man{#1}} } 
\newcommand{\barman}[1]{ \bar{\man{#1}} }
\newcommand{\bman}[1]{ \bar{\man{#1}} }
\newcommand{\hman}[1]{ \hat{\man{#1}} }
\newcommand{\whman}[1]{ \widehat{\man{#1}} }
\newcommand{\sman}[1]{ {#1} } 
\newcommand{\tint}[0]{\man{I}} 

\newcommand{\chart}[2]{ \man{#1}_{\hspace{-0.2mm}\textup{\tiny(} {\scriptscriptstyle{#2}} \textup{\tiny)}} }
\newcommand{\rchart}[2]{ \mathbb{#1}_{\hspace{-0.2mm}\textup{\tiny(} {\scriptscriptstyle{#2}} \textup{\tiny)}} }
\newcommand{\chrt}[2]{ {#1}_{\hspace{-0.2mm}\textup{\tiny(} {\scriptscriptstyle{#2}} \textup{\tiny)}} }

\newcommand{\vsp}[1]{ \mathbb{#1} } 
\newcommand{\aff}[1]{\man{#1}} 
\newcommand{\tvsp}[1]{ \tilde{\vsp{#1}} } 
\newcommand{\bvsp}[1]{ \bar{\vsp{#1}} } 
\newcommand{\taff}[1]{ \tilde{\aff{#1}} } 
\newcommand{\baff}[1]{ \bar{\aff{#1}} } 

\newcommand{\affE}[0]{\man{E}}
\newcommand{\vecE}[0]{\vsp{E}}
\newcommand{\Eaf}[0]{\man{E}}
\newcommand{\Evec}[0]{\vsp{E}}
\newcommand{\emet}[0]{I} 
\newcommand{\bemet}[0]{\sfb{I}}
\newcommand{\rfun}[0]{{r}} 
\newcommand{\lang}[0]{{\ell}} 
\newcommand{\Lang}[0]{{L}}
\newcommand{\blang}[0]{\bm{\ell}} 
\newcommand{\bLang}[0]{\sfb{L}} 
\newcommand{\slang}[0]{{\ldash}} 
\newcommand{\Slang}[0]{{\Ldash}} 
\newcommand{\bslang}[0]{\ldashb} 
\newcommand{\bSlang}[0]{{\sfb{\L}}}

\newcommand*{\fun}[0]{{\scalebox{0.85}[1]{$\mathcal{F}$}}}  
\newcommand*{\tens}[0]{ \mathscr{T} } 
\newcommand*{\forms}[0]{ \Lambdaup }
\newcommand{\formsex}[0]{ \Lambdaup_\ii{\mrm{ex}\!}}
\newcommand{\formscl}[0]{ \Lambdaup_\ii{\mrm{cl}\!}}

\newcommand*{\veckl}[0]{ \mathfrak{X}_{\ii{\!\mathfrak{i\!s}}\!} }
\newcommand*{\vect}[0]{ \mathfrak{X} } 
\newcommand*{\vecsp}[0]{ \mathfrak{X}_{\ii{\!\mathfrak{s\!p}}\!\!} } 
\newcommand*{\vechm}[0]{ \mathfrak{X}_{\ii{\!\mathfrak{h\!\!m}}\!\!} } 
\newcommand{\ham}[1]{{\mathscr{#1}}}
\newcommand{\tham}[1]{\widetilde{\mathscr{#1}}}

 \newcommand{\Tspacefont}[1]{ {\mathrm{#1}} } 
\newcommand{\Tan}[0]{ \Tspacefont{T}   }
\newcommand{\Tanh}[0]{  \Tspacefont{H}  }
\newcommand{\Tanv}[0]{  \Tspacefont{V}  }
\newcommand{\tsp}[1][\mkern2mu]{ \Tspacefont{T}_{\mkern-2mu #1} }
\newcommand{\tspv}[1][\mkern2mu]{\Tspacefont{V}_{\!#1}}
\newcommand{\tsph}[1][]{\Tspacefont{H}_{#1}}
\newcommand{\tspn}[1][\hspace{.5mm}]{\Tspacefont{N}_{\hspace{-.2mm}#1}}
\newcommand{\cotsp}[1][\hspace{1mm}]{ {\Tspacefont{T}^* {\hspace{-1ex}}_{\hspace{-.4mm}#1\hspace{0.5mm}}} } 
\newcommand{\cotspv}[1][\hspace{1mm}]{ \Tspacefont{V}^* {\hspace{-1ex}}_{\!#1\hspace{0.5mm}} }
\newcommand{\cotsph}[1][\hspace{1mm}]{ \Tspacefont{H}^* {\hspace{-1ex}}_{#1\hspace{0.1mm}} }
\newcommand{\cotspn}[1][\hspace{1mm}]{ \Tspacefont{N}^* {\hspace{-1ex}}_{#1\hspace{0.1mm}} }

\newcommand{\prj}[0]{ \scalebox{0.7}{$\Pi$} }  
 \newcommand{\tpr}[0]{ \scalebox{0.5}[0.68]{$\bs{\mcal{T}}$} }
\newcommand{\copr}[0]{{ \mkern1mu\hat{\mkern-1.5mu\smash{\tpr}} }}

\newcommand{\vfun}[1][]{{V^{#1}}}
\newcommand{\pfun}[1][]{{P^{#1}}}
\newcommand{\varpfun}[1][]{{l^{#1}}}

\newcommand*{\tlift}[0]{ T\hspace{-.3mm} }
\newcommand*{\colift}[0]{ \hat{T}\hspace{-.3mm} }
\newcommand*{\uptlift}[0]{ \mrm{T}\hspace{-.3mm} }
\newcommand*{\upcolift}[0]{ \hat{\mrm{T}}\hspace{-.3mm} }

\newcommand{\formsh}[0]{\Lambdaup_\ii{\mrm{h}\!}} 
\newcommand{\formsv}[0]{\Lambdaup_\ii{\mrm{v}\!}} 
\newcommand{\vectv}[0]{\mathfrak{X}_\ii{\mrm{v}\!}} 
\newcommand{\vecth}[0]{\mathfrak{X}_\ii{\mrm{h}\!}} 
\newcommand{\formsbh}[0]{\Lambdaup_\ii{\mrm{bh}\!}} 
\newcommand{\vectbv}[0]{\mathfrak{X}_\ii{\mrm{bv}\!}}

\newcommand{\lft}[1]{ #1^{\hspace{-0.2mm}\scriptscriptstyle\upharpoonright} } 
\newcommand{\lift}[1]{ #1^{\hspace{-0.2mm}\scriptscriptstyle\uparrow} } 
\newcommand{\cotlft}[1]{  #1^{\hspace{-0.2mm}\hat{\scriptscriptstyle{\uparrow}}} }
\newcommand{\invlift}[1]{ #1^{\hspace{-0.2mm}\scriptscriptstyle\downarrow} }

\newcommand{\vlift}[1]{ #1^{\hspace{-0.2mm}\scriptscriptstyle\uparrow} }
\newcommand{\Vlift}[1]{  #1^{\scriptscriptstyle\Uparrow} }
\newcommand{\coVlift}[1]{  #1^{\hat{\scriptscriptstyle{\Uparrow}}} }%

\newcommand{\covlift}[1]{ #1^{\hat{\textup{\tiny{\textsf{V}}}}}}
\newcommand{\verlift}[1]{ #1^{\textup{\tiny{\textsf{V}}}} }%
\newcommand{\coverlift}[1]{ #1^{\hat{\textup{\tiny{\textsf{V}}}}}}

\newcommand{\hlift}[1]{ #1^{\textup{\tiny{\textsf{H}}}} }%
\newcommand{\horlift}[1]{ #1^{\textup{\tiny{\textsf{H}}}} }%
\newcommand{\cohlift}[1]{ #1^{\hat{\textup{\tiny{\textsf{H}}}}} }%

 \newcommand{\dubuparrow}[0]{\rotatebox[origin=c]{90}{$\ii{\twoheadrightarrow}$}}
\newcommand{\tailuparrow}[0]{\rotatebox[origin=c]{90}{$\ii{\rightarrowtail}$}}
\newcommand{\lftt}[1]{#1^{\dubuparrow}}
\newcommand{\llft}[1]{#1^{\tailuparrow}}

\newcommand{\rhk}[0]{\hookrightarrow}
\newcommand{\lhk}[0]{\hooklefttarrow}
\newcommand{\subman}[0]{ \,\accentset{\scriptstyle{\subset}}{\scriptstyle{\hookrightarrow}}\, }

\newcommand{\pbrak}[2]{\{#1  ,  #2 \}}
\newcommand{\Pbrak}[2]{\big\{#1\;,\;#2\big\}}
\newcommand{\ppbrak}[2]{\{ \hspace{-0.85mm} \{#1 ,  #2 \} \hspace{-0.85mm}\} }

\newcommand{\lbrak}[2]{ [ #1 , #2 ] }
\newcommand{\Lbrak}[2]{ \big[ #1\;,\;#2 \big] }
\newcommand{\llbrak}[2]{ [\hspace{-0.8mm}[#1 ,  #2 ]\hspace{-0.8mm}] }

\newcommand{\altbrak}[2]{ {\lfloor #1 ,  #2 \rfloor} }
\newcommand{\aaltbrak}[2]{ {\lfloor \hspace{-0.8mm} \lfloor #1 ,  #2 \rfloor \hspace{-0.8mm} \rfloor} }
\newcommand{\lagbrak}[2]{ {\lfloor #1 ,  #2 \rfloor}  }
\newcommand{\llagbrak}[2]{ {\lfloor \hspace{-0.8mm} \lfloor #1 ,  #2 \rfloor \hspace{-0.8mm} \rfloor} }

\newcommand{\inner}[2]{\langle#1  , #2\rangle}
\newcommand{\iinner}[2]{\langle \hspace{-0.8mm} \langle #1,#2\rangle \hspace{-0.8mm}\rangle}
\newcommand{\pprent}[2]{ (\hspace{-0.8mm}(#1,#2) \hspace{-0.8mm}) }

\newcommand{\ang}[1]{\langle #1 \rangle}
\newcommand{\Ang}[1]{\big\langle #1 \big\rangle }


\newcommand{\abs}[1]{ {\lvert #1 \rvert} }
\newcommand{\Abs}[1]{ \left| #1 \right| }
\renewcommand{\mag}[1]{ |\!| #1 |\!| }
\newcommand{\nrm}[1]{{\lvert #1 \rvert} }
\renewcommand{\det}[0]{ {\textrm{\textup{\footnotesize{det}}}\mkern1mu} }
\renewcommand{\dim}[0]{ {\textrm{\textup{\footnotesize{dim}}}\mkern1mu} }
\newcommand{\codim}[0]{ {\textrm{\textup{\footnotesize{codim}}}\mkern1mu} } %
\newcommand{\rnk}[0]{ {\textrm{\textup{\footnotesize{rnk}}}\mkern1mu} }
\newcommand{\sgn}[0]{ {\textrm{\textup{\footnotesize{sgn}}}\mkern1mu} }
\newcommand{\tr}[0]{ {\textrm{\textup{\footnotesize{tr}}}\mkern1mu} }
\renewcommand{\ker}[0]{ {\textrm{\textup{\footnotesize{ker}}}\mkern1mu} }
\newcommand{\img}[0]{ {\textrm{\textup{\footnotesize{img}}}\mkern1mu} }
\newcommand{\spn}[0]{ {\textrm{\textup{\footnotesize{span}}}\mkern1mu} }
\renewcommand{\div}[1][\,]{ {{\textrm{\textup{\footnotesize{div}}}}_{\!#1}}}

\newcommand{\snn}[1][]{\sin{{\mkern-2mu}#1}}
\newcommand{\csn}[1][]{\cos{{\mkern-2mu}#1}}

\newcommand{\crd}[2]{ \tensor*[^{\ss{#2}}]{[#1]}{} }
\newcommand{\crdl}[2]{ \tensor*[^{\ss{#2}}]{#1}{} }
\newcommand{\cordl}[2]{  #1_{\scriptscriptstyle{\!#2}}  } 
\newcommand{\cord}[2]{  #1_{\scalebox{0.7}{$#2$}}  }

\newcommand{\eval}[2]{\left.{#1}\right|_{#2}}

\newcommand{\pdup}[0]{\uppartial} %
\newcommand{\upd}[0]{{\rotatebox[origin=t]{10}{$\partial$} \mkern-2mu}}
\newcommand{\pd}[0]{{\rotatebox[origin=t]{15}{$\partial$} \mkern-2mu}}


\newcommand{\mypd}[0]{\rotatebox[origin=t]{10}{$\partial$}} 

\newcommand*{\ffsize}[1]{{\scalebox{0.7}{$#1$}}} 

\newcommand*{\bpart}[1]{\bs{\mypd}_{\hspace{-.3mm}#1}}
\newcommand*{\bpartup}[1]{\bs{\mypd}^{#1}}
\newcommand*{\tbpart}[1]{\tilde{\bs{\mypd}}_{\hspace{-.3mm}#1}}
\newcommand*{\tbpartup}[1]{{\tilde{\bs{\mypd}}\vphantom{l}^{#1}}}
\newcommand*{\hbpart}[1]{\hbs{\mypd}_{\mkern-1.5mu #1}}
\newcommand*{\hbpartup}[1]{{\hbs{\mypd}\vphantom{l}^{#1}}}

\newcommand*{\pdii}[1]{ \bs{\mypd}_{\ffsize{\!#1}}}
\newcommand*{\pdupii}[1]{ \bs{\mypd}^{\ffsize{#1}} }
\newcommand*{\pdiiup}[1]{ \bs{\mypd}^{\ffsize{#1}} }
\newcommand*{\tpdii}[1]{ \tilde{\bs{\mypd}}_{\hspace{-.5mm}\ffsize{#1}} }
\newcommand*{\tpdupii}[1]{ \tilde{\bs{\mypd}}{}^{\ffsize{#1}} }
\newcommand*{\hpdii}[1]{ \hbs{\mypd}_{\ffsize{\!#1}} }
\newcommand*{\hpdiiup}[1]{ \hbs{\mypd}\vphantom{l}^{\ffsize{#1}} }
\newcommand*{\barpdii}[1]{ \bar{\bs{\mypd}}_{\hspace{-.6mm}\ffsize{#1}} }
\newcommand*{\barpdiiup}[1]{ {\bs{\bar{\mypd}}}^{\ffsize{#1}} }

\newcommand{\bpd}[1][]{ {\bs{\mypd}_{\ffsize{\!#1}}} }%
\newcommand{\bpdup}[1][]{ {\bs{\mypd}^{\ffsize{#1}}} }
\newcommand{\hbpd}[1][]{ {\hbs{\mypd}_{\ffsize{\!#1}}} }%
\newcommand{\hbpdup}[1][]{ {\hbs{\mypd}^{\ffsize{#1}}} }
\newcommand{\tbpd}[1][]{ {\tbs{\mypd}_{\ffsize{\!#1}}} }%
\newcommand{\tbpdup}[1][]{ {\tbs{\mypd}^{\ffsize{#1}}} }
\newcommand{\barbpd}[1][]{ {\barbs{\mypd}_{\ffsize{\!#1}}} }%
\newcommand{\barbpdup}[1][]{ {\barbs{\mypd}^{\ffsize{#1}}} }

\newcommand*{\bpdh}[1]{ {\bs{\mypd}_{\mkern-1mu\hat{#1}}} }
\newcommand*{\bpdhup}[1]{ {\bs{\mypd}^{\hat{#1}}} }
\newcommand*{\bpdt}[1]{ {\bs{\mypd}_{\mkern-1mu\til{#1}}} }
\newcommand*{\bpdtup}[1]{ {\bs{\mypd}^{\til{#1}}} }
\newcommand*{\bpdbar}[1]{ {\bs{\mypd}_{\mkern-1mu\bar{#1}}} }
\newcommand*{\bpdbarup}[1]{ {\bs{\mypd}^{\bar{#1}}} }
\newcommand*{\bdelh}[1]{ \bs{\delta}^{\hat{#1}}\vphantom{l} }
\newcommand{\bdelhdn}[1][]{ {\bs{\delta}_{\mkern-1mu\hat{#1}}\vphantom{l}} }
\newcommand*{\bdelt}[1]{ \bs{\delta}^{\til{#1}}\vphantom{l} }
\newcommand{\bdeltdn}[1][]{ {\bs{\delta}_{\mkern-1mu\til{#1}}\vphantom{l}} }
\newcommand*{\bdelbar}[1]{ \bs{\delta}^{\bar{#1}}\vphantom{l} }
\newcommand{\bdelbardn}[1][]{ {\bs{\delta}_{\mkern-1mu\bar{#1}}\vphantom{l}} }



\newcommand{\bdel}[1][]{ {\bs{\delta}^{\ffsize{#1}}} }%
\newcommand{\bdeldn}[1][]{ {\bs{\delta}_{\ffsize{\!#1}}} }
\newcommand{\hbdel}[1][]{ \hbs{\delta}\vphantom{l}^{\ffsize{#1}}\vphantom{l} }%
\newcommand{\hbdeldn}[1][]{ \hbs{\delta}_\ffsize{\mkern-1.5mu #1}\vphantom{l} } 
\newcommand{\tbdel}[1][]{ \tbs{\delta}^{\ffsize{#1}}\vphantom{l} }%
\newcommand{\tbdeldn}[1][]{ \tbs{\delta}_\ffsize{\mkern-1.5mu #1}\vphantom{l} }
\newcommand{\barbdel}[1][]{ \barbs{\delta}^{\ffsize{#1}}\vphantom{l} }%
\newcommand{\barbdeldn}[1][]{ \barbs{\delta}_\ffsize{\mkern-1.5mu #1}\vphantom{l} }


\newcommand{\bD}[1][]{{ {\sfb{D}}_{\ffsize{\!#1}}} }
\newcommand{\bDel}[1][]{{ {\bs{\Delta}}^{\mkern-1mu\ffsize{#1}}}}%
\newcommand{\bDeldn}[1][]{{ {\bs{\Delta}}_{\ffsize{#1}}}}%

\newcommand{\ff}[2]{\sfb{#1}_{\ffsize{\!#2}}} 
\newcommand{\ffup}[2]{\sfb{#1}^{\ffsize{#2}}} 
\newcommand{\coff}[2]{\bs{#1}^{\ffsize{#2}}}
\newcommand{\coffdn}[2]{\bs{#1}_{\ffsize{\!#2}}}

\newcommand{\bi}[1][]{{\sfb{i}_{\ffsize{\!#1}}}}
\newcommand{\bio}[1][]{{\bs{\iota}^{\ffsize{#1}}} \vphantom{\iota}}
\newcommand{\hbi}[1][]{{\hsfb{\i}_{\ffsize{\!#1}}}} 
\newcommand{\hbio}[1][]{{\hbs{\iota}^{\ffsize{#1}}} \vphantom{\iota}}

\newcommand{\be}[1][]{{\sfb{e}_{\ffsize{\mkern-2mu #1}}}}
\newcommand{\beup}[1][]{{\sfb{e}^{\ffsize{#1}}}} 
\newcommand{\tbe}[1][]{ {\tsfb{e}_{\ffsize{\mkern-2mu #1}}} }
\newcommand{\hbe}[1][]{ {\hsfb{e}_{\ffsize{\mkern-2mu #1}}} }
\newcommand{\bep}[1][\!]{{\bs{\epsilon}^{\ffsize{#1}}}}
\newcommand{\bepdn}[1][]{{\bs{\epsilon}_{\ffsize{\!#1}}}}
\newcommand{\tbep}[1][\!]{\tbs{\epsilon}^{\ffsize{#1}} \vphantom{\epsilon}}
\newcommand{\hbep}[1][\!]{{\hbs{\epsilon}^{\ffsize{#1}}} \vphantom{\epsilon}}

\newcommand{\bt}[1][]{ \sfb{t}_{\ffsize{\!#1}} \vphantom{t}}
\newcommand{\tbt}[1][]{\tsfb{t}_{\ffsize{\!#1}} \vphantom{t}}
\newcommand{\hbt}[1][]{\hsfb{t}_{\ffsize{\!#1}} \vphantom{t}}
\newcommand{\btup}[1][\,]{ \sfb{t}^{\ffsize{#1}\!} \vphantom{t}}
\newcommand{\tbtup}[1][\,]{\tsfb{t}^{\ffsize{#1}\!} \vphantom{t}}
\newcommand{\hbtup}[1][\,]{\hsfb{t}^{\ffsize{#1}\!} \vphantom{t}}

\newcommand{\btau}[1][\!]{\bm{\tau}^{\ffsize{#1}}\vphantom{\tau}}
\newcommand{\tbtau}[1][\!]{{\tbm{\tau}^{\ffsize{#1}}}}
\newcommand{\hbtau}[1][\!]{{\hbm{\tau}^{\ffsize{#1}}}}
\newcommand{\btaudn}[1][]{{\bm{\tau}_{\ffsize{\!#1}}}}
\newcommand{\tbtaudn}[1][]{{\tbm{\tau}_{\ffsize{\!#1}}}}
\newcommand{\hbtaudn}[1][]{{\hbm{\tau}_{\ffsize{\!#1}}}}


\newcommand{\Zvar}{{ \text{\ooalign{\hidewidth\raisebox{0.25ex}{--}\hidewidth\cr$Z$\cr}}\vphantom{Z}} }

\newcommand{\zvar}{%
  \text{\ooalign{\hidewidth -\kern-.3em-\hidewidth\cr$z$\cr}} \vphantom{z}%
}

\renewcommand{\Zbar}{{\textnormal{\ooalign{\hidewidth\raisebox{0.3ex}{\footnotesize{--}}\hidewidth\cr$Z$\cr}} \vphantom{Z}} }

\newcommand{\zbar}{ \mkern0.5mu
  {\text{\ooalign{\hidewidth\raisebox{0.1ex}{\scriptsize{--}}\hidewidth\cr$\mkern-0.5muz$\cr}} \vphantom{z}}%
} 

\newcommand{\Jbar}{ \mkern2mu {\textnormal{\ooalign{\hidewidth\raisebox{0.4ex}{\scriptsize{--}}\hidewidth\cr$\mkern-2mu J$\cr}} \vphantom{J}} }

\newcommand{\Ibar}{{\textnormal{\ooalign{\hidewidth\raisebox{0.4ex}{\scriptsize{--}}\hidewidth\cr$I$\cr}} \vphantom{I}} }

\newcommand{\Sbar}{{\mkern0mu \textnormal{\ooalign{\hidewidth\raisebox{0.4ex}{\scriptsize{\textbf{---}}}\hidewidth\cr$\mkern-0mu S$\cr}} \vphantom{S}} }

\newcommand{\sbar}{ \mkern-0.5mu
  {\text{\ooalign{\hidewidth\raisebox{0.1ex}{\scriptsize{--}}\hidewidth\cr$\mkern0.5mu s$\cr}} \vphantom{s}}%
}

\newcommand{\Cbar}{{\mkern-7mu \textnormal{\ooalign{\hidewidth\raisebox{0.35ex}{\footnotesize{--}}\hidewidth\cr$\mkern7muC$\cr}} \vphantom{C}} }

\newcommand{\cbar}{%
  {\mkern-3mu\text{\ooalign{\hidewidth\raisebox{0.1ex}{\scriptsize{--}}\hidewidth\cr$\mkern3muc$\cr}} \vphantom{c}}%
}

\newcommand{\ldash}[0]{{\mkern-2mu \textit{\l}\mkern1mu}} 
\newcommand{\Ldash}[0]{{\textit{\L}}} 
\newcommand{\ldashb}[0]{{\itsb{\l}}}
\newcommand{\Ldashb}[0]{{\txsfb{\L}}} 
\newcommand{\Lbar}{ {\mkern-2mu \textnormal{\ooalign{\hidewidth\raisebox{0.35ex}{\scriptsize{--}}\hidewidth\cr$\mkern2mu L$\cr}} \vphantom{L}}  }

\newcommand{\lbar}{%
    {\textnormal{\ooalign{\hidewidth\raisebox{0.3ex}{-}\hidewidth\cr$l$\cr}} \vphantom{l}}%
} 

\newcommand{\blbar}{%
    {\textbf{\ooalign{\hidewidth\raisebox{0.4ex}{\scriptsize{--}}\hidewidth\cr$\bs{l}$\cr}} \vphantom{l}}%
} 

\newcommand{\kbar}{%
   {\mkern-1mu \text{\ooalign{\hidewidth\raisebox{0.8ex}{\scriptsize{--}}\hidewidth\cr$\mkern1mu k$\cr}} \vphantom{k} }%
}
\newcommand{\kbarup}{%
   {\mkern-4mu \textnormal{\ooalign{\hidewidth\raisebox{0.8ex}{\scriptsize{--}}\hidewidth\cr$\mkern4mu \mrm{k}$\cr}} \vphantom{k} }%
}
\newcommand{\txkbar}[0]{\kbarup}

\newcommand{\hvar}{%
   {\mkern-1mu \text{\ooalign{\hidewidth\raisebox{0.8ex}{\scriptsize{--}}\hidewidth\cr$\mkern1mu h$\cr}} \vphantom{h} }%
}
\newcommand{\hbarup}{%
   {\mkern-4mu \textnormal{\ooalign{\hidewidth\raisebox{0.8ex}{\scriptsize{--}}\hidewidth\cr$\mkern4mu \mrm{h}$\cr}} \vphantom{k} }%
}
\newcommand{\txhbar}[0]{\hbarup}

\newcommand{\bbar}{%
   {\mkern-1mu \text{\ooalign{\hidewidth\raisebox{0.8ex}{\scriptsize{--}}\hidewidth\cr$\mkern1mu b$\cr}} \vphantom{b} }%
}
\newcommand{\bbarup}{%
   {\mkern-4mu \textnormal{\ooalign{\hidewidth\raisebox{0.8ex}{\scriptsize{--}}\hidewidth\cr$\mkern4mu \mrm{b}$\cr}} \vphantom{b} }%
} 
\newcommand{\txbbar}[0]{\bbarup}

\newcommand{\dbar}{%
   {\mkern4.5mu \text{\ooalign{\hidewidth\raisebox{0.8ex}{\scriptsize{--}}\hidewidth\cr$\mkern-4.5mu d$\cr}} \vphantom{d} }%
}
\newcommand{\dbarup}{%
   {\mkern3mu \textnormal{\ooalign{\hidewidth\raisebox{0.8ex}{\scriptsize{--}}\hidewidth\cr$\mkern-3mu \mrm{d}$\cr}} \vphantom{d} }%
} 
\newcommand{\txdbar}[0]{\dbarup}

\newcommand{\Gambar}{%
   {\mkern-4mu \text{\ooalign{\hidewidth\raisebox{0.4ex}{\scriptsize{--}}\hidewidth\cr$\mkern4mu \Gamma$\cr}} \vphantom{h} }%
}

\newcommand{\mathsc}[1]{\scalebox{0.7}{$#1$}}

\renewcommand{\v}[0]{{\nu}} 
\newcommand{\txv}[0]{{\txi{v}}}
\newcommand{\altv}[0]{{\mathsc{\mathcal{V}}}}
\newcommand{\newv}[0]{{\mathsc{\mathscr{V}}}}

\newcommand{\one}[0]{{\ss{1}}}
\newcommand{\two}[0]{{\ss{2}}}
\newcommand{\three}[0]{{\ss{3}}}
\newcommand{\four}[0]{{\ss{4}}}
\newcommand{\six}[0]{{\ss{6}}}
\newcommand{\eight}[0]{{\ss{8}}}


\renewcommand{\a}[0]{{\alpha}}
\renewcommand{\b}[0]{{\ii{\beta}}} 
\newcommand{\g}[0]{{\ss{\gamma}}} 
\newcommand{\gam}[0]{{\gamma}}
\newcommand{\Gam}[0]{{\Gamma}}
\newcommand{\y}[0]{{\ss{\lambda}}} 
\newcommand{\lam}[0]{{\lambda}}
\newcommand{\Lam}[0]{{\Lambda}}
\newcommand{\sig}[0]{{\sigma}}
\newcommand{\Sig}[0]{{\Sigma}}
\newcommand{\ep}[0]{{\epsilon}}
\newcommand{\omg}[0]{{\omega}}
\newcommand{\Omg}[0]{{\Omega}}
\newcommand{\kap}[0]{{\kappa}}
\renewcommand{\th}[0]{{\theta}}
\newcommand{\Th}[0]{{\Theta}}
\newcommand{\varth}[0]{{\vartheta}}
\newcommand*{\varep}[0]{{\varepsilon}}
\newcommand{\varsig}[0]{{\varsigma}}
\newcommand{\varkap}[0]{{\varkappa}}
\newcommand{\ro}[0]{{\varrho}}
\newcommand{\varpsi}[0]{\txi{\textpsi}} 

\DeclareRobustCommand{\rchi}{{\mathpalette\irchi\relax}}
\newcommand{\irchi}[2]{\raisebox{\depth}{$#1\chi$}} 
\DeclareRobustCommand{\pup}{{\mathpalette\newp\relax}}
\newcommand{\newp}[2]{\raisebox{0.8\depth}{$#1p$}}
\DeclareRobustCommand{\jup}{{\mathpalette\newj\relax}}
\newcommand{\newj}[2]{\raisebox{0.8\depth}{$#1\jmath$}}

\newcommand{\upa}[0]{{\upalpha}}
\newcommand{\upb}[0]{{\upbeta}}
\newcommand{\upgam}[0]{{\upgamma}}
\newcommand{\upGam}[0]{{\upGamma}}
\newcommand{\Upgam}[0]{{\Upgamma}}
\newcommand{\uplam}[0]{{\uplambda}}
\newcommand{\upLam}[0]{{\upLambda}}
\newcommand{\Uplam}[0]{{\Uplambda}}
\newcommand{\updel}[0]{{\updelta}}
\newcommand{\upsig}[0]{{\upsigma}}
\newcommand{\upSig}[0]{{\upSigma}} 
\newcommand{\Upsig}[0]{{\Upsigma}}
\newcommand{\upomg}[0]{{\upomega}}
\newcommand{\upOmg}[0]{{\upOmega}}
\newcommand{\Upomg}[0]{{\Upomega}}
\newcommand{\upep}[0]{\upepsilon}
\newcommand{\upth}[0]{\uptheta}
\newcommand{\upvarth}[0]{\upvartheta}
\newcommand{\upTh}[0]{\upTheta}
\newcommand{\Upth}[0]{\Uptheta}
\newcommand{\upkap}[0]{\upkappa}
\newcommand{\upvark}[0]{\upvarkappa}
\newcommand{\upn}[0]{\upeta}
\newcommand{\upi}[0]{\upiota}

\newcommand{\aup}[0]{{\alphaup}}
\newcommand{\bup}[0]{{\betaup}}
\newcommand{\gamup}[0]{{\gammaup}}
\newcommand{\Gamup}[0]{{\Gammaup}}
\newcommand{\lamup}[0]{{\lambdaup}}
\newcommand{\Lamup}[0]{{\Lambdaup}}
\newcommand{\delup}[0]{{\deltaup}}
\newcommand{\sigup}[0]{{\sigmaup}}
\newcommand{\Sigup}[0]{{\Sigmaup}}
\newcommand{\wup}[0]{{\omegaup}} 
\newcommand{\omgup}[0]{{\omegaup}}
\newcommand{\Omgup}[0]{{\Omegaup}}
\newcommand{\epup}[0]{\epsilonup}
\newcommand{\thup}[0]{\thetaup}
\newcommand{\varthup}[0]{\varthetaup}
\newcommand{\kup}[0]{\kappaup}
\newcommand{\kapup}[0]{\kappaup}
\newcommand{\varkup}[0]{\varkappaup}
\newcommand{\nup}[0]{\etaup}
\newcommand{\iup}[0]{\iotaup}

\newcommand{\txa}[0]{{\text{\textalpha}}}
\newcommand{\txb}[0]{{\text{\textbeta}}}
\newcommand{\txbet}[0]{{\text{\textbeta}}}
\newcommand{\txg}[0]{{\text{\textgamma}}}
\newcommand{\txgam}[0]{{\text{\textgamma}}}
\newcommand{\txGam}[0]{{\text{\textGamma}}}
\newcommand{\txy}[0]{{\text{\textlambda}}}
\newcommand{\txlam}[0]{{\text{\textlambda}}}
\newcommand{\txLam}[0]{{\text{\textLambda}}}
\newcommand{\txr}[0]{{\text{\textrho}}}
\newcommand{\txrho}[0]{{\text{\textrho}}}
\newcommand{\txsig}[0]{{\text{\textsigma}}}
\newcommand{\txSig}[0]{{\text{\textSigma}}}
\newcommand{\txw}[0]{{\text{\textomega}}}
\newcommand{\txomg}[0]{{\text{\textomega}}}
\newcommand{\txOmg}[0]{{\text{\textOmega}}}
\newcommand{\txep}[0]{{\text{\textepsilon}}}
\newcommand{\txth}[0]{{\text{\texttheta}}}
\newcommand{\txTh}[0]{{\text{\textTheta}}}
\newcommand{\txvarth}[0]{{\text{\textvartheta}}}
\newcommand{\txphi}[0]{{\text{\textphi}}}
\newcommand{\txPhi}[0]{{\text{\textPhi}}}
\newcommand{\txvarphi}[0]{{\text{\textvarphi}}}
\newcommand{\txpsi}[0]{{\text{\textpsi}}}
\newcommand{\txPsi}[0]{{\text{\textPsi}}}
\newcommand{\txd}[0]{{\text{\textdelta}}}
\newcommand{\txdel}[0]{{\text{\textdelta}}}
\newcommand{\txiota}[0]{{\text{\textiota}}}
\newcommand{\txn}[0]{{\text{\texteta}}}
\newcommand{\txeta}[0]{{\text{\texteta}}}
\newcommand{\txxi}[0]{{\text{\textxi}}}
\newcommand{\txchi}[0]{{\text{\textchi}}}
\newcommand{\txpi}[0]{{\text{\textpi}}}
\newcommand{\txtau}[0]{{\text{\texttau}}}
\newcommand{\txz}[0]{{\text{\textzeta}}}
\newcommand{\txzeta}[0]{{\text{\textzeta}}}
\newcommand{\txnu}[0]{{\text{\textnu}}}
\newcommand{\txmu}[0]{{\text{\textmugreek}}} 
\newcommand{\txups}[0]{{\text{\textupsilon}}} 
\newcommand{\txk}[0]{{\text{\textkappa}}}
\newcommand{\txkap}[0]{{\text{\textkappa}}}

\newcommand{\tximu}[0]{ {\textit{\textmugreek}} } %
\newcommand{\txiu}[0]{ {\textit{\textupsilon}} } 
\newcommand{\txitau}[0]{ {\textit{\texttau}} } %
\newcommand{\txipi}[0]{ {\textit{\textpi}} }%
\newcommand{\txirho}[0]{ {\textit{\textrho}} } 
\newcommand{\txilam}[0]{ {\textit{\textlambda}} }

\newcommand{\wtxt}[0]{{\textomega}}
\newcommand{\ytxt}[0]{{\textlambda}}
\newcommand{\thtxt}[0]{{\texttheta}}
\newcommand{\btxt}[0]{{\textbeta}}
\newcommand{\atxt}[0]{{\textalpha}}
\newcommand{\gtxt}[0]{{\textgamma}}
\newcommand{\tautxt}[0]{{\texttau}}

\newcommand{\gvgr}[1]{\vec{\tbf{#1}}}
\newcommand{\uvgr}[1]{\hat{\tbf{#1}}}
\newcommand{\tvgr}[1]{\widetilde{\tbf{#1}}}
\newcommand{\bvgr}[1]{\bar{\tbf{#1}}}
\newcommand{\gvbb}[1]{\vec{\pmb{#1}}}
\newcommand{\uvbb}[1]{\hat{\pmb{#1}}}
\newcommand{\tvbb}[1]{\widetilde{\pmb{#1}}}
\newcommand{\bvbb}[1]{\bar{\pmb{#1}}}

\newcommand{\sigvec}[0]{\vec{\pmb{\upsigma}} }
 \newcommand{\sigtvec}[0]{\widetilde{\pmb{\upsigma}} }

\newcommand{\gvec}[1]{\vec{\bs{\mathrm{#1}}}}
\newcommand{\gv}[1]{\vec{\bm{#1}}}
\newcommand{\gvb}[1]{\vec{\bs{#1}}}
\newcommand*{\bsvec}[1]{\vec{\bs{#1}}}
\newcommand*{\sfvec}[1]{\,\vec{\!\sfb{#1}}}

\newcommand{\uvec}[1]{\hat{\bs{#1}}}
\newcommand{\uv}[1]{\hat{\bs{#1}}}
\newcommand{\ihat}[0]{\hat{\bs{\iota}}}
\newcommand{\ehat}[0]{\hat{\bs{e}}}
\newcommand{\bhat}[0]{\hat{\bs{b}}}
\newcommand{\shat}[0]{\hat{\bs{s}}}
\newcommand{\nhat}[0]{\uvec{n}}
\newcommand{\evec}[0]{\vec{\bs{e}}}
\newcommand{\epvec}[0]{\vec{\bs{\epsilon}}}

\newcommand{\uvecb}[1]{ {\hbs{#1}} }
\newcommand{\uvb}[1]{ {\hbs{#1}} }
\newcommand{\ihatb}[0]{\hat{\pmb{\upiota}}} 
\newcommand{\ehatb}[0]{\uvecb{e}} 
\newcommand{\bhatb}[0]{\!\!\uvecb{\;b}}
\newcommand{\hhatb}[0]{\!\!\uvecb{\;h}}
\newcommand{\shatb}[0]{\uvecb{s}}
\newcommand{\nhatb}[0]{\uvecb{n}}
\newcommand{\ohatb}[0]{\uvecb{o}}
\newcommand{\ahatb}[0]{\uvecb{a}}
\newcommand{\epvecb}[0]{\vec{\bs{\epsilon}}}
\newcommand{\evecb}[0]{ \hspace{0.3mm}\vec{\mbf{e}}\hspace{0.1mm} }

\newcommand{\tvb}[1]{\widetilde{\mbf{#1}}}
\newcommand{\tvecb}[1]{\bar{\bs{#1}}}
\newcommand{\bv}[1]{\bar{\bs{#1}}}
\newcommand{\bvb}[1]{\bar{\bs{\mathrm{#1}}}}


\newcommand{\rnote}[1]{\noindent{\footnotesize\textit{{\color{red}${\circ}$}  #1}}}
\newcommand{\red}[1]{{\color{red}#1}}

\newcommand{\bnote}[1]{\noindent{\footnotesize\textit{{\color{blue}${\circ}$} #1}}}

\newcommand{\gnote}[1]{\noindent{\footnotesize\textit{$\circ$  #1}}}

 \newcommand{\note}[1]{\noindent{\footnotesize\textit{{\color{darkgray} #1}}}}



\theoremstyle{plain} 
\newtheorem{thrm}{Theorem}[section]
\newtheorem{defn}{Def.}[section]

\theoremstyle{plain} 
\newtheorem{remit}{Remark}[section]

\theoremstyle{definition} 
\newtheorem{remark}{Remark}[section]


\newtheoremstyle{remsans}
{2ex} 
{2ex} 
{\rmfamily\slshape\small}
{}
{\sffamily\bfseries\small}
{.}
{.3em}
{}

\theoremstyle{remsans}
\newtheorem{remsf}{$\iisqr\,$remark}[section]

\newtheoremstyle{MYremark}
{2ex} 
{2ex} 
{\rmfamily\itshape}
{}
{\rmfamily\mdseries\upshape\scshape}
{.}
{.3em}
{}

\theoremstyle{MYremark}
\newtheorem{remrm}{$\iisqr\,$Remark}[section]
\newtheorem{noether}[remark]{\textit{Noether}}

\newtheoremstyle{remslant}
{8pt} 
{12pt} 
{\rmfamily\slshape\small}
{}
{\rmfamily\slshape\bfseries\small}
{.}
{.3em}
{}

\theoremstyle{remslant} 
\newtheorem{remsl}{Remark}[section]

\newtheoremstyle{remnopunc}
{8pt} 
{8pt} 
{\rmfamily\small}
{}
{\bfseries\small}
{}
{.2em}
{}
\theoremstyle{remnopunc} 
\newtheorem*{noteblt}{$\nmblt$}
\newtheorem*{notestr}{\raisebox{0.1ex}{$\star$}}
\newtheorem*{noteast}{$\bs{*}$}

\newenvironment{notesq}
    {
     \begin{itemize}[left=1mm,labelsep=0.3ex,topsep=6pt]
     \item[$\iisqr$] }
    {\end{itemize}
     }

\newtheoremstyle{remslantnopunc}
{8pt} 
{12pt} 
{\slshape\rmfamily\small}
{}
{\slshape\bfseries\small}
{}
{.2em}
{}
\theoremstyle{remslantnopunc} 
\newtheorem*{notesl}{$\iisqr$}

\newtheoremstyle{remrmnonbold}
{2ex} 
{2ex} 
{\rmfamily\small}
{}
{\rmfamily\itshape\small}
{.}
{.2em}
{}
\theoremstyle{remrmnonbold}
\newtheorem*{notation}{Notation}



\newenvironment{smol}
    {}
    {%
    }





\vspace{-1.5\baselineskip}
\begin{abstract}
This note derives the full Hamiltonian equations of motion on extended phase space for nonconservative and/or nonautonomous dynamical systems,
with particular attention given to 
the momentum coordinate and generalized force associated with time.
Time reparameterization is also considered, clarifying when it does or does not preserve Hamiltonian structure.
The analytical mechanics framework is used throughout, with all developments formulated in terms of \textit{arbitrary} symplectic coordinates; 
the main developments are not predicated on a Lagrangian description. 
\end{abstract}

\renewcommand{\contentsname}{}
\vspace{-2\baselineskip}
\tableofcontents 
\vspace{-1\baselineskip}
\setcounter{section}{-1} 

 \section{Introduction}\label{sec:intro}

Although nonconservative forces and extended phase space are, individually, established topics in Hamiltonian dynamics, their combination is seldom developed explicitly.
Nonconservative forces are routinely incorporated as generalized forces through virtual work and the Hamilton--d'Alembert principle, as treated in classical texts such as \cite{lanczos1986variational,whittaker1937treatise,greenwood2003advanced,papastavridis2014analytical}.
Independently, the (symplectic) extended phase space formulation, in which time is appended as a generalized coordinate paired with its conjugate momentum coordinate, 
is developed notably by Lanczos \cite{lanczos1986variational}, Greenwood \cite{greenwood1997classical}, Johns \cite{johns2011analytical}, and Struckmeier \cite{struckmeier2005hamiltonian,struckmeier2009extended}. 
Although some sources address both nonconservative forces and extended Hamiltonian dynamics (e.g., \cite{lanczos1986variational,greenwood1997classical,johns2011analytical}), 
 their combination within the extended Hamiltonian framework does not appear to be developed, leaving the two subjects largely separate. We aim to fill that gap.

 Analytical mechanics precedents to this work's Hamiltonian construction are found mainly on the Lagrangian side.
Sharma et al.~\cite{SHARMA2018energyInt} formulate an extended Lagrange-d'Alembert principle for time-dependent systems with nonconservative forces and account for virtual work associated with variations in time. 
Johns \cite{johns2011analytical} likewise develops extended Lagrangian equations with nonconservative forces, including the additional generalized force and virtual displacement associated with the time coordinate.
Yet, such developments have not been carried over to the Hamiltonian description in arbitrary symplectic coordinates.


\paragraph*{Why extended phase space?}
While the consideration of nonconservative forces requires little explanation, the role of extended phase space is less immediately apparent and warrants brief motivation. 
Two common situations motivate the use of extended phase space:

\begin{enumerate}[left=0pt]
   \item \emph{Nonautonomous systems.}
    Nonautonomous Hamiltonian systems arise whenever the Hamiltonian function itself depends explicitly on \eq{t}
    (time\footnote{Although we often refer to \eq{t} as time throughout, it may represent whatever evolution parameter (i.e., independent variable) is relevant to the system under consideration.}), 
    as occurs throughout dynamical systems and optimal control problems. A standard ``trick'' to render such a system as autonomous is to simply append \eq{t} as an additional dependent state variable, with governing equation \eq{\diff{t}{t}=1}, and thereby conveniently circumvent the additional complications associated with explicit \eq{t}-dependence.
     Yet, this typically produces a state space of \textit{odd} dimension  \eq{2n{+}1} and therefore does not permit a canonical Hamiltonian formulation.
    Although such a \eq{(2n{+}1)}-dim formulation may be treated elegantly using cosymplectic structures or related presymplectic formulations 
    \cite{Munoz1991TimeDependent, deLeonSardon2017Cosymplectic, MunozRoy2025geometrymechanics}, 
    doing so requires familiarity with concepts and notation specialized to the differential geometric formalism (avoided in this work). 
    An alternative presentation, accessible within the analytical mechanics framework, is to introduce an additional momentum coordinate, \eq{\pi_t}, conjugate to \eq{t}. The resulting \eq{(2n{+}2)}-dim \textit{extended phase space} again admits the familiar Hamiltonian formulation \cite{lanczos1986variational,johns2011analytical,struckmeier2005hamiltonian}.

    \item \emph{Evolution parameter transformations.}
    Extended phase space also arises naturally whenever one wishes to replace some Hamiltonian system's evolution parameter, \eq{t}, by some other parameter, \eq{\varep}, prescribed via \eq{\mrm{d}t=\cf \mrm{d}\varep} for some function \eq{\cf > 0} (a conformal scaling factor). 
    At the level of first-order ODEs, such transformations can be easily implemented by directly multiplying the original ODEs by \eq{\diff{t}{\varep} =\cf}. 
    Unfortunately, for a general phase-space-dependent \eq{\cf}, such direct conformal scaling does \textit{not} preserve the Hamiltonian structure of the original system (apart from certain special cases) \cite{carinena2022infinitesimal}.
    Fortunately, passing to extended phase space provides a systematic implementation of the same evolution parameter transformation while retaining the canonical Hamiltonian form of the original system, albeit on a higher-dimensional phase space \cite{carinena1988timeScaling,CARINENA1991,struckmeier2005hamiltonian}.

    Such transformations of the evolution parameter in Hamiltonian systems arise prominently in regularization and linearization methods in celestial mechanics \cite{stiefel1971linear,bond1981canonical,peterson2026projective}. 
\end{enumerate}

\noindent
The above scenarios are not mutually exclusive and the extended phase space formulation accommodates either or both as standard, not exceptional. 
Conversely, neither scenario is required for its use. 

\paragraph*{This work.}
We develop the extended phase space construction of Hamiltonian dynamics in the presence of general nonconservative forces using the framework of classic analytical mechanics. Developments are formulated in arbitrary symplectic coordinates \eq{(\tup{x},\tup{\piup})}, which need \textit{not} be configuration and momentum coordinates in the physical sense. 
Three points receive particular attention. 
(1) The complete forced extended equations of motion are derived, including the generalized force associated with \eq{t}. 
(2) The often-seen relation \eq{\pi_t=-\mscr{H}} is clarified:~it characterizes an invariant level set of the extended Hamiltonian which selects the family of extended solution curves that reproduce the original physical dynamics. 
(3) The Hamiltonian structure (or lack thereof) of the original, directly reparameterized, fully extended, and restricted extended dynamics is compared. 
If the nonconservative forces are omitted, the same developments give a concise account of the extended phase space construction of classical conservative Hamiltonian systems.

\subparagraph*{Organization.}
We first review nonconservative Hamiltonian dynamics and its Hamilton-d'Alembert derivation on ordinary phase space (\cref{sec:non_conserv}).
Transforming the evolution parameter by direct conformal scaling is then considered to motivate the extended construction (\cref{sec:confHam}). 
The full nonconservative dynamics are subsequently derived on extended phase space (\cref{sec:ext_coord_EOMs}) and it is shown that the extended Hamiltonian is \textit{always} conserved
(\cref{sec:extPbrak_Hconst}).
The physical significance of the extended Hamiltonian's zero level set is discussed and the relation \eq{\pi_t=-\mscr{H}} is clarified (\cref{sec:projToActualDyn}). 
Finally, we summarize how the Hamiltonian structure of these various presentations of the dynamics compare (\cref{sec:extHam_compare}). 
In Appx.~\ref{sec:extLagNote}, the extended Lagrangian and its Legendre transformation reaffirm the interpretation of \eq{\pi_t} and its relation with the Hamiltonian.



\paragraph*{Notation and conventions.}
This work is written in the coordinate-dominant manner of classic analytical dynamics 
with all developments formulated in terms of local coordinates or functions thereof. Our mathematical arena is only ever regarded as some finite-dimensional real coordinate vector space, \eq{\mbb{R}^\emm}, equipped with the usual structures, operations, and abuses of notation.
\begin{small}
\begin{itemize}[left=4pt]
\item Expressions such as \eq{\tup{u}\in \mbb{R}^\en} are often an abuse of notation indicating that \eq{\tup{u}} is an object that \textit{takes values in} \eq{\mbb{R}^\en}. For any such \eq{\tup{u}}, we make no distinction between \eq{\tup{u}} as an ordered \eq{\en}-tuple, \eq{(u_1,\dots,u_\en)}, and \eq{\tup{u}} as a column vector, \eq{\trn{[u_1 \dots u_\en]}}.

\item  The ``dot product'' denotes contraction in the simplest sense. For \eq{\tup{u},\tup{v}\in \mbb{R}^\en}, then \eq{\tup{u}\cdot\tup{v} = u_i \v_i} (summation implied over \eq{i=1\dots \en}). Contraction with some  \eq{A \in\mbb{R}^{\en \times \en}} is denoted either with or without a dot. That is, 
\eq{A \tup{u} = A \cdot\tup{u}=A_{ij}u_j \ibase_i} and \eq{\trn{A}\tup{u} = \trn{A}\cdot\tup{u}=\tup{u}\cdot A = A_{ij}u_i \ibase_j = A_{ji}u_j \ibase_i } (with \eq{\ibase_i} the ``standard basis'' for \eq{\mbb{R}^\en}).

\item 
We (ab)use notation \eq{y=y(x)}, or just \eq{y(x)}, to indicate something called \eq{y} is a ``function of'' something called \eq{x}. Similarly, given some implied transformation \eq{x\leftrightarrow \til{x}}, then \eq{y(x)} and \eq{y(\til{x})} should be interpreted, respectively, as ``\eq{y} expressed as a function of \eq{x}'' and ``\eq{y} expressed as a function of \eq{\til{x}}''.

\item  \eq{\pd_{x}y \equiv \pderiv{y}{x}} are equivalent notations for the partial derivative of some \eq{x}-dependent \eq{y} with respect to \eq{x}.
If \eq{\tup{y}\in\mbb{R}^{\emm}} and \eq{\tup{x}\in\mbb{R}^{\en}}, then \eq{\pd_{\tup{x}}\tup{y} = \pderiv{\tup{y}}{\tup{x}}\in\mbb{R}^{\emm \times \en}} has components \eq{(\pd_{\tup{x}}\tup{y})_{aj}  = \pderiv{y_a}{x_j}} (for \eq{a=1\dots \emm} and \eq{j=1\dots \en}). 
For any \eq{\tup{z}\in\mbb{R}^{\emm}}:
\begin{smol}
\begin{align}
      \tup{z}\cdot \pd_{\tup{x}}\tup{y} \,\equiv\, \trn{\pderiv{\tup{y}}{\tup{x}}}\cdot \tup{z}
\end{align}
\end{smol}
\end{itemize}
\end{small}%
\Verx[-2]%
More specific notation used in this work includes:
\begin{small}
\begin{itemize}[left=4pt]
\item 
\eq{\tup{r}=(r_1,\dots,r_{\en})\in\mbb{R}^{\en}} denote cartesian position coordinates associated with an orthonormal inertial basis for \eq{\en}-dim Euclidean vector space, and \eq{\tup{F}^{\nc} \in\mbb{R}^{\en}} denotes components of any/all nonconservative forces in that same basis.

\item 
\eq{(\tup{x},\tup{\piup})\in\mbb{R}^{2n}} denote arbitrary symplectic coordinates (i.e., canonical coordinates), where \eq{n\leq \en} is the degrees of freedom (DOF) of the system. 
By \textit{arbitrary} symplectic coordinates we mean that \eq{(\tup{x},\tup{\piup})} satisfy Hamilton's equations of motion in their canonical form, but that \eq{\tup{x}} and \eq{\tup{\piup}} need \textit{not} be configuration and momentum coordinates, respectively, 
  in the physical sense.\footnote{E.g., \eq{(\tup{x},\tup{\piup})} could be action-angle coordinates such as, say, the Delaunay or Poincaré orbit elements for two-body orbital dynamics.
  }
  That is, we allow that \eq{\tup{r}=\tup{r}(\tup{x},\tup{\piup},t)} may be a ``function of'' all \eq{2n} coordinates \eq{(\tup{x},\tup{\piup})}, and possibly also explicitly \eq{t}-dependent.

\item \eq{(\tup{q},\tup{p})\in\mbb{R}^{2n}} denote symplectic coordinates where \eq{\tup{q}} and \eq{\tup{p}} \textit{are} configuration-level and momentum-level coordinates, respectively, in the typical physical sense such that \eq{\tup{r}=\tup{r}(\tup{q},t)} does not depend on momentum coordinates \eq{\tup{p}}.

\item  
\eq{\pbrak{\slot}{\slot}} denotes the canonical symplectic Poisson bracket, expressed in arbitrary symplectic coordinates \eq{(\tup{x},\tup{\piup})} as follows for any scalar-valued functions \eq{f} and \eq{h}:
\begin{align} \label{pbrak_NOext}
    \pbrak{f}{h} \,=\, \pdiv{f}{\tup{x}} \cdot \pdiv{h}{\tup{\piup}} \,-\, \pdiv{f}{\tup{\piup}} \cdot \pdiv{h}{\tup{x}}
\;,\qquad\quad
    \pbrak{f}{h} \,=\, -\pbrak{h}{f}
\end{align}

\item \eq{\dot{\square}:=\diff{\square}{t}} and \eq{\rng{\square}:=\diff{\square}{\varep}} denote, respectively, differentiation with respect to evolution parameters \eq{t} and \eq{\varep}.


\item \textit{Acronyms.} DOF = degree(s) of freedom; 
ODE = ordinary differential equation(s); 
EoM = equation(s) of motion. 
\end{itemize}
\end{small}%

\section{Review of classical Hamiltonian dynamics with nonconservative forces} \label{sec:non_conserv}


Before considering extended phase space, we first summarize the canonical Hamiltonian equations of motion for systems subject to nonconservative forces. We further allow for the system to be nonautonomous.

\paragraph*{The equations of motion.}
Consider some \eq{n}-DOF system described in arbitrary symplectic coordinates, \eq{(\tup{x},\tup{\piup})\in\mbb{R}^{2n}}, by some Hamiltonian function \eq{\mscr{H}(\tup{x},\tup{\piup},t)}. 
If nonconservative forces also act on the system then
Hamilton's equations of motion incorporating such forces are stated as follows 
(derived below):
\begin{smol}
\begin{flalign} \label{Dqp_nonExt}
&&
\boxed{
 \begin{array}{lllll}
     \dot{\tup{x}} \,=\, \pdiv{\mscr{H}}{\tup{\piup}} \,-\, \tup{\alphaup}_{\nbtup{\piup}} 
\\[4pt]
      \dot{\tup{\piup}} \,= -\pdiv{\mscr{H}}{\tup{x}} \,+\, \tup{\alphaup}_{\nbtup{x}} 
 \end{array} }
\;,\qquad
\begin{array}{llll}
     \tup{\alphaup}_{\nbtup{x}} :=\, \tup{F}^{\nc} \cdot \pdiv{\tup{r}}{\tup{x}}
  \\[4pt]
     \tup{\alphaup}_{\nbtup{\piup}} :=\, \tup{F}^{\nc} \cdot \pdiv{\tup{r}}{\tup{\piup}}  
 \end{array}
\,,
&&
\end{flalign} 
\end{smol}%
where \eq{\dot{\square}:=\diff{\square}{t}} and where \eq{\tup{r},\tup{F}^{\nc} \in \mbb{R}^\en} are
cartesian coordinate vectors of position and force as previously described.
It follows from \cref{Dqp_nonExt} that the equation governing the evolution of any function \eq{f(\tup{x},\tup{\piup},t)} is then
\begin{smol}
\begin{flalign}\label{eom_pbrack1}
&&
&&
\begin{array}{rlll}
    \dot{f}
     \,=\, \pbrak{f}{\mscr{H}} 
     \,+\, \pdiv{f}{t}
     \,+\,  \tup{\alphaup}_{\nbtup{x}} {\cdot} \pdiv{f}{\tup{\piup}} \,-\, \tup{\alphaup}_{\nbtup{\piup}} \cdot \pdiv{f}{\tup{x}}    
\end{array}
&&
     \smsz{ =\, \pbrak{f}{\mscr{H}} 
     + \pdiv{f}{t}
     + \tup{F}^{\nc} \cdot \pbrak{\tup{r}}{f} }
     \,,
\end{flalign}
\end{smol}%
from which we see that \eq{\mscr{H}} itself is generally not conserved by the dynamics in \cref{Dqp_nonExt}:
\begin{smol}
\begin{flalign} \label{Hdot_nonExt}
&&
&&
    \dot{\mscr{H}} \,=\,
   \pdiv{\mscr{H}}{t}
     \,+\,  \tup{\alphaup}_{\nbtup{x}} {\cdot} \pdiv{\mscr{H}}{\tup{\piup}} \,-\, \tup{\alphaup}_{\nbtup{\piup}} \cdot \pdiv{\mscr{H}}{\tup{x}}  
 &&
\smsz{ =\, \pdiv{\mscr{H}}{t} + \tup{F}^{\nc} \cdot \pbrak{\tup{r}}{\mscr{H}} 
     \,.
     \phantom{\pbrak{\mscr{H}}{\mscr{H}}} }
\end{flalign}
\end{smol}

\begin{notesq}
\textit{Note.} If \eq{(\tup{q},\tup{p})\in\mbb{R}^{2n}} are
typical configuration-level and momentum-level coordinates in the physical sense such that \eq{\tup{r}=\tup{r}(\tup{q},t)}, then \eq{\tup{\alphaup}_\ss{\nbtup{p}} =0} and the ODEs in \cref{Dqp_nonExt}--\cref{eom_pbrack1} simplify as follows:
\begin{smol}
\begin{align} \label{dQP_nc_2}
\begin{array}{lll}
       \dot{\tup{q}} \,=\, \pdiv{H}{\tup{p}} 
    \;,\qquad 
      \dot{\tup{p}} \,=\, -\pdiv{H}{\tup{q}} + \tup{\alphaup}_\ss{\nbtup{q}}
  \;,\qquad
  \dot{f} 
     \,=\, \pbrak{f}{H} 
     \,+\,  \pdiv{f}{t}
     \,+\, \tup{\alphaup}_\ss{\nbtup{q}} \cdot \pdiv{f}{\tup{p}} 
      \,,
\end{array} 
\end{align}
\end{smol}%
with \eq{H(\tup{q},\tup{p},t)} being the relevant Hamiltonian 
with governing equation \eq{\dot{H} = \pdiv{H}{t} +  \tup{\alphaup}_{\nbtup{q}} {\cdot} \pdiv{H}{\tup{p}}}. 
\end{notesq}

\noindent 
 Forced Hamiltonian dynamics in the form of \cref{dQP_nc_2} are standard \cite{greenwood2003advanced,johns2011analytical,schaub2026analytical}. The more general form given in \cref{Dqp_nonExt} is less common, though Tveter's derivations are found in \cite{tveter1994hamilton,Tveter1998HamNoncon2}. 
 We nevertheless provide our own derivation of \cref{Dqp_nonExt} in order to establish a reference point for subsequent developments in \cref{sec:ext_coord}. 
\verx
\begin{small}
\begin{itemize}[left=3ex]
    \item[] \bemph{Derivation of \cref{Dqp_nonExt}.}   
Classically, the equations of motion may be derived using Hamilton's principle; 
of all admissible paths between given endpoints, the physical trajectories are those that render the action integral stationary:
\begin{flalign} \label{HamDeAlam_conservative}
&&
\begin{array}{llll}
    \vard I \,=\, \vard \int_{t_\iio}^{t_f}  (\tup{\piup}\cdot\dot{\tup{x}} - \mscr{H} )\,\mrm{d}t
    \,=\, 0
    \,.
\end{array}
&&
\scrsize{(conservative)}
\end{flalign}
The above assumes purely conservative dynamics. 
To account for nonconservative forces (with cartesian components \eq{\tup{F}^{\nc}}), we add a virtual work term, \eq{\vard W^\nc}, such that the above is replaced by:
\begin{align} \label{HamDeAlam_gen}
 \begin{array}{llll}
   \vard I \,+\, \int_{t_\iio}^{t_f} \vard W^{\nc} \mrm{d}t  \,=\, 0
   \;,
   \qquad 
   \vard W^{\nc} = \tup{F}^{\nc}\cdot\vard\tup{r} 
\end{array}
 \,,
\end{align}
where, despite our notation, \eq{\vard W^{\nc}} is \textit{not} the variation of a smooth function called \eq{W^{\nc}}. Rather, \eq{\vard W^{\nc}} is merely an abuse of notation for the virtual work, expressed more properly in cartesian coordinates as above
(see footnote\footnote{The notation \eq{\vard W^{\nc}}, though not uncommon, can be misleading; it is \textit{not} the variation of a scalar-valued function \eq{W^{\nc}}. 
It may be better  to think of \eq{\vard W^{\nc}} merely as shorthand for \eq{\tup{F}^{\nc}\cdot\vard\tup{r}} (where \eq{\vard\tup{r}} \textit{is} a variation). 
Likewise, the entire expression \eq{\vard I + \int_{t_\iio}^{t_f} \vard W^{\nc} \mrm{d}t } is \textit{not} the variation of a genuine action functional (only the \eq{\vard I} term is). It would then be more appropriate, strictly speaking, to refer to \cref{HamDeAlam_gen} as the Hamilton-d’Alembert principle, rather than Hamilton’s principle in its usual conservative form.}).

Let us start with the variation of action integral, \eq{\vard I}, in \cref{HamDeAlam_gen}. Carrying out the variation leads to:
\begin{align} \label{dIQP}
\begin{array}{rlllll}
   \vard I
   \,=\, \vard \int_{t_\iio}^{t_f} (\tup{\piup}\cdot\dot{\tup{x}} - \mscr{H} ) \mrm{d}t  
   &=\, (\tup{\piup}\cdot\dot{\tup{x}} - \mscr{H})  \cancel{ \vard t \big|}_{t_\iio}^{t_f}  \;+\, 
   \int_{t_\iio}^{t_f} \vard (\tup{\piup}\cdot\dot{\tup{x}} - \mscr{H} ) \mrm{d}t 
   \,=\,
   \int_{t_\iio}^{t_f} ( \tup{\piup}\cdot\vard \dot{\tup{x}} + \dot{\tup{x}}\cdot\vard \tup{\piup} 
    - \vard \mscr{H} 
    ) \mrm{d} t 
\\[6pt] 
    &=\,  \tup{\piup}\cdot \cancel{ \vard \tup{x}\big|}_{t_\iio}^{t_f}  \;+\, 
    \int_{t_\iio}^{t_f} ( - \dot{\tup{\piup}}\cdot\vard \tup{x}
    + \dot{\tup{x}}\cdot\vard \tup{\piup}
    - \vard \mscr{H} 
    ) \mrm{d} t
\\[6pt]
    &=\, \int_{t_\iio}^{t_f} \big[ {-}(\dot{\tup{\piup}} + \pdiv{\mscr{H}}{\tup{x}})\cdot\vard \tup{x} 
    \,+\, (\dot{\tup{x}}-\pdiv{\mscr{H}}{\tup{\piup}})\cdot\vard \tup{\piup} 
    \big] \mrm{d} t 
    \,,
\end{array}
\end{align}
where the above has used the following steps. 
1\textsuperscript{st} line:~the variation is brought inside the integral using the fact \eq{\vard t_\iio = \vard t_\ii{f}=0}. 
2\textsuperscript{nd} line:~integration by parts is applied to the term with \eq{\vard \dot{\tup{x}}}, along with the fact that \eq{(\vard \tup{x},\vard \tup{\piup})} 
vanish at \eq{t_\iio} and \eq{t_\ii{f}}.\footnote{Though we take the admissible variations \eq{(\vard \tup{x},\vard \tup{\piup})} to vanish at \eq{t_\iio} and \eq{t_\ii{f}}, note that only \eq{\vard \tup{x}|_{t_\iio}=\vard \tup{x}|_{t_f}=0} is required for the derivation.}
3\textsuperscript{rd} line:~simply use \eq{\vard \mscr{H} = \pdiv{\mscr{H}}{\tup{x}} \cdot \vard \tup{x} + \pdiv{\mscr{H}}{\tup{\piup}} \cdot \vard \tup{\piup} }.
These steps are common in standard textbooks. 

Next, the nonconservative virtual work term in \cref{HamDeAlam_gen} is given in cartesian coordinates by \eq{\vard W^{\nc}=\tup{F}^{\nc}\cdot\vard \tup{r}}. We need it expressed in phase space coordinates \eq{(\tup{x},\tup{\piup})}. For the most general case that \eq{\tup{r}=\tup{r}(\tup{x},\tup{\piup},t)}:
\begin{align} \label{dW_QP}
    \vard W^\nc \!=\, \tup{F}^{\nc}\cdot\vard \tup{r} 
    \,=\,
    \tup{F}^{\nc}\cdot(\pderiv{\tup{r}}{\tup{x}} \cdot \vard \tup{x} + \pderiv{\tup{r}}{\tup{\piup}} \cdot \vard \tup{\piup})
    \;\;=\;
    \tup{\alphaup}_{\nbtup{x}} \cdot \vard \tup{x} \,+\,  \tup{\alphaup}_{\nbtup{\piup}} \cdot \vard \tup{\piup}
    \,,
\end{align}
with  \eq{\tup{\alphaup}_{\nbtup{x}}:=\tup{F}^{\nc}\cdot  \pdiv{\tup{r}}{\tup{x}}} and \eq{\tup{\alphaup}_{\nbtup{\piup}}:=\tup{F}^{\nc}\cdot  \pdiv{\tup{r}}{\tup{\piup}}}. 
 Substitution of \cref{dW_QP} and \cref{dIQP} into the Hamilton–d’Alembert condition of  \cref{HamDeAlam_gen} finally leads to:
\begin{align} \label{dIQP2}
\begin{array}{llll}
   0 \,=\,
   \vard I +  \int_{t_\iio}^{t_f} \vard W^{\nc} \mrm{d}t
    \;=\;
   \int_{t_\iio}^{t_f} \Big( 
    {-}(\dot{\tup{\piup}} + \pdiv{\mscr{H}}{\tup{x}} -  \tup{\alphaup}_{\nbtup{x}} ) \cdot\vard \tup{x} 
    \,+\, 
    (\dot{\tup{x}} - \pdiv{\mscr{H}}{\tup{\piup}}+ \tup{\alphaup}_{\nbtup{\piup}}  )\cdot\vard \tup{\piup} 
    \Big)\mrm{d} t 
    \,.
\end{array}
\end{align}
For the above to hold for any and all \eq{(\vard \tup{x},\vard \tup{\piup})}, it is required that their coefficients vanish, thus verifying the governing equations of motion claimed in \cref{Dqp_nonExt}.
\hfill \eq{\sbx}
\end{itemize}
\end{small}%

\paragraph*{Canonical transformation of generalized forces.}
Consider any, possibly \eq{t}-dependent,  transformation of arbitrary symplectic coordinates (i.e., a canonical transformation), \eq{(\tup{x},\tup{\piup},t) \leftrightarrow (\tiltup{x},\tiltup{\piup},t)}. 
The dynamics for \eq{(\tiltup{x},\tiltup{\piup})} then, by definition, have the same structure as in \cref{Dqp_nonExt}:
\begin{smol}
\begin{align} \label{dqp_nc}
      \dot{\tiltup{x}} \,=\, \pdiv{\tscr{H}}{\tiltup{\piup}} - \tup{\alphaup}_{\nbtiltup{\piup}}
     \;,\qquad \dot{\tiltup{\piup}} \,=\, -\pdiv{\tscr{H}}{\tiltup{x}} + \tup{\alphaup}_{\nbtiltup{x}}
\;,&&
    \tup{\alphaup}_{\nbtiltup{x}} :=\, \tup{F}^{\nc} \cdot \pdiv{\tup{r}}{\tiltup{x}}
 \;,\qquad 
       \tup{\alphaup}_{\nbtiltup{\piup}} :=\, \tup{F}^{\nc} \cdot \pdiv{\tup{r}}{\tiltup{\piup}}
      \,,
\end{align}
\end{smol}
with \eq{\tscr{H}(\tiltup{x},\tiltup{\piup},t)} the new Hamiltonian. 
What then is the relation between \eq{(\tup{\alphaup}_{\nbtiltup{x}}, \tup{\alphaup}_{\nbtiltup{\piup}})} above and \eq{(\tup{\alphaup}_{\nbtup{x}},\tup{\alphaup}_{\nbtup{\piup}})} in \cref{Dqp_nonExt}?
It follows from their definitions in \cref{dqp_nc} and \cref{Dqp_nonExt}:%
\footnote{\textit{Derivation:} It follows from the chain rule that 
    \eq{ \tup{\alphaup}_{\nbtiltup{x}} :=\tup{F}^{\nc} \cdot\pderiv{\tup{r}}{\tiltup{x}} = \tup{F}^{\nc} \cdot  \big(\pderiv{\tup{r}}{\tup{x}} \cdot\pderiv{\tup{x}}{\tiltup{x}} +\pderiv{\tup{r}}{\tup{\piup}} \cdot\pderiv{\tup{\piup}}{\tiltup{x}} \big) 
    = \tup{\alphaup}_{\nbtup{x}} \cdot \pderiv{\tup{x}}{\tiltup{x}}   + \tup{\alphaup}_{\nbtup{\piup}} \cdot\pderiv{\tup{\piup}}{\tiltup{x}} },
and likewise for \eq{\tup{\alphaup}_{\nbtiltup{\piup}} := \tup{F}^{\nc} \cdot\pderiv{\tup{r}}{\tiltup{\piup}}  }.
}
\begin{smol}
\begin{align} \label{genForce_CT}
\begin{array}{llllll}
 \tup{\alphaup}_{\nbtiltup{x}}
      \,=\, \tup{\alphaup}_{\nbtup{x}} \cdot \pdiv{\tup{x}}{\tiltup{x}}   \,+\, \tup{\alphaup}_{\nbtup{\piup}} \cdot\pdiv{\tup{\piup}}{\tiltup{x}}
\;,\qquad 
        \tup{\alphaup}_{\nbtiltup{\piup}}
    \,=\, \tup{\alphaup}_{\nbtup{x}} \cdot\pdiv{\tup{x}}{\tiltup{\piup}}   \,+\, \tup{\alphaup}_{\nbtup{\piup}} \cdot\pdiv{\tup{\piup}}{\tiltup{\piup}}
\end{array}
\,.
\end{align}
\end{smol}%
The above is unaffected by whether or not the transformation between \eq{(\tup{x},\tup{\piup})} and \eq{(\tiltup{x},\tiltup{\piup})} is explicitly \eq{t}-dependent. Such \eq{t}-dependence matters only for the relation between \eq{\mscr{H}} and \eq{\tscr{H}}.



\section{Transforming the evolution parameter via direct conformal scaling} \label{sec:confHam}

As in \cref{sec:non_conserv}, we consider a system described in arbitrary symplectic coordinates \eq{(\tup{x},\tup{\piup})\in\mbb{R}^{2n}} by a Hamiltonian \eq{\mscr{H}(\tup{x},\tup{\piup},t)}. 
The usual \eq{t}-parameterized Hamiltonian dynamics (plus nonconservative forces) are given in \cref{Dqp_nonExt}.  
Suppose that, instead of time \eq{t}, we wish to use some other evolution parameter, \eq{\varep}, that is related to \eq{t} by some given differential relation:
\begin{smol}
\begin{align}
   \mrm{d} t \,=\, \cf(\tup{x},\tup{\piup},t) \mrm{d} \varep
    \qquad\quad \fnsize{i.e.,}\quad \rng{t} := \diff{t}{\varep}  = \cf(\tup{x},\tup{\piup},t)
    \,,
\end{align}
\end{smol}
where \eq{\rng{\square}:=\diff{\square}{\varep}} and  \eq{\cf \neq 0} may be any smooth, nowhere-vanishing, possibly \eq{t}-dependent, function on phase space. Often, one further requires that \eq{\cf >0} be strictly positive everywhere. 
We will call \eq{\cf} the \textit{conformal factor}.

If one regards the \eq{t}-parameterized equations of motion in \cref{Dqp_nonExt} simply as first-order ODEs, without regard to their Hamiltonian/symplectic structure, then their reparameterization using \eq{\varep} is not complicated; 
simply scale the original ODEs from \cref{Dqp_nonExt} by  \eq{\rng{t} \equiv \cf}:\footnote{This is simply the chain rule \eq{\diff{}{\varep} = \diff{t}{\varep}\diff{}{t} = \cf\diff{}{t}}.} 
\begin{smol}
\begin{align} \label{HnoExt_conf}
 \begin{array}{lllll}
     \dot{\tup{x}} \,=\, \pdiv{\mscr{H}}{\tup{\piup}} \,-\, \tup{\alphaup}_{\nbtup{\piup}}  
\\[4pt]
      \dot{\tup{\piup}} \,= -\pdiv{\mscr{H}}{\tup{x}} \,+\, \tup{\alphaup}_{\nbtup{x}}
 \end{array}
 \qquad\quad
 \xRightarrow[]{\mrm{d} t \,=\, \cf \mrm{d} \varep}
 \qquad\quad
 \begin{array}{lllll}
     \rng{\tup{x}} \,=\,  \rng{t} \dot{\tup{x}}
     & \equiv \, \cf(\pdiv{\mscr{H}}{\tup{\piup}} \,-\, \tup{\alphaup}_{\nbtup{\piup}} ) 
\\[4pt]
      \rng{\tup{\piup}} \,=\, \rng{t} \dot{\tup{\piup}}
       & \equiv\, \cf (-\pdiv{\mscr{H}}{\tup{x}} \,+\, \tup{\alphaup}_{\nbtup{x}}) 
\end{array}
\,.
\end{align}
\end{smol}%
However, the canonical Hamiltonian structure of the conservative part of the dynamics generally does \textit{not} survive the above conformal scaling by \eq{\cf} (i.e., the evolution parameter transformation \eq{\mrm{d} t = \cf \mrm{d} \varep}).

\paragraph*{Conformally Hamiltonian dynamics.}
To isolate the effect of the evolution parameter transformation in \cref{HnoExt_conf} on the Hamiltonian structure --- without the separate issue of tracking the nonconservative forces --- we momentarily limit consideration to the conservative case \eq{\tup{\alphaup}_{\nbtup{x}}=\tup{\alphaup}_{\nbtup{\piup}} =\tup{0}}.  \cref{HnoExt_conf} simplifies to
\begin{smol}
\begin{align} \label{HnoExt_conf_conservative}
 \begin{array}{lllll}
     \dot{\tup{x}} = \pdiv{\mscr{H}}{\tup{\piup}} 
      \;,\quad 
      \dot{\tup{\piup}} = -\pdiv{\mscr{H}}{\tup{x}}
 \end{array}
 \qquad\quad
 \xRightarrow[]{\mrm{d} t \,=\, \cf \mrm{d} \varep}
 \qquad\quad
 \begin{array}{lllll}
     \rng{\tup{x}} 
     = \cf \pdiv{\mscr{H}}{\tup{\piup}} 
      \;,\quad 
      \rng{\tup{\piup}} 
       = -\cf \pdiv{\mscr{H}}{\tup{x}} 
\end{array}
\,.
\end{align}
\end{smol}%
where the original \eq{t}-parameterized system on the left has the canonical Hamiltonian structure. 
However, for a general conformal factor \eq{\cf}, the \eq{\varep}-parameterized system on the right does \textit{not} necessarily retain that structure.  
That is, for general choices of
\eq{\cf(\tup{x},\tup{\piup},t)}, there does \textit{not} always exist a new Hamiltonian \eq{\mscr{M}(\tup{x},\tup{\piup},t)} such that the conformally scaled  equations
in \cref{HnoExt_conf_conservative} are equivalently written as a canonical Hamiltonian system:
\begin{smol}
\begin{align} \label{confHam_condition}
    \rng{\tup{x}}
     = \cf\pdiv{\mscr{H}}{\tup{\piup}}
    = \pdiv{\mscr{M}}{\tup{\piup}}
\;,\qquad
    \rng{\tup{\piup}}
    =-\cf\pdiv{\mscr{H}}{\tup{x}}
    =-\pdiv{\mscr{M}}{\tup{x}}
    \,.
\end{align}
\end{smol}%
The \eq{\cf}-scaled system (i.e., \eq{\varep}-parameterized system) is thus generally not Hamiltonian in the original phase-space coordinates. It is sometimes said to be a conformally
\footnote{The term ``conformal'' can have different meanings in the context
of Hamiltonian systems. Our use here agrees with some, though certainly not all, sources.}
Hamiltonian system (with conformal factor \eq{\cf}); 
it need not preserve the
canonical symplectic form or associated phase space volume.

\subparagraph*{Special cases.}
The loss of Hamiltonian structure of the \eq{\varep}-parameterized dynamics in \cref{HnoExt_conf_conservative} is generic, but not guaranteed; 
for certain special cases of \eq{\cf}, a new Hamiltonian \eq{\mscr{M}} satisfying \cref{confHam_condition} can indeed be found. 
In particular,
suppose that \eq{\cf=\cf(\mscr{H},t)} depends on  \eq{(\tup{x},\tup{\piup})} only through the original Hamiltonian: 
\begin{smol}
\begin{align}
    \cf
    =
     \cf \big(
        \mscr{H}(\tup{x},\tup{\piup},t),t \big)
    \,.
\end{align}
\end{smol}%
In this case, there exists a function \eq{\mscr{M}} that satisfies \cite{carinena2022infinitesimal}:%
\footnote{One may then choose \eq{\mscr{M}} to be an antiderivative of the function \eq{\cf(\mscr{H},t)} with respect to \eq{\mscr{H}}, holding \eq{t} fixed.}
\begin{smol}
\begin{align}
    \pdiv{\mscr{M}}{\tup{\piup}}
    =
    \cf\pdiv{\mscr{H}}{\tup{\piup}}
    \;,\qquad
    \pdiv{\mscr{M}}{\tup{x}}
    =
    \cf\pdiv{\mscr{H}}{\tup{x}}
    \,,
\end{align}
\end{smol}%
such that the conformally scaled dynamics in \cref{HnoExt_conf_conservative} retain Hamiltonian form with
Hamiltonian \eq{\mscr{M}}, and can be written as in \cref{confHam_condition}. 
Three simple examples are (for \eq{k\neq-1}):\footnote{These examples are understood on regions where the indicated scaling function \eq{\cf} is smooth and positive.}
\begin{small}
\begin{flalign} \label{confH_sp_examp}
\quad
     \cf=\cf(t)
    \;\;\Rightarrow\;\;
    \mscr{M}=\cf \mscr{H}
&&\Big|&& 
    \cf=\mscr{H}^{k} 
   \;\;\Rightarrow\;\;
    \mscr{M}
    =\tfrac{1}{k+1}\mscr{H}^{k+1}
    =\tfrac{1}{k+1}\cf \mscr{H}
&&\Big|&& 
    \cf=\mscr{H}^{-1} 
\;\;\Rightarrow\;\;
    \mscr{M}
    = \ln|\mscr{H}|
    \;.
\;\;
\end{flalign}
\end{small}%
Such cases require a particular relation between \eq{\cf} and \eq{\mscr{H}} in order for Hamiltonian structure of the \eq{\varep}-parameterized dynamics to be preserved in the sense of \cref{confHam_condition}. 
 For a general conformal scaling function \eq{\cf(\tup{x},\tup{\piup},t)}, the Hamiltonian structure cannot generally be preserved on standard (nonextended) phase space; however, it can indeed be retained systematically by instead making use of the \textit{extended} phase space, as detailed in the following.

\section{Nonconservative Hamiltonian dynamics on extended phase space} \label{sec:ext_coord}

The preceding section shows that direct conformal scaling generally does not preserve Hamiltonian structure on nonextended phase space (with certain special exceptions). We now show how a general prescribed evolution-parameter transformation can instead be implemented within the extended Hamiltonian framework in order to preserve this structure. 
The arbitrary nonconservative forces of 
\cref{sec:non_conserv} are incorporated throughout.

\subsection{The general extended equations of motion.} \label{sec:ext_coord_EOMs}
We first state the extended phase space dynamics, then derive them. To start, define extended phase space coordinates, \eq{\whtup{x} := (\tup{x},t)} and \eq{\whtup{\piup} := (\tup{\piup},\pi_t)},
\begin{smol}
\begin{flalign} \label{qp_ext_def}
&&
(\whtup{x},\whtup{\piup}):=(\tup{x},t,\tup{\piup},\pi_t)\in\mbb{R}^{2n+2}
 &&
\end{flalign}
\end{smol}
where phase space is extended to include \eq{t} as a generalized configuration coordinate with conjugate momentum coordinate \eq{\pi_t}.
A physical interpretation of \eq{\pi_t} is unclear at the moment, but will soon emerge in \cref{sec:projToActualDyn}. 

The \eq{\varep}-parameterized canonical equations of motion for \eq{(\whtup{x},\whtup{\piup})\in\mbb{R}^{2n+2}}, including nonconservative forces, are then obtained not from the original Hamiltonian \eq{\mscr{H}} but, rather, from its corresponding \textit{extended Hamiltonian}, 
\eq{\whscr{H} := \cf (\mscr{H}+ \pi_t)}
--- whose definition is motivated in the subsequent derivation ---
as follows:
\begin{smol}
 \begin{flalign} \label{Dqp_ext}
\boxed{
\begin{array}{cccc}
    \whscr{H}(\whtup{x},\whtup{\piup}) :=\, 
     \cf (\mscr{H}+ \pi_t)
\\[6pt]
    \begin{array}{rlrllll}
        \rng{\tup{x}} \,=& \pdiv{\whscr{H}}{\tup{\piup}} - \tup{A}_{\nbtup{\piup}}    
        &,\qquad 
         \rng{t} \,=& \pdiv{\whscr{H}}{\pi_t} 
       \;\;\;= \cf
    \\[4pt]
       \rng{\tup{\piup}} \,=& -\pdiv{\whscr{H}}{\tup{x}} +  \tup{A}_{\nbtup{x}}  
   &,\qquad 
          \rng{\pi}_t =& -\pdiv{\whscr{H}}{t} + A_t
    \end{array}
\end{array}
}
\;,
\qquad
\begin{array}{rllll}
    \tup{A}_{\nbtup{x}}  :=&  \cf \tup{\alphaup}_{\nbtup{x}}  
    = \cf\tup{F}^{\nc} \cdot \pdiv{\tup{r}}{\tup{x}}
\\[4pt]
      \tup{A}_{\nbtup{\piup}}  :=& \cf \tup{\alphaup}_{\nbtup{\piup}}  
      = \cf\tup{F}^{\nc} \cdot \pdiv{\tup{r}}{\tup{\piup}} 
\\[4pt]
    A_t \,=& 
    \fnsize{\cref{Ftgen_ext_def} }
\\[2pt]
    A_{\pi_t}  =& 0
 \end{array}
\,,
\end{flalign}
\end{smol}
with \eq{ \tup{\alphaup}_{\nbtup{x}}  = \tup{F}^{\nc} \cdot \pd_{\tup{x}} \tup{r}   }  and \eq{\tup{\alphaup}_{\nbtup{\piup}}  = \tup{F}^{\nc} \cdot \pd_{\tup{\piup}} \tup{r}} the usual generalized forces on nonextended space as in \cref{Dqp_nonExt}, 
and where we have introduced a new generalized ``force'' \eq{A_t} which we soon show is given as follows:
\begin{smol}
\begin{flalign} \label{Ftgen_ext_def}
&&
&&
    \boxed{ 
    A_t 
    \,=\, -(\tup{\alphaup}_{\nbtup{x}}  \cdot \pdiv{\whscr{H}}{\tup{\piup}} \,-\, \tup{\alphaup}_{\nbtup{\piup}}  \cdot \pdiv{\whscr{H}}{\tup{x}} ) 
    }
&&
    \smsz{
    =\, -\tup{F}^{\nc} \cdot \pbrak{\tup{r}}{\whscr{H}}
    \;\;=\, -( \tup{\alphaup}_{\nbtup{x}}  \cdot  \rng{\tup{x}}   + \tup{\alphaup}_{\nbtup{\piup}}  \cdot  \rng{\tup{\piup}})
    }
    \,. 
\end{flalign}
\end{smol}%
Note the dynamics for \eq{(\whtup{x},\whtup{\piup})=(\tup{x},t,\tup{\piup},\pi_t)} in \cref{Dqp_ext} are also written in the familiar form,
\begin{smol}
\begin{flalign} \label{Dqp_ext_canon}
   \rng{\whtup{x}}
   \,=\, \pdiv{\whscr{H}}{\whtup{\piup}}
   \,-\, \whtup{A}_{\nbwhtup{x}} 
\;,\qquad
   \rng{\whtup{\piup}}
   \,=\, -\pdiv{\whscr{H}}{\whtup{x}} \,+\, \whtup{A}_{\nbwhtup{\piup}} 
   \;,
\end{flalign}
\end{smol}%
with 
\begin{smol}
\begin{align} \label{Fgen_ext1}
    (\whtup{A}_{\nbwhtup{x}} ,\whtup{A}_{\nbwhtup{\piup}} ) 
    :=\, ( \tup{A}_{\nbtup{x}} ,A_t,\tup{A}_{\nbtup{\piup}} ,A_{\pi_t}) 
    \,=\, (\cf \tup{\alphaup}_{\nbtup{x}} ,A_t,\cf \tup{\alphaup}_{\nbtup{\piup}} ,0)
    \,\in \mbb{R}^{2n+2}
    \,.
\end{align}
\end{smol}%

\verx 
\begin{small}
\begin{itemize}
\item[] \bemph{Derivation of \cref{Dqp_ext}.} 
In \cref{sec:non_conserv}, nonconservative Hamiltonian dynamics in standard, nonextended phase space coordinates were derived using the Hamilton-d'Alembert principle, \eq{\vard I + \int_{t_\iio}^{t_f} \vard W^{\nc} \mrm{d}t  = 0} (see \cref{HamDeAlam_gen}). 
For extended phase space, we begin by reparameterizing the integrals in \cref{HamDeAlam_gen} using \eq{\mrm{d} t = \rng{t} \mrm{d} \varep}:
\begin{flalign} \label{HamDeAlam_gen_pre}
\;\;
\begin{array}{lllll}
    I \,=\,  \int_{t_\iio}^{t_f} (\tup{\piup}\cdot\dot{\tup{x}} - \mscr{H} ) \mrm{d}t  
   \,=
    \int_{\varep_\iio}^{\varep_f} \rng{t}(\tup{\piup}\cdot\dot{\tup{x}} - \mscr{H} ) \mrm{d}\varep 
    \,=
     \int_{\varep_\iio}^{\varep_f} \big(\whtup{\piup}\cdot\rng{\whtup{x}} - \rng{t}(\mscr{H} + \pi_t) \big) \mrm{d}\varep 
\end{array}
\;,&&
\begin{array}{lllll}
     \int_{t_\iio}^{t_f} \vard W^{\nc} \mrm{d}t
     \;=\; 
      \int_{\varep_\iio}^{\varep_f} \rng{t}\vard W^{\nc}\mrm{d}\varep
      \,,
\end{array}
\;
\end{flalign}
where we have also expressed \eq{I} in terms of the extended coordinates, \eq{(\whtup{x},\whtup{\piup})=(\tup{x},t,\tup{\piup},\pi_t)}, and have used the relation \eq{\rng{t}\diff{}{t} = \diff{t}{\varep}\diff{}{t} = \diff{}{\varep}}.\footnote{Explicitly: \eq{\rng{t}(\tup{\piup}\cdot\dot{\tup{x}} - \mscr{H} ) =\tup{\piup}\cdot\rng{\tup{x}} - \rng{t}\mscr{H} = \tup{\piup}\cdot\rng{\tup{x}} + \pi_t \rng{t} - \rng{t}(\mscr{H} + \pi_t)  = \whtup{\piup}\cdot\rng{\whtup{x}} - \rng{t}(\mscr{H} + \pi_t) }. }
This expression for \eq{I} motivates defining an extended Hamiltonian \eq{\whscr{H}} 
and action integral \eq{\wh{I}} as:
\begin{align}
\begin{array}{llll}
    \whscr{H}:=\cf (\mscr{H} + \pi_t)
\;, \qquad
    \wh{I} :=\, \int_{\varep_\iio}^{\varep_f} (\whtup{\piup}\cdot \rng{\whtup{x}} - \whscr{H}) \mrm{d}\varep
    \,.
\end{array}
\end{align}
Note \eq{\wh{I}} is not simply another expression for the reparameterized \eq{I} in \cref{HamDeAlam_gen_pre}; their integrands are related by:\footnote{Explicitly: \eq{\whtup{\piup}\cdot\rng{\whtup{x}} - \whscr{H}
    \,=\, \tup{\piup}\cdot\rng{\tup{x}} + \pi_t \rng{t} - \cf(\mscr{H} + \pi_t) 
    \,=\, \tup{\piup}\cdot\rng{\tup{x}} - \rng{t}\mscr{H} + (\rng{t} - \cf)(\mscr{H} + \pi_t)
    \,=\, \rng{t}(\tup{\piup}\cdot\dot{\tup{x}} - \mscr{H}) + (\rng{t} - \cf)(\mscr{H} + \pi_t) }.
    }
\begin{align} \label{integrand_rel}
 \whtup{\piup}\cdot\rng{\whtup{x}} - \whscr{H}
    \,=\, \rng{t}(\tup{\piup}\cdot\dot{\tup{x}} - \mscr{H}) \,+\, (\rng{t} - \cf)(\mscr{H} + \pi_t) 
    \,,
\end{align}
which relies only on  \eq{\diff{t}{\varep}\diff{}{t} = \diff{}{\varep}} (i.e., \eq{\rng{t}\dot{\square} = \rng{\square}}) and the definition of \eq{\whscr{H}}, but where we cannot yet assume that \eq{\rng{t} = \cf}.\footnote{In the case that \eq{\rng{t}=l} does hold (or, for that matter, in the case \eq{\pi_t=-\mscr{H}}), then we see from \cref{HamDeAlam_gen_pre}-\cref{integrand_rel} that the integrands of \eq{\wh{I}} and \eq{I} coincide along such trajectories.}

Motivated by \cref{HamDeAlam_gen_pre}-\cref{integrand_rel}, we formulate the Hamilton-d'Alembert principle on extended phase space, compared to the classic formulation in \cref{HamDeAlam_gen}, as follows:
\begin{flalign} \label{HamDeAlam_gen_ext}
&&
\begin{array}{lllll}
   \vard I +  \int_{t_\iio}^{t_f} \vard W^{\nc} \mrm{d}t \,=\, 0
\\[8pt]
    \scrsize{with }
    \left\{\begin{array}{rlll}
         I &=  \int_{t_\iio}^{t_f} (\tup{\piup}\cdot\dot{\tup{x}} - \mscr{H} )\,\mrm{d}t 
     \\[5pt]
         \vard W^{\nc} \!\!&= \tup{F}^{\nc}\cdot\vard\tup{r} 
    \end{array}\right.
\end{array}
&& \rightsquigarrow &&
\boxed{\begin{array}{lllll}
   \vard \wh{I}
    \,+\, 
    \int_{\varep_\iio}^{\varep_f} \wh{\vard W}^{\nc} \mrm{d}\varep
    \,=\, 0
\\[8pt]
    \scrsize{with }
    \left\{\begin{array}{rllll}
         \wh{I} &:=  \int_{\varep_\iio}^{\varep_f} (\whtup{\piup}\cdot \rng{\whtup{x}} - \whscr{H}) \mrm{d}\varep
     \\[5pt]
         \wh{\vard W}^{\nc} \!\!\!&:= \rng{t}\vard W^{\nc} 
   =\, \tup{F}^{\nc}\cdot (\rng{t}\vard \tup{r} - \rng{\tup{r}}\vard t ) 
    \end{array}\right.
\end{array} }
 \,,
 &&
\end{flalign}
where we have assigned the notation \eq{\wh{\vard W}^{\nc} \!\!:= \rng{t}\vard W^{\nc}}. This virtual work term will require careful handling in the extended setting. 
Above, note we have claimed, without explanation, that this is given by \eq{\tup{F}^{\nc}\cdot (\rng{t}\vard \tup{r} - \rng{\tup{r}}\vard t ) }. This will be justified shortly. For the moment, we proceed with the derivation taking this relation for granted. 

Let us first address the \eq{\vard \wh{I}} term in \cref{HamDeAlam_gen_ext}, which is more straightforward and handled the same as the classic variation \eq{\vard I} in the nonextended setting. 
Carrying out the variation yields:
\begin{align} \label{dI_ext}
\!\!
\begin{array}{rlllll} 
 \vard \wh{I} 
       \,=\, \vard \int_{\varep_\iio}^{\varep_f} (\whtup{\piup}\cdot \rng{\whtup{x}} - \whscr{H}) \mrm{d}\varep  
      &=\, (\whtup{\piup}\cdot \rng{\whtup{x}} - \whscr{H})  \cancel{ \vard \varep \big|}_{\varep_\iio}^{\varep_f} 
       \,+
       \int_{\varep_\iio}^{\varep_f} \vard (\whtup{\piup}\cdot \rng{\whtup{x}} - \whscr{H} ) \mrm{d}\varep 
       \;=\;
       \int_{\varep_\iio}^{\varep_f}  ( \whtup{\piup} \cdot \vard \rng{\whtup{x}}
       + \rng{\whtup{x}}\cdot \vard \whtup{\piup} 
       - \vard \whscr{H} 
        ) \mrm{d}\varep 
\\[5pt] 
      &=\, \whtup{\piup}\cdot \cancel{ \vard \whtup{x}\big|}_{\varep_\iio}^{\varep_f}  
      \,+\,
      \int_{\varep_\iio}^{\varep_f} ( 
      {-}\rng{\whtup{\piup}} \cdot  \vard \whtup{x} 
      + \vard \whtup{\piup}\cdot \rng{\whtup{x}} 
      - \vard \whscr{H}  ) \mrm{d}\varep 
\\[5pt] 
      &= \int_{\varep_\iio}^{\varep_f} 
      \big[ {-}(\rng{\whtup{\piup}} + \pdiv{\whscr{H}}{\whtup{x}})\cdot\vard \whtup{x} 
    \,+\, (\rng{\whtup{x}}-\pdiv{\whscr{H}}{\whtup{\piup}})\cdot\vard \whtup{\piup} 
     \big] \mrm{d}\varep 
     \\[5pt] 
      &=
     \int_{\varep_\iio}^{\varep_f} 
      \big[ {-}(\rng{\tup{\piup}} + \pdiv{\whscr{H}}{\tup{x}})\cdot\vard \tup{x} 
    + (\rng{\tup{x}}-\pdiv{\whscr{H}}{\tup{\piup}})\cdot\vard \tup{\piup} 
    - (\rng{\pi}_t + \pdiv{\whscr{H}}{t})\cdot\vard t + (\rng{t}-\pdiv{\whscr{H}}{\pi_t})\cdot\vard \pi_t 
     \big] \mrm{d}\varep 
    \,,
\end{array} 
\end{align}
where the following steps were used. 
1\textsuperscript{st} line:~the variation is taken inside the integral as usual using the fact \eq{\vard \varep_\iio = \vard \varep_\ii{f}=0}. 
2\textsuperscript{nd} line:~integration by parts is used on the term with \eq{\vard \rng{\whtup{x}}}, along with the fact that
\eq{\vard \whtup{x}} vanish at \eq{\varep_\iio} and \eq{\varep_\ii{f}}.
3\textsuperscript{rd} and 4\textsuperscript{th} lines:~we simply write out \eq{\vard \whscr{H} = \pdiv{\whscr{H}}{\whtup{x}} \cdot \vard \whtup{x} + \pdiv{\whscr{H}}{\whtup{\piup}} \cdot \vard \whtup{\piup} }.

Next, momentarily taking the expression for virtual work \eq{\wh{\vard W}^{\nc}} in \cref{HamDeAlam_gen_ext} for granted (it is derived further down), and considering the general case that \eq{\tup{r}=\tup{r}(\tup{x},\tup{\piup},t)}, we find this term is expressed in  coordinates \eq{(\whtup{x},\whtup{\piup})} as:
\begin{align} \label{virtWork_qp_ext}
\begin{array}{rllll}
       \wh{\vard W}^{\nc} \!:=\, \rng{t}\vard W^{\nc} 
       \;=\; \tup{F}^{\nc}\cdot (\rng{t}\vard \tup{r} - \rng{\tup{r}}\vard t )
    &=\,
    \tup{F}^{\nc} \cdot \big[  \rng{t}\big( \pderiv{\tup{r}}{\whtup{x}} \cdot \vard \whtup{x}  +  \pderiv{\tup{r}}{\whtup{\piup}} \cdot \vard \whtup{\piup} \big)
    \,-\,
    \big( \pderiv{\tup{r}}{\whtup{x}} \cdot \rng{\whtup{x}}  +  \pderiv{\tup{r}}{\whtup{\piup}} \cdot \rng{\whtup{\piup}} \big) \vard t \big]
\\[4pt]
     &=\,
    \tup{F}^{\nc}\cdot \big[ \rng{t} \big( \pderiv{\tup{r}}{\tup{x}} \cdot \vard \tup{x} + \pderiv{\tup{r}}{t} \vard t + \pderiv{\tup{r}}{\tup{\piup}} \cdot \vard \tup{\piup} \big)
    \,-\, 
    \big( \pderiv{\tup{r}}{\tup{x}} \cdot \rng{\tup{x}} + \pderiv{\tup{r}}{t}\rng{t} +  \pderiv{\tup{r}}{\tup{\piup}} \cdot \rng{\tup{\piup}}  \big) \vard t \big]
\\[4pt]
 &=\,
    \tup{F}^{\nc}\cdot \big[ \rng{t} \big( \pderiv{\tup{r}}{\tup{x}} \cdot \vard \tup{x} + \pderiv{\tup{r}}{\tup{\piup}} \cdot \vard \tup{\piup} \big)
    \,-\, 
    \big( \pderiv{\tup{r}}{\tup{x}} \cdot \rng{\tup{x}}  +  \pderiv{\tup{r}}{\tup{\piup}} \cdot \rng{\tup{\piup}}  \big) \vard t \big]
\\[4pt] 
     &=\,  \rng{t}(\tup{\alphaup}_{\nbtup{x}} \cdot  \vard \tup{x}  \,+\,  \tup{\alphaup}_{\nbtup{\piup}} \cdot \vard \tup{\piup})  \,-\,(\tup{\alphaup}_{\nbtup{x}}  \cdot  \rng{\tup{x}}   \,+\,  \tup{\alphaup}_{\nbtup{\piup}}  \cdot  \rng{\tup{\piup}})\vard t
\end{array}
\,,
\end{align}
with the usual \eq{\tup{\alphaup}_{\nbtup{x}} :=\tup{F}^{\nc}\cdot  \pdiv{\tup{r}}{\tup{x}}} and \eq{\tup{\alphaup}_{\nbtup{\piup}} :=\tup{F}^{\nc}\cdot  \pdiv{\tup{r}}{\tup{\piup}}}. 
Substitution of \cref{dI_ext} and \cref{virtWork_qp_ext} into  \cref{HamDeAlam_gen_ext} then gives:
\begin{align} \nonumber 
  0 \,=\,  \int_{\varep_\iio}^{\varep_f} 
    \Big( {-}(\rng{\tup{\piup}} + \pdiv{ \whscr{H}}{\tup{x}} - \rng{t}\tup{\alphaup}_{\nbtup{x}} )\cdot \vard \tup{x} 
    \,+\, (\rng{\tup{x}}-\pdiv{\whscr{H}}{\tup{\piup}} + \rng{t}\tup{\alphaup}_{\nbtup{\piup}} )\cdot \vard \tup{\piup}
    \,+\,  (\rng{t} - \pdiv{\whscr{H}}{\pi_t})\vard \pi_t 
    -  ( \rng{\pi}_t + \pdiv{\whscr{H}}{t} + \tup{\alphaup}_{\nbtup{x}}  \cdot  \rng{\tup{x}}   +  \tup{\alphaup}_{\nbtup{\piup}}  \cdot  \rng{\tup{\piup}} ) \vard t
    \Big) \,\mrm{d}\varep   
\,.
\end{align}
For the above to hold for all arbitrary \eq{(\vard \whtup{x},\vard \whtup{\piup})= (\vard \tup{x}, \vard t, \vard \tup{\piup}, \vard \pi_t)} requires that their coefficients vanish, leading to:
\begin{align}
     \rng{\tup{x}} = \pdiv{\whscr{H}}{\tup{\piup}} - \rng{t}\tup{\alphaup}_{\nbtup{\piup}}   
 \;,\qquad  
   \rng{\tup{\piup}} = -\pdiv{\whscr{H}}{\tup{x}} + \rng{t}\tup{\alphaup}_{\nbtup{x}}  
\;,\qquad
    \rng{t} = \pdiv{\whscr{H}}{\pi_t} = \cf
\;,\qquad 
      \rng{\pi}_t = -\pdiv{\whscr{H}}{t} - (\tup{\alphaup}_{\nbtup{x}}  \cdot  \rng{\tup{x}}   +  \tup{\alphaup}_{\nbtup{\piup}}  \cdot  \rng{\tup{\piup}} )
  \,.
\end{align}
Note that \eq{\pdiv{\whscr{H}}{\pi_t} = \cf} always holds (assuming \eq{\pdiv{\cf}{\pi_t}=0}) such that any solution curve must automatically satisfy the desired relation \eq{\rng{t}=\cf}. This fact allows for \eq{\rng{t}\tup{\alphaup}_{\nbtup{x}} } and \eq{\rng{t}\tup{\alphaup}_{\nbtup{\piup}} } above to be replaced by \eq{\tup{A}_{\nbtup{x}} =\cf\tup{\alphaup}_{\nbtup{x}} } and \eq{\tup{A}_{\nbtup{\piup}} =\cf\tup{\alphaup}_{\nbtup{\piup}} }, as in \cref{Dqp_ext}. 
We have, therefore, now verified the ODEs for \eq{\rng{\tup{x}}}, \eq{\rng{\tup{\piup}}}, and \eq{\rng{t}} that were claimed in \cref{Dqp_ext}. 
For the remaining ODE for \eq{\rng{\pi}_t}, note that the term \eq{\tup{\alphaup}_{\nbtup{x}}  \cdot  \rng{\tup{x}}   +  \tup{\alphaup}_{\nbtup{\piup}}  \cdot  \rng{\tup{\piup}}}, after substitution of the above dynamics for \eq{(\rng{\tup{x}},\rng{\tup{\piup}})}, is given by:
\begin{align} \label{f_t_deriv_gen}
     \tup{\alphaup}_{\nbtup{x}}  \cdot  \rng{\tup{x}}   + \tup{\alphaup}_{\nbtup{\piup}}  \cdot  \rng{\tup{\piup}}
    \,=\, \tup{\alphaup}_{\nbtup{x}}  \cdot \pdiv{\whscr{H}}{\tup{\piup}} - \tup{\alphaup}_{\nbtup{\piup}}  \cdot \pdiv{\whscr{H}}{\tup{x}}  
    \,-\,  \rng{t} ( \cancel{ \tup{\alphaup}_{\nbtup{x}}  \cdot \tup{\alphaup}_{\nbtup{\piup}}  -  \tup{\alphaup}_{\nbtup{\piup}}  \cdot \tup{\alphaup}_{\nbtup{x}}   } )
    \;=:  -A_t 
    && \Rightarrow &&
    \rng{\pi}_t = -\pdiv{\whscr{H}}{t} + A_t
\end{align}
which verifies the ODE for \eq{\rng{\pi}_t}
that was  claimed in \cref{Dqp_ext}. 

Additionally, upon imposing the equations of motion in \cref{Dqp_ext}, note that the extended virtual work in \cref{virtWork_qp_ext} is then expressed as follows (along solution curves of \cref{Dqp_ext}):
\begin{align}
\begin{array}{lllll}
 \wh{\vard W}^{\nc} 
    &=\; \rng{t}\tup{\alphaup}_{\nbtup{x}} \cdot  \vard \tup{x}  \,+\,  \rng{t}\tup{\alphaup}_{\nbtup{\piup}} \cdot \vard \tup{\piup}  
    \,-\, (\tup{\alphaup}_{\nbtup{x}}  \cdot  \rng{\tup{x}}   \,+\,  \tup{\alphaup}_{\nbtup{\piup}}  \cdot  \rng{\tup{\piup}})\vard t
\\[4pt]
    & 
    \!\!\! \xlongequal{\!\!\!\text{\cref{Dqp_ext}}\!\!\!} 
    \;\;
     \whtup{A}_{\nbwhtup{x}}  \cdot \vard \whtup{x} \,+\,  \whtup{A}_{\nbwhtup{\piup}}  \cdot \vard \whtup{\piup}
      \;=\; 
     \tup{A}_{\nbtup{x}}  \cdot \vard \tup{x} \,+\,  \tup{A}_{\nbtup{\piup}}  \cdot \vard \tup{\piup} \,+\,  A_{t} \cdot \vard t 
     \,+\, \cancel{A_{\pi_t}} \cdot \vard \pi_t
\end{array}
\end{align}
The first line is the general kinematic expression for the extended virtual work, whereas the second line expresses the same quantity assuming the equations of motion in \cref{Dqp_ext} are satisfied (with \eq{(\whtup{A}_{\nbwhtup{x}} ,\whtup{A}_{\nbwhtup{\piup}} )} defined in \cref{Fgen_ext1}). 
\end{itemize}
\end{small}


\begin{small}
\begin{itemize}
\item[]  \bemph{On the virtual work in \cref{HamDeAlam_gen_ext}.} 
The above derivation is not quite complete. 
It remains for us to justify the expression for the nonconservative virtual work that we claimed in \cref{HamDeAlam_gen_ext}:
\begin{align} \label{W_ext_1}
    \wh{\vard W}^{\nc} \!:=\, \rng{t}\vard W^{\nc} 
   =\, \tup{F}^{\nc}\cdot (\rng{t}\vard \tup{r} - \rng{\tup{r}}\vard t )  
   \qquad 
   \fnsize{i.e.,}
   \quad
    \vard W^{\nc} =\,\tup{F}^{\nc}\cdot (\vard \tup{r} - \dot{\tup{r}}\vard t )
\end{align}
where \eq{\wh{\vard W}^{\nc} \!\!:= \rng{t}\vard W^{\nc}} is an assignment of notation that arises from the reparameterization in \cref{HamDeAlam_gen_pre}. What needs justification is our claimed expression for \eq{\vard W^{\nc}} in terms of \eq{\tup{F}^{\nc}} and \eq{(\tup{r},t)} seen above.

Note that \eq{\vard W^{\nc}} in \cref{W_ext_1} appears, at first glance, to have a different meaning than it did in the nonextended setting in \cref{HamDeAlam_gen} where we said \eq{\vard W^{\nc} = \tup{F}^{\nc}\cdot \vard \tup{r}}. 
There is, in fact, no contradiction; it appears that there is because \eq{\vard} now implies something different on extended space than it did previously in \cref{sec:non_conserv}.
In \cref{sec:non_conserv}, the variation \eq{\vard} had its typical interpretation with \eq{t} as the evolution parameter; let us denote it \eq{\vard_t} (variation at fixed \eq{t}). 
However, in the present context of extended space, the variation \eq{\vard} is now with \eq{\varep} as the evolution parameter and \eq{t} a coordinate like any other; let us denote it \eq{\vard_{\varep}} (variation at fixed \eq{\varep}). 
On extended space, \eq{\vard \tup{r}} always implies that at fixed \eq{\varep}, \eq{\vard_{\varep} \tup{r}}; it consists of the displacement at fixed \eq{t}, \eq{\vard_t \tup{r}}, \textit{and} the displacement during the accompanying change in time, \eq{\dot{\tup{r}}\vard_{\varep} t}, such that:
\begin{align}
    \vard_{\varep} \tup{r} \,=\, \vard_t \tup{r} + \dot{\tup{r}}\vard_{\varep} t
\;,\qquad 
   \vard_t \tup{r} \,=\, \vard_{\varep} \tup{r} - \dot{\tup{r}}\vard_{\varep} t
\end{align}
Thus, \eq{\vard W^{\nc}} represents the same quantity, given by \eq{\tup{F}^{\nc}\cdot \vard_t \tup{r}}, on both nonextended and extended space. Yet, its expression differs on extended space (where \eq{\vard \equiv \vard_{\varep}\neq \vard_t}):
\begin{align}
\begin{array}{rllll}
    \scrsize{nonext.~phase space, }\; \vard \equiv \vard_t :  &\quad \vard W^{\nc} =\, \tup{F}^{\nc}\cdot \vard_t \tup{r}
\\[5pt]
    \scrsize{ext.~phase space, }\; \vard \equiv \vard_{\varep} : 
    &\quad \vard W^{\nc} 
    =\, \tup{F}^{\nc}\cdot \vard_t \tup{r}
    =\,\tup{F}^{\nc}\cdot (\vard_{\varep} \tup{r} - \dot{\tup{r}}\vard_{\varep} t )
    & \quad (\neq \tup{F}^{\nc}\cdot \vard_{\varep} \tup{r})
\end{array}
\,.
\end{align}
When writing \eq{ \wh{\vard W}^{\nc} \!\!:= \rng{t}\vard W^{\nc}} in \cref{HamDeAlam_gen_ext} or \cref{W_ext_1}, it is the second expression above, with \eq{\vard \equiv \vard_{\varep}},  that is relevant:
\begin{align}
    \wh{\vard W}^{\nc} =\, \rng{t}\vard W^{\nc}
    =\, \rng{t} \tup{F}^{\nc}\cdot \vard_t \tup{r}
    \,=\,\tup{F}^{\nc}\cdot (\rng{t}\vard_{\varep} \tup{r} - \rng{\tup{r}}\vard_{\varep} t )
    \;\equiv\, \tup{F}^{\nc}\cdot (\rng{t}\vard \tup{r} - \rng{\tup{r}}\vard t )
\end{align}
where we have used the purely kinematic relation \eq{\rng{t}\dot{\tup{r}} = \rng{\tup{r}}}.  
\end{itemize}
\end{small}

\subsection{Extended Poisson bracket \& conservation of the extended Hamiltonian.}
\label{sec:extPbrak_Hconst}

Let \eq{\pbrak{\slot}{\slot}} be the canonical Poisson bracket on nonextended \eq{2n}-dim phase space as in \cref{pbrak_NOext}. 
We define an analogous Poisson bracket, \eq{\ppbrak{\slot}{\slot}}, on extended \eq{(2n{+}2)}-dim phase space as follows for any scalar-valued \eq{f} and \eq{h}:
\begin{smol}
\begin{align} \label{pbrak_ext}
    \ppbrak{f}{h} :=\, \pbrak{f}{h} \,+\,  \pderiv{f}{t}\pderiv{h}{\pi_t} - \pderiv{f}{\pi_t}\pderiv{h}{t} 
    \;.
\end{align}
\end{smol}%
The ODE governing the evolution of any function \eq{f(\whtup{x},\whtup{\piup})} is then given in terms of the above, and the extended generalized nonconservative forces \eq{(\whtup{A}_{\nbwhtup{x}} ,\whtup{A}_{\nbwhtup{\piup}} ) := (\cf \tup{\alphaup}_{\nbtup{x}} ,A_t,\cf \tup{\alphaup}_{\nbtup{\piup}} ,0)} from \cref{Dqp_ext}, as:
\begin{smol}
\begin{align} \label{pbrak_ext_fdot}
\begin{array}{rllll}
       \rng{f}(\whtup{x},\whtup{\piup})
       \,=\,  \ppbrak{f}{\whscr{H}} 
     + 
    \whtup{A}_{\nbwhtup{x}}  \cdot \pderiv{f}{\whtup{\piup}} -  \whtup{A}_{\nbwhtup{\piup}}  \cdot \pderiv{f}{\whtup{x}} 
    \quad
    &=\, 
    \ppbrak{f}{\whscr{H}} 
    \,+\,   \tup{A}_{\nbtup{x}}  \cdot \pderiv{f}{\tup{\piup}} -   \tup{A}_{\nbtup{\piup}}  \cdot \pderiv{f}{\tup{x}} 
    \,+\,  A_t \pderiv{f}{\pi_t} 
 \\[5pt]
 &=\,  \ppbrak{f}{\whscr{H}} 
     \,+\, \cf \big(\tup{\alphaup}_{\nbtup{x}}  \cdot \pderiv{f}{\tup{\piup}} - \tup{\alphaup}_{\nbtup{\piup}}  \cdot \pderiv{f}{\tup{x}} \big)  
     \,-\,  \pderiv{f}{\pi_t} \big( \tup{\alphaup}_{\nbtup{x}}  \cdot \pderiv{\whscr{H}}{\tup{\piup}} - \tup{\alphaup}_{\nbtup{\piup}}  \cdot \pderiv{\whscr{H}}{\tup{x}} \big)
    \,,
\end{array}
\end{align}
\end{smol}%

(\smsize{equivalently, $\rng{f} = \ppbrak{f}{\whscr{H}} + \cf \tup{F}^{\nc} \cdot \pbrak{\tup{r}}{f} - \pderiv{f}{\pi_t} \tup{F}^{\nc} \cdot \pbrak{\tup{r}}{\whscr{H}}\,$}).

\verx
\noindent
It follows immediately 
that \eq{\whscr{H}} itself is \textit{always} a constant of motion:

\begin{remrm} \label{rem:Hext_IOM}
    The extended Hamiltonian, \eq{\whscr{H}=\cf(\mscr{H}+\pi_t)}, is \emph{always} --- even in the presence of nonconservative forces --- a constant of motion of the extended phase space dynamics in \cref{Dqp_ext}:
    \footnote{\textit{Derivation:}~using \eq{(\whtup{A}_{\nbwhtup{x}} ,\whtup{A}_{\nbwhtup{\piup}} ) := (\cf \tup{\alphaup}_{\nbtup{x}} ,A_t,\cf \tup{\alphaup}_{\nbtup{\piup}} ,0)}, with \eq{A_t= -( \tup{\alphaup}_{\nbtup{x}}  \cdot \pdiv{\whscr{H}}{\tup{\piup}} - \tup{\alphaup}_{\nbtup{\piup}}  \cdot \pdiv{\whscr{H}}{\tup{x}}) } given in \cref{Ftgen_ext_def}, then \cref{pbrak_ext_fdot} leads to:
     \begin{align} \nonumber
    \begin{array}{rllll}
           \rng{\whscr{H}}
           \,=\, 
           \cancel{ \ppbrak{\whscr{H}}{\whscr{H}} } \,+\,
          \whtup{A}_{\nbwhtup{x}} \cdot \pderiv{\whscr{H}}{\whtup{\piup}} 
         \,-\, \whtup{A}_{\nbwhtup{\piup}}  \cdot \pderiv{\whscr{H}}{\whtup{x}}
        \,=\, 
         \cf \tup{\alphaup}_{\nbtup{x}}  \cdot \pderiv{\whscr{H}}{\tup{\piup}}  
         \,-\,
         \cf \tup{\alphaup}_{\nbtup{\piup}}  \cdot \pderiv{\whscr{H}}{\tup{x}}
         \,+\, A_t  \pderiv{\whscr{H}}{\pi_t} 
        \,=\,  \cf ( \tup{\alphaup}_{\nbtup{x}}  \cdot \pderiv{\whscr{H}}{\tup{\piup}} 
        -
        \tup{\alphaup}_{\nbtup{\piup}}  \cdot \pderiv{\whscr{H}}{\tup{x}} )
         + \cf  A_t
        \,=\,  - \cf  A_t
         + \cf  A_t \,=\, 0
         \,.
    \end{array}
    \end{align}%
    Thus, \eq{\whscr{H}} not just a constant of motion, but an \emph{integral} of motion (i.e., evolution-parameter-independent constant of motion).
    }
    \begin{smol}
    \begin{flalign}
    \begin{array}{cc}
        \scrsize{integral}  \\[-2pt]
        \scrsize{of motion} 
    \end{array}\!\!:
    &&
    \begin{array}{rllll}
           \rng{\whscr{H}}
          \,=\, 
          \tup{A}_{\nbtup{x}}  \cdot \pderiv{\whscr{H}}{\tup{\piup}} -   \tup{A}_{\nbtup{\piup}}  \cdot \pderiv{\whscr{H}}{\tup{x}} \,+\,  A_t \pderiv{\whscr{H}}{\pi_t} 
         \,=\, 0
    \end{array}
    \qquad \Rightarrow \qquad
      \whscr{H} = \whscr{H}_\zr = \fnsize{ const.}
    &&
    \end{flalign}
    \end{smol}%
This holds regardless of whether or not \eq{\mscr{H}} is conserved under the nonextended phase space dynamics of \cref{Dqp_nonExt}.
\end{remrm}

Now, most quantities of physical interest are functions \eq{f(\tup{x},\tup{\piup})} on nonextended phase space, or, more generally, functions \eq{f(\tup{x},\tup{\piup},t)} with explicit \eq{t}-dependence. 
Such expressions can be viewed as functions on extended phase space  with no explicit \eq{\varep}-dependence and which are further independent of \eq{\pi_t}. 
For such functions, \cref{pbrak_ext_fdot} with \eq{\pdiv{f}{\pi_t}=0} (and \eq{\pdiv{\whscr{H}}{\pi_t}=\cf}) leads to: 
\begin{smol}
\begin{align} \label{pbrak_ext_fdot_lifted}
\begin{array}{rllll}
       \rng{f}(\tup{x},\tup{\piup},t)
 \,=\,
    \ppbrak{f}{\whscr{H}} 
    +   \tup{A}_{\nbtup{x}}  \cdot \pdiv{f}{\tup{\piup}} -   \tup{A}_{\nbtup{\piup}}  \cdot \pdiv{f}{\tup{x}} 
    \;\;
 &=\,  \ppbrak{f}{\whscr{H}} 
     + \cf \big( \tup{\alphaup}_{\nbtup{x}}  \cdot \pdiv{f}{\tup{\piup}} - \tup{\alphaup}_{\nbtup{\piup}}  \cdot \pdiv{f}{\tup{x}} \big)  
\\[5pt]
    &=\,  \pbrak{f}{\whscr{H}} 
     \,+\,  \cf \pdiv{f}{t}
        \,+\, \cf \big( \tup{\alphaup}_{\nbtup{x}}  \cdot \pdiv{f}{\tup{\piup}} - \tup{\alphaup}_{\nbtup{\piup}}  \cdot \pdiv{f}{\tup{x}} \big) 
\,,
\end{array}
\end{align}
\end{smol}%

(\smsize{equivalently, $\rng{f} = \pbrak{f}{\whscr{H}} +  \cf \pdiv{f}{t} + \cf \tup{F}^{\nc} \cdot \pbrak{\tup{r}}{f}\,$}).

\verx
\noindent
Expanding \eq{\pbrak{f}{\whscr{H}}} using \eq{\whscr{H}=\cf(\mscr{H}+\pi_t)}, then  \cref{{pbrak_ext_fdot_lifted}} yields:%
\footnote{With \eq{\whscr{H}=\cf(\mscr{H}+\pi_t)}, then \eq{\pbrak{f}{\whscr{H}} = \cf \pbrak{f}{\mscr{H}} + \tfrac{1}{\cf}\whscr{H}\pbrak{f}{\cf} }. Similarly, \eq{\ppbrak{f}{\whscr{H}} = \cf \pbrak{f}{\mscr{H}} + \cf \pdiv{f}{t} + \tfrac{1}{\cf}\whscr{H}\pbrak{f}{\cf} } (since \eq{\pdiv{\cf}{\pi_t}=\pdiv{f}{\pi_t}=0}).}
\begin{smol}
\begin{align} \label{pbrak_fdot_ext_ohNeat}
    \rng{f}(\tup{x},\tup{\piup},t)
    \,=\, \cf \Big( \pbrak{f}{\mscr{H}}
     +  \pdiv{f}{t}
     + \tup{\alphaup}_{\nbtup{x}}  \cdot \pdiv{f}{\tup{\piup}} - \tup{\alphaup}_{\nbtup{\piup}}  \cdot \pdiv{f}{\tup{x}}  \Big)
     + \tfrac{\whscr{H}}{\cf}\pbrak{f}{\cf}
     \qquad=\, \cf \dot{f}
     \,+\, \tfrac{\whscr{H}}{\cf}\pbrak{f}{\cf}
     \,,
\end{align}
\end{smol}%
with \eq{\dot{f}= \pbrak{f}{\mscr{H}} + \pdiv{f}{t} + \tup{\alphaup}_{\nbtup{x}}  \cdot \pdiv{f}{\tup{\piup}} - \tup{\alphaup}_{\nbtup{\piup}}  \cdot \pdiv{f}{\tup{x}}}
 being the usual \eq{t}-parameterized governing ODE for any \eq{f(\tup{x},\tup{\piup},t)} 
on nonextended space, as per
\cref{eom_pbrack1}. 
The implications of \cref{pbrak_fdot_ext_ohNeat} and \cref{rem:Hext_IOM} are considered next.

\subsection{The physical extended dynamics:~restriction to the zero level set.}
\label{sec:projToActualDyn}

For arbitrary initial conditions \eq{(\whtup{x}_\zr,\whtup{\piup}_\zr)\in\mbb{R}^{2n+2}}, there exists a unique solution curve, \eq{(\whtup{x}_\varep,\whtup{\piup}_\varep)}, satisfying the general extended dynamics \cref{Dqp_ext}. However, the projection of such solutions back to the standard, nonextended phase space --- that is, the ``\eq{(\tup{x},\tup{\piup})}-part'' of the solution --- does \textit{not} necessarily agree with the true physical dynamics on nonextended phase space given in \cref{Dqp_nonExt}, even for the same \eq{(\tup{x}_\zr,\tup{\piup}_\zr)\in\mbb{R}^{2n}}.

\paragraph*{The zero level set, $\Sigma_{0}$.}
As established in \cref{rem:Hext_IOM}, \eq{\whscr{H}} is conserved and, therefore, for any value \eq{\kappa \in \mbb{R}}, the level set \eq{\Sigma_{\kappa} = \{ \whscr{H}=\kappa \}} is invariant under the extended dynamics of \cref{Dqp_ext}.
More specifically, each \eq{\Sigma_{\kappa}}
\begin{smol}
\begin{align} \label{Hext_levelsets}
\Sigma_{\kappa}
:=\,
\big\{\,
(\whtup{x},\whtup{\piup})\in\mbb{R}^{2n+2}
\;\big|\;
\whscr{H}(\whtup{x},\whtup{\piup})=\kappa
\,\big\} \,\subset \mbb{R}^{2n+2}
\;,
\end{align}
\end{smol}%
is a \eq{(2n{+}1)}-dim embedded hypersurface in \eq{(2n{+}2)}-dim extended phase space that is invariant under the full extended dynamics --- including nonconservative forces --- of \cref{Dqp_ext}.
The extended dynamics admit solution curves on any/every such surface \eq{\Sigma_{\kappa}}. Yet, we will show that one of them, the zero level set (\eq{\kappa=0}),
\begin{smol}
\begin{align} \label{Hext_0_set}
    \Sigma_{0}
    \,=\,
    \big\{\,\whscr{H}=0\, \big\}
    \,=\,
    \big\{\,\pi_t=-\mscr{H}\, \big\}
\;,
\end{align}
\end{smol}%
is distinguished:
\textit{when restricted to \eq{\Sigma_{0}}, the extended dynamics universally reproduce (up to reparameterization) the original nonextended dynamics in \cref{Dqp_nonExt}, 
and the additional momentum \eq{\pi_t} satisfies \eq{\pi_t=-\mscr{H}}}.\footnote{
The relations \eq{\pi_t=-\mscr{H}} and \eq{\whscr{H}=0} are equivalent; this follows from \eq{\whscr{H}:=\cf(\mscr{H}+\pi_t)} and the assumption that \eq{\cf} is
nowhere vanishing. 
The extended Lagrangian construction in \cref{sec:extLagNote} later provides an independent derivation of the relation \eq{\pi_t=-\mscr{H}}.
}
This is verified starting around \cref{fdot_ext_difference} below. First, let us clarify the relations in \cref{Hext_0_set}:


Since \eq{\Sigma_{0}\subset\mbb{R}^{2n+2}} is invariant under the extended dynamics, every solution curve starting in \eq{\Sigma_{0}} remains in \eq{\Sigma_{0}} and, consequently, every such solution curve \eq{(\whtup{x}_{\varep},\whtup{\piup}_{\varep})\in\Sigma_{0}} satisfies, 
\begin{smol}
\begin{flalign} \label{pt_H_specific_curve}
    &&
    \whscr{H}(\whtup{x}_{\varep},\whtup{\piup}_{\varep}) = 0
    \qquad \Leftrightarrow \qquad
    (\pi_t)_{\varep}
    =-\mscr{H}(\tup{x}_{\varep},\tup{\piup}_{\varep},t_{\varep})
    \;,
&&
\end{flalign}
\end{smol}%
for all \eq{\varep}. 
When we use the casual notation \eq{\whscr{H}=0} or \eq{\pi_t=-\mscr{H}}, it should be interpreted as above. 
It does \textit{not} mean, for instance, that \eq{\pi_t} should be treated as a function, \eq{-\mscr{H}(\tup{x},\tup{\piup},t)}, of the other coordinates. This distinction is important when differentiating \eq{\whscr{H}} to obtain the extended ODEs in \cref{Dqp_ext}; when doing so, \eq{\pi_t} must be treated as an independent coordinate. 
However, \textit{after} obtaining the extended ODEs this way, we may then restrict them to \eq{\Sigma_{0}} by substituting into their right-hand sides
the relation \eq{\whscr{H}=0} (equivalently, \eq{\pi_t=-\mscr{H}}). 


We note that restriction of the extended dynamics to \eq{\Sigma_{0}} imposes no physical constraints on the original nonextended dynamics. More specifically, it imposes no restriction on the admissible initial data of the nonextended system:~for any initial condition \eq{(\tup{x}_{\zr},\tup{\piup}_{\zr})\in\mbb{R}^{2n}} at time \eq{t_{\zr}}, there is a unique choice of the initial value of the additional momentum coordinate, 
\eq{(\pi_t)_{\zr}=-\mscr{H}(\tup{x}_{\zr},\tup{\piup}_{\zr},t_{\zr})},
that places the corresponding extended initial condition \eq{(\whtup{x}_{\zr},\whtup{\piup}_{\zr})} in \eq{\Sigma_{0}\subset\mbb{R}^{2n+2}}. Equivalently, \eq{\whscr{H}(\whtup{x}_{\zr},\whtup{\piup}_{\zr})=0}.
Since \eq{\Sigma_{0}} is invariant, the relation \eq{\whscr{H}=0}, and the equivalent relation \eq{\pi_t=-\mscr{H}}, remains satisfied along the resulting extended solution curve. Thus, the restriction constrains only the additional coordinate \eq{\pi_t}; it does not further constrain or exclude any initial condition of 
--- and hence no corresponding evolution of --- 
the original system. Rather, it selects an invariant class of extended solution curves that are guaranteed to agree, up to the prescribed reparameterization, with the original nonextended dynamics, as shown below.

\paragraph*{Recovering the nonextended dynamics.}
The role of the zero level set is already visible in
\cref{pbrak_fdot_ext_ohNeat}. For any observable/function \eq{f(\tup{x},\tup{\piup},t)} "lifted" to a function on extended phase space:
\begin{smol}
\begin{align} \label{fdot_ext_difference}
    \rng{f}
    \,=\,
    \cf\dot{f} + \tfrac{\whscr{H}}{\cf}\pbrak{f}{\cf}
    \;,
\end{align}
\end{smol}%
with \eq{\dot{f}} as in \cref{eom_pbrack1}. 
Thus, restriction to \eq{\Sigma_{0}} guarantees
the \eq{\varep}-parameterized dynamics agree with the original \eq{t}-parameterized dynamics (up to reparameterization):
\begin{smol}
\begin{align} \label{fdot_ext_restricted}
    \rng{f}\big|_{\Sigma_{0}}
    =\,
    \cf\dot{f}
    =\,
    \rng{t}\dot{f}
    &&
    \fnsize{i.e.,}
    \quad 
     \whscr{H}=0 \quad\Rightarrow\quad 
     \rng{f} \simeq
    \cf\dot{f} =
    \rng{t}\dot{f}
    \;,
\end{align}
\end{smol}%
where we will use the shorthand notation "\eq{\simeq}" to indicate relations that hold when restricting consideration to \eq{\Sigma_{0}}. 
The above \cref{fdot_ext_difference}-\cref{fdot_ext_restricted} imply that:

\begin{remrm} \label{rem:Hext_0}
  \eq{\Sigma_{0}\subset\mbb{R}^{2n+2}} is the invariant level set on which the extended dynamics
    universally reproduce the original, nonextended dynamics (up to the reparameterization prescribed by \eq{\mrm{d} t = \cf \mrm{d} \varep}). 
\end{remrm}


\begin{remrm} \label{rem:lifted_IOM}
If a function \eq{f(\tup{x},\tup{\piup},t)} is a constant of motion of the nonextended
dynamics, then it is also a constant of motion of the extended
dynamics restricted to \eq{\Sigma_{0}}:\footnote{
In fact, it becomes an \emph{integral} of motion of the restricted
extended dynamics, even if it was only a constant of motion of
the nonextended system (its lift has no explicit \eq{\varep}-dependence).}
\begin{smol}
\begin{align} \label{lifted_IOM_ext}
\dot{f}=0
\qquad\Longrightarrow\qquad
\rng{f}\big|_{\Sigma_{0}} = \cf \dot{f} = 0
\;.
\end{align}
\end{smol}%
\end{remrm}

The word ``universally'' in \cref{rem:Hext_0} warrants clarification. For a particular function \eq{f}, the term \eq{\tfrac{1}{\cf}\whscr{H}\pbrak{f}{\cf}}
in \cref{fdot_ext_difference} may also vanish if \eq{\pbrak{f}{\cf}=0}, which \textit{can} happen even off of restriction to
\eq{\Sigma_{0}} (i.e., even if \eq{\whscr{H}\neq 0}). For example, if \eq{\cf} is
independent of \eq{(\tup{x},\tup{\piup})} --- including the case
\eq{\cf=\cf(t)} --- all such correction terms vanish on every level set. Therefore,
\eq{\whscr{H}=0} is always sufficient, but not always necessary. Yet, it often \textit{is} necessary in the case of a
nontrivial conformal scaling function \eq{\cf(\tup{x},\tup{\piup},t)} for which \eq{\pbrak{f}{\cf}\neq 0} for general observables \eq{f}.

\paragraph*{Equations of motion on $\Sigma_{0}$.}
To illustrate \cref{rem:Hext_0}, we will show that the full \eq{\varep}-parameterized extended dynamics for \eq{(\whtup{x},\whtup{\piup})\in\mbb{R}^{2n+2}} in \cref{Dqp_ext} do not necessarily agree with the original \eq{t}-parameterized nonextended dynamics for \eq{(\tup{x},\tup{\piup})\in\mbb{R}^{2n}} in \cref{Dqp_nonExt}, but that they \textit{do} agree when the former is restricted to \eq{\Sigma_{0}}.

Let \eq{z_i} denote any extended coordinate other than
\eq{\pi_t} and note that
\begin{smol}
\begin{flalign} \label{qpdot_ext_alt_deriv}
\qquad
\begin{array}{llll}
    \pderiv{\whscr{H}}{z_i} = \pderiv{}{z_i} \big( \cf (\mscr{H}+\pi_t) \big) 
    \,=\, \cf  \pderiv{\mscr{H}}{z_i} +  (\mscr{H}+\pi_t) \pderiv{\cf}{z_i}  
    \,=\,
   \cf  \pderiv{\mscr{H}}{z_i}  
   +\tfrac{\whscr{H}}{\cf} \pderiv{\cf}{z_i}
\end{array}
 && \xRightarrow{\whscr{H}\,=\,0} &&
 \pderiv{\whscr{H}}{z_i} \simeq \cf  \pderiv{\mscr{H}}{z_i}
 \,,
 \qquad
\end{flalign}
\end{smol}%
such that, for the \eq{2n} coordinates \eq{(\tup{x},\tup{\piup})}, the extended dynamics in \cref{Dqp_ext} are equivalently:
\begin{smol}
\begin{align}\label{qpdot_ext_offsurface}
 \begin{array}{llllllll}
      \rng{\tup{x}} \,=\, \pdiv{\whscr{H}}{\tup{\piup}} - \cf\tup{\alphaup}_{\nbtup{\piup}}   
      &\;\;=\, \cf (\pdiv{\mscr{H}}{\tup{\piup}} - \tup{\alphaup}_{\nbtup{\piup}} ) 
      \,+\, \tfrac{\whscr{H}}{\cf} \pdiv{\cf}{\tup{\piup}} 
      &\;\;=\, \cf \dot{\tup{x}} 
      \,+\, \tfrac{\whscr{H}}{\cf} \pdiv{\cf}{\tup{\piup}} 
\\[4pt]
    \rng{\tup{\piup}} \,= -(\pdiv{\whscr{H}}{\tup{x}} -  \cf\tup{\alphaup}_{\nbtup{x}} )
    &\;\;= - \cf (\pdiv{\mscr{H}}{\tup{x}} - \tup{\alphaup}_{\nbtup{x}} ) 
      \,-\, \tfrac{\whscr{H}}{\cf} \pdiv{\cf}{\tup{x}} 
    &\;\;=\,  \cf \dot{\tup{\piup}} 
      \,-\, \tfrac{\whscr{H}}{\cf} \pdiv{\cf}{\tup{x}} 
\end{array} 
\end{align}
\end{smol}%
with \eq{(\dot{\tup{x}},\dot{\tup{\piup}})} being the original \eq{t}-parameterized nonextended dynamics from \cref{Dqp_nonExt}. 
Consequently, these same dynamics are recovered if 
--- but not quite only if\footnote{Off of \eq{\Sigma_{0}} (i.e., if \eq{\whscr{H}\neq 0}), the extended dynamics can still reproduce the nonextended dynamics if \eq{\pdiv{\cf}{\tup{x}} = \pdiv{\cf}{\tup{\piup}} =0 }. This corresponds to the special case that \eq{\cf=\cf(t)}, including constant \eq{\cf}.} ---
the extended dynamics are restricted to \eq{\Sigma_{0}},
\begin{smol}
\begin{flalign} \label{qpdot_ext_alt_apx}
&&
    \whscr{H}=0 
    \qquad\Rightarrow\qquad 
\begin{array}{rllll}
    \rng{\tup{x}}
    &\simeq\,
    \cf (
    \pdiv{\mscr{H}}{\tup{\piup}}-\tup{\alphaup}_{\nbtup{\piup}}  )
    =\cf\dot{\tup{x}}
    =\rng{t}\dot{\tup{x}}
    \;,
    \\[5pt]
    \rng{\tup{\piup}}
    &\simeq\,
    \cf (
    -\pdiv{\mscr{H}}{\tup{x}}+\tup{\alphaup}_{\nbtup{x}}  )
    =\cf\dot{\tup{\piup}}
    =\rng{t}\dot{\tup{\piup}}
    \;.
\end{array}
&&
\end{flalign}
\end{smol}%

The remaining extended equation, the ODE for \eq{\rng{\pi}_t} in \cref{Dqp_ext}, also reduces consistently.
Using \cref{qpdot_ext_alt_deriv},
this equation is equivalently written as
\begin{smol}
\begin{align}
\begin{array}{rllllll}
      \rng{\pi}_t =& -\pdiv{\whscr{H}}{t} + A_t 
\\[3pt]
      =&
      -\cf \pdiv{\mscr{H}}{t} + A_t - \tfrac{\whscr{H}}{\cf}\pdiv{\cf}{t}
\end{array}
\qquad
\begin{array}{rllllll}
   \scrsize{with:}\quad  
    A_t :=&  -(\tup{\alphaup}_{\nbtup{x}}  \cdot \pdiv{\whscr{H}}{\tup{\piup}} \,-\, \tup{\alphaup}_{\nbtup{\piup}}  \cdot \pdiv{\whscr{H}}{\tup{x}} )
\\[3pt]
    =&
    -\cf (\tup{\alphaup}_{\nbtup{x}}  \cdot \pdiv{\mscr{H}}{\tup{\piup}} - \tup{\alphaup}_{\nbtup{\piup}}  \cdot \pdiv{\mscr{H}}{\tup{x}} ) 
    -
    \tfrac{\whscr{H}}{\cf} (\tup{\alphaup}_{\nbtup{x}}  \cdot \pdiv{\cf}{\tup{\piup}} - \tup{\alphaup}_{\nbtup{\piup}}  \cdot \pdiv{\cf}{\tup{x}} )
\end{array}
\end{align}
\end{smol}%
such that restriction to \eq{\Sigma_{0}} yields:
\begin{smol}
\begin{align}
   \whscr{H}=0 
   \qquad\Rightarrow\qquad 
    \rng{\pi}_t \simeq -\cf(\pdiv{\mscr{H}}{t} - \alpha_t)
    \;,\qquad 
    \scrsize{with:} \quad
    \alpha_t :=\,  -(\tup{\alphaup}_{\nbtup{x}}  \cdot \pdiv{\mscr{H}}{\tup{\piup}} \,-\, \tup{\alphaup}_{\nbtup{\piup}}  \cdot \pdiv{\mscr{H}}{\tup{x}} )
    \,
\end{align}
\end{smol}%
\smsize{(equivalently:~$\alpha_t = -\tup{F}^{\nc}\cdot \pbrak{\tup{r}}{\mscr{H}}
= -\tup{F}^{\nc}\cdot  (\dot{\tup{r}} - \pd_t \tup{r} ) 
$.)}\footnote{Note \eq{\alpha_t} has physical units of power. Also, if \eq{\pd_t \tup{r}=0}, then \eq{\alpha_t = -\tup{F}^{\nc}\cdot \dot{\tup{r}}} is the (negative) rate at which nonconservative forces do work on the system. 
}\\
Note from the above that, on \eq{\Sigma_{0}}, the ODE for \eq{\rng{\pi}_t} precisely reproduces (up to reparameterization) the governing equation for 
\eq{-\mscr{H}} on the nonextended space from \cref{Hdot_nonExt}:
\begin{smol}
\begin{align}
    \rng{\pi}_t \simeq
    -\cf \big( \pdiv{\mscr{H}}{t} \,+\, \tup{\alphaup}_{\nbtup{x}}  \cdot \pdiv{\mscr{H}}{\tup{\piup}} \,-\, \tup{\alphaup}_{\nbtup{\piup}}  \cdot \pdiv{\mscr{H}}{\tup{x}} \big)
    \;=\; -\cf \dot{\mscr{H}}
    \;.
\end{align}
\end{smol}%

Altogether, the complete extended dynamics from \cref{Dqp_ext}, and their equivalent expressions when restricted to \eq{\Sigma_{0}}, are summarized as:
\begin{smol}
\begin{align} \label{Dqp_Hext0}
\boxed{\begin{array}{rlllllllllll}
      \rng{\tup{x}} 
      \,=& \pdiv{\whscr{H}}{\tup{\piup}} - \tup{A}_{\nbtup{\piup}}  
      &\quad \simeq&
      \cf ( \pdiv{\mscr{H}}{\tup{\piup}}-\tup{\alphaup}_{\nbtup{\piup}}  )
    &=\cf\dot{\tup{x}}
\\[5pt]
     \rng{\tup{\piup}} 
     \,=& -(\pdiv{\whscr{H}}{\tup{x}} -  \tup{A}_{\nbtup{x}} )
     &\quad \simeq&
     -\cf ( \pdiv{\mscr{H}}{\tup{x}}-\tup{\alphaup}_{\nbtup{x}}  )
    &=\cf\dot{\tup{\piup}}
 \\[5pt]
     \rng{t} 
     \,=& \pdiv{\whscr{H}}{\pi_t} 
       \;\;= \cf
       &\quad =&
       \quad \scrsize{(no change)}
       &= \cf \dot{t}
 \\[5pt]
     \rng{\pi}_t 
     \,=& -(\pdiv{\whscr{H}}{t} - A_t)
     &\quad \simeq&
     -\cf(\pdiv{\mscr{H}}{t} - \alpha_t)
    &= -\cf\dot{\mscr{H}}
\end{array} }
\end{align}
\end{smol}
where "\eq{\simeq}" indicates relations that hold when restricting consideration to \eq{\Sigma_{0}}.
That is, 
\begin{smol}
\begin{align} \label{full_ext_restricted}
( \rng{\tup{x}},\rng{t},\rng{\tup{\piup}},\rng{\pi}_t )
\big|_{\Sigma_{0}}
=\,
\cf \cdot ( \dot{\tup{x}},1,\dot{\tup{\piup}},-\dot{\mscr{H}} )
=\,
\diff{t}{\varep} \cdot ( \dot{\tup{x}},1,\dot{\tup{\piup}},-\dot{\mscr{H}} )
\;.
\end{align}
\end{smol}%
Note the \eq{2n}-dim ``\eq{(\tup{x},\tup{\piup})}-part'' of
\cref{Dqp_Hext0} is precisely the directly reparameterized,
conformally Hamiltonian system in \cref{HnoExt_conf}.

\subsection{Keeping track of Hamiltonian structure.} \label{sec:extHam_compare}

We have now encountered four closely related forms of the equations of motion for a dynamical system described by a Hamiltonian function \eq{\mscr{H}(\tup{x},\tup{\piup},t)}:
(1) the original nonextended dynamics in \cref{Dqp_nonExt}, (2) their directly
reparameterized form in \cref{HnoExt_conf}, (3) the extended dynamics in
\cref{Dqp_ext}, and (4) the restriction of the extended dynamics to the zero level set
\eq{\Sigma_{0}=\{\whscr{H}=0\}} in \cref{Dqp_Hext0}. 
We pause to summarize their structural
differences since they are not all canonical Hamiltonian systems. 

To more clearly track --- without the separate complication introduced by nonconservative forces --- which aspects of the Hamiltonian structure are retained or lost among these similar presentations of the equations of motion, we momentarily restrict consideration to the conservative case, \eq{\tup{F}^{\nc}= 0}, so that all nonconservative generalized forces vanish.\footnote{
Such nonconservative forces can always be restored in the dynamics as previously described. We omit them here because their presence generally corrupts Hamiltonian structure (the matter at hand) from the start and they will only distract from the present discussion.
}


\begin{enumerate}
\item
\emph{The original \eq{t}-parameterized nonextended dynamics on \eq{\mbb{R}^{2n}}}
(\cref{Dqp_nonExt}):
\begin{small}
\begin{flalign} \label{Dqp_compare1}
&&
   \dot{\tup{x}}
   = \pdiv{\mscr{H}}{\tup{\piup}}
   \;,\qquad
   \dot{\tup{\piup}}
   =-\pdiv{\mscr{H}}{\tup{x}} 
   \;,
   && \phantom{\big(\scrsize{with}\;\rng{t}=\cf\big)}
\end{flalign}
\end{small}%
are Hamiltonian equations in standard canonical form on the original
\eq{2n}-dim phase space. 

\item
\emph{The \eq{\varep}-parameterized nonextended dynamics on \eq{\mbb{R}^{2n}} by direct conformal scaling}
(\cref{HnoExt_conf}):
\begin{small}
\begin{flalign} \label{Dqp_compare2}
&&
   \rng{\tup{x}}
   =\cf\pdiv{\mscr{H}}{\tup{\piup}}
   \;,\qquad
   \rng{\tup{\piup}}
   =-\cf\pdiv{\mscr{H}}{\tup{x}}
   \;,
&&
   \big(\scrsize{with}\;\rng{t}=\cf\big)
\end{flalign}
\end{small}%
are obtained simply by scaling
the original ODEs by \eq{\cf=\mrm{d}t/\mrm{d}\varep}.
For a general scaling function
\eq{\cf = \cf(\tup{x},\tup{\piup},t)}, the resulting equations above are only conformally
Hamiltonian but are generally \textit{not} themselves Hamiltonian with
respect to the original canonical symplectic structure. Yet, certain special choices of \eq{\cf} may nevertheless preserve Hamiltonian form, as discussed around \cref{confH_sp_examp}. 

If \eq{t} is to be recovered in the solution, its ODE \eq{\rng{t}=\cf} must also be included, giving a
\eq{(2n{+}1)}-dim system.

\item
\emph{The \eq{\varep}-parameterized extended dynamics on \eq{\mbb{R}^{2n+2}}}
(\cref{Dqp_ext}), generated by
\eq{\whscr{H}=\cf(\mscr{H}+\pi_t)}:
\begin{small}
\begin{align} \label{Dqp_compare3a}
   \rng{\tup{x}}
   &=\pdiv{\whscr{H}}{\tup{\piup}}
   \;,\qquad
   \rng{\tup{\piup}}
   =-\pdiv{\whscr{H}}{\tup{x}}
   \;,\qquad
   \rng{t}
   =\pdiv{\whscr{H}}{\pi_t}
   =\cf
   \;,\qquad
   \rng{\pi}_t
   =-\pdiv{\whscr{H}}{t}
   \;,
\end{align}
\end{small}%
are equivalently written in the familiar canonical form,
\begin{small}
\begin{flalign} \label{Dqp_compare3b}
&&
   \rng{\whtup{x}}
   =\pdiv{\whscr{H}}{\whtup{\piup}}
   \;,\qquad
   \rng{\whtup{\piup}}
   =-\pdiv{\whscr{H}}{\whtup{x}}
   \;.
 && \phantom{\big(\scrsize{with}\;\rng{t}=\cf\big)}
\end{flalign}
\end{small}%
Unlike the direct conformal scaling in \cref{Dqp_compare2}, the above \textit{are} canonical Hamiltonian
equations on the full \eq{(2n{+}2)}-dim extended phase space. The
additional coordinate pair \eq{(t,\pi_t)} therefore allows the evolution parameter transformation to be implemented while retaining canonical
Hamiltonian structure fully analogous to \cref{Dqp_compare1}. 

\item
\emph{The \eq{\varep}-parameterized extended dynamics restricted to \eq{\Sigma_{0}\subset \mbb{R}^{2n+2}}}
(\cref{Dqp_Hext0}):
\begin{small}
\begin{align} \label{Dqp_compare4}
      \rng{\tup{x}} \,= \pdiv{\whscr{H}}{\tup{\piup}} 
      \,\simeq \cf\pdiv{\mscr{H}}{\tup{\piup}} 
\;,\qquad 
      \rng{\tup{\piup}} \,= -\pdiv{\whscr{H}}{\tup{x}} 
      \,\simeq -\cf\pdiv{\mscr{H}}{\tup{x}} 
\;,\qquad 
    \rng{t} \,= \pdiv{\whscr{H}}{\pi_t} \,= \cf
\;,\qquad 
        \rng{\pi}_t = -\pdiv{\whscr{H}}{t}
        \,\simeq -\cf\pdiv{\mscr{H}}{t}
 \;,
\end{align}
\end{small}%
are not actually a fourth, separate formulation; they are the equivalent form of the full extended dynamics in  \cref{Dqp_compare3b} when restricted to their invariant level set \eq{\Sigma_{0}} 
--- that is, simplifying the ODEs using \eq{\whscr{H}=0} ---
on which the physically relevant extended solution curves reside. 
However,
\eq{\Sigma_{0}} has odd dimension \eq{2n{+}1} and therefore cannot itself carry a symplectic structure: the canonical symplectic structure belongs to the
ambient \eq{(2n{+}2)}-dim extended phase space and becomes degenerate
when restricted to \eq{\Sigma_{0}}. Correspondingly, the above
\eq{2n{+}2} ODEs evolve subject to the conserved relation \eq{\pi_t=-\mscr{H}} (i.e., \eq{\whscr{H}=0}), so only \eq{2n{+}1} coordinates are independent on \eq{\Sigma_{0}}. 


We reiterate that such restriction to \eq{\Sigma_{0}} does \textit{not} imply any physical constraints; nothing need be done to ``enforce'' the restriction aside from limit consideration of initial conditions to those that satisfy \eq{\whscr{H}(\whtup{x}_{\zr},\whtup{\piup}_{\zr}) = 0}. Equivalently, for any given \eq{(\tup{x}_{\zr},\tup{\piup}_{\zr},t_{\zr})}, one simply selects \eq{(\pi_t)_{\zr} = -\mscr{H}(\tup{x}_{\zr},\tup{\piup}_{\zr},t_{\zr})}.
\textit{This is not a constraint relation; it is a conservation property}. 
It imposes no restriction on the nonextended phase space. 
The dynamics on \eq{\Sigma_{\zr}} are merely a description of the physical dynamics on the subspace where they naturally evolve.
\end{enumerate}

\noindent
Thus, direct conformal scaling provides the simplest transformation of the
evolution parameter but generally sacrifices canonical Hamiltonian
structure. Extension to \eq{\mbb{R}^{2n+2}} with the additional coordinate pair \eq{(t,\pi_t)} retains that
structure on the full extended phase space. Subsequent restriction to the surface
\eq{\Sigma_{0}\subset \mbb{R}^{2n+2}} singles out the extended solution curves that reproduce the
original physical dynamics. 
The restricted equations remain the restriction of the ambient extended Hamiltonian system. 


\begin{appendices}
\renewcommand{\appendixname}{}

\switchToTenPoint

\section{Appendix}


\subsection{Relation to the Lagrangian \& the Legendre transformation.}
\label{sec:extLagNote}

The preceding Hamiltonian developments need not be predicated on an accompanying Lagrangian description. Still, to complete the picture, 
we will briefly show how the extended Hamiltonian and momentum coordinate conjugate to \eq{t} are recovered by considering an extended Lagrangian and the corresponding Legendre transformation. We refer the reader to \cite{johns2011analytical,struckmeier2005hamiltonian,struckmeier2009extended} for a more complete treatment. 
Nonconservative forces play no part in the relations considered here and will therefore not appear. 


In contrast to the arbitrary symplectic coordinates of preceding developments,
we now specialize to typical configuration coordinates \eq{\tup{q}\in\mbb{R}^{n}}
(in the physical sense\footnote{That is, the cartesian coordinate position vector is some mapping  \eq{\tup{r}=\tup{r}(\tup{q},t)}, independent of \eq{\tup{v}} (or of \eq{\tup{p}}). }
)
with corresponding generalized velocity coordinates \eq{\tup{v}\in\mbb{R}^{n}}. 
I.e., \eq{\tup{v}_t=\dot{\tup{q}}_t} when evaluated along a curve \eq{\tup{q}_t\equiv \tup{q}(t)}.
The corresponding conjugate momenta will be denoted \eq{\tup{p}\in\mbb{R}^{n}}.

\paragraph*{Extended Lagrangian.}
Let \eq{\noscr{L}(\tup{q},\tup{v},t)} be the usual nonextended Lagrangian function for the system of interest;  we assume it permits such a description and also that the associated nonextended Legendre transformation is regular. 
As before, suppose we wish to exchange \eq{t} for some other evolution parameter \eq{\varep}. We begin by reparameterizing the Lagrangian action integral: 
\begin{smol}
\begin{align} \label{extLag_action}
\begin{array}{lllll}
    I_\ss{\noscr{L}}
    \,=\, \int_{t_\iio}^{t_f}\noscr{L} \,\mrm{d}t
     \,=\, \int_{\varep_\iio}^{\varep_f}
       \rng{t}\noscr{L} \,\mrm{d}\varep
    \,.
\end{array}
\end{align}
\end{smol}%
where we have used only the kinematic relation \eq{\mrm{d} t = \rng{t} \mrm{d} \varep}. 
The above motivates defining an extended Lagrangian function \eq{\whnoscr{L}} of extended configuration and velocity coordinates \eq{(\whtup{q},\whtup{u})\in\mbb{R}^{2n+2}}:
\begin{smol}
\begin{flalign} \label{extLag_def}
&&
  \whnoscr{L}(\whtup{q},\whtup{u})
    :=
    u_t\noscr{L} 
\;,\qquad
    \whtup{q}:=(\tup{q},t)
\,,\quad
    \whtup{u}:=(\tup{u},u_t) 
\;, &&
\end{flalign}
\end{smol}%
with \eq{\whtup{u}=(\tup{u},u_t)} related to the usual nonextended velocity coordinates \eq{\tup{v}} as follows (for \eq{u_t \neq 0}):
\begin{smol}
\begin{align}
    \tup{v}= \tfrac{1}{u_t} \tup{u}
\quad \leftrightarrow\quad 
    \tup{u} = u_t \tup{v}  
\;.
\end{align}
\end{smol}%
The extended velocity coordinates \eq{\whtup{u}=(\tup{u},u_t)} are understood as follows:~when evaluated along any extended configuration coordinate curve, \eq{\whtup{q}_{\varep}=(\tup{q}_{\varep},t_{\varep})}, they satisfy:
\begin{smol}
\begin{flalign} \label{extVel_meaning}
\;\;\fnsize{along a curve:}
&&
    \whtup{u}_{\varep} = \rng{\whtup{q}}_{\varep}
\qquad\quad
\fnsize{i.e.,}
\quad
    \tup{u}_{\varep} = \rng{\tup{q}}_{\varep} = \rng{t}_{\varep} \dot{\tup{q}}_{t}
\;,\quad 
    (u_t)_{\varep} = \rng{t}_{\varep}
\;.
&&
\end{flalign}
\end{smol}%
Thus, along such curves, \eq{\whnoscr{L}} in \cref{extLag_def} reproduces the integrand of the action integral in \cref{extLag_action}.

\subparagraph*{Note.}
For \eq{ \whnoscr{L} := u_t\noscr{L}} given in \cref{extLag_def},  the original \eq{\noscr{L}(\tup{q},\tup{v},t)} is regarded as a function \eq{\noscr{L}(\tup{q},\tup{v}(\whtup{u}),t)} of \eq{(\whtup{q},\whtup{u})}
---that is, the original expression for \eq{\noscr{L}(\tup{q},\tup{v},t)} rewritten using direct substitution of \eq{\tup{v}=\tup{u}/u_t}.

\paragraph*{Extended Legendre transformation.}
The Legendre transformation associated with the extended Lagrangian \eq{\whnoscr{L}} proceeds much as it does in the nonextended setting. 
We introduce extended momentum coordinates \eq{\whtup{p}=(\tup{p},p_t)\in\mbb{R}^{n+1}} using a familiar definition:
\begin{smol}
\begin{align} \label{extLag_pext_def}
     \whtup{p} :=  \pdiv{\whnoscr{L}}{\whtup{u}}
     \;.
\end{align}
\end{smol}%
The first \eq{n} components, \eq{ \tup{p} := \pdiv{\whnoscr{L}}{\tup{u}} \in \mbb{R}^n}, coincide 
with the usual nonextended relation \eq{ \tup{p} = \pdiv{\noscr{L}}{\tup{v}}}:
    \footnote{\textit{Derivation:} using \eq{\whnoscr{L}=u_t \noscr{L}} and \eq{\tup{v} = \tup{u}/u_t} we find that  \eq{\pdiv{\whnoscr{L}}{\tup{u}}  =  u_t \pdiv{\noscr{L}}{\tup{u}} = u_t \pdiv{\noscr{L}}{\tup{v}} \cdot \pdiv{\tup{v}}{\tup{u}} = u_t \pdiv{\noscr{L}}{\tup{v}} \cdot (u_t^{-1} \imat_n) = \pdiv{\noscr{L}}{\tup{v}}}. 
    }
\begin{smol}
\begin{align} \label{extLag_p_def}
    \tup{p}
    &:=
    \pdiv{\whnoscr{L}}{\tup{u}}
    \,=\,
    \pdiv{\noscr{L}}{\tup{v}}
    \,.
\end{align}
\end{smol}%
\cref{extLag_pext_def} also defines the momentum coordinate conjugate to \eq{t} as 
\eq{p_t := \pdiv{\whnoscr{L}}{u_t}}, which turns out to be \eq{-\noscr{H}}:
    \footnote{\textit{Derivation:} using \eq{\whnoscr{L}= u_t \noscr{L}} with the original \eq{\noscr{L}} now a function \eq{\noscr{L}(\tup{q},\tup{v}(\whtup{u}),t)} of \eq{(\whtup{q},\whtup{u})} 
    --- i.e., the original expression for \eq{\noscr{L}(\tup{q},\tup{v},t)} rewritten using direct substitution of \eq{\tup{v}(\whtup{u})=\tup{u}/u_t}. Then, noting that \eq{\pdiv{\tup{v}}{u_t} = -u_t^{-2} \tup{u} = -u_t^{-1}\tup{v}},
    we verify \cref{extLag_pt_def} as follows:\\
    \eq{\qquad p_t := \pdiv{\whnoscr{L}}{u_t}  
    \,=\, \pdiv{(u_t\noscr{L})}{u_t} = \noscr{L} + u_t \pdiv{\noscr{L}}{u_t} 
    \,=\, \noscr{L} + u_t \pdiv{\noscr{L}}{\tup{v}}\cdot \pdiv{\tup{v}}{u_t} 
    \,=\,
    \noscr{L} + u_t \pdiv{\noscr{L}}{\tup{v}}\cdot(-u_t^{-1} \tup{v}) 
    \,=\, \noscr{L} -  \pdiv{\noscr{L}}{\tup{v}}\cdot\tup{v}
    \,=\, \noscr{L} -  \tup{p}\cdot\tup{v}
    } .\\
    where the final equality has made use of \cref{extLag_p_def}. 
    }
\begin{smol}
\begin{align} \label{extLag_pt_def}
    p_t
    :=
    \pdiv{\whnoscr{L}}{u_t}
 \;=\; 
    \noscr{L} -  \pdiv{\noscr{L}}{\tup{v}} \cdot \tup{v}
    \,=\, \noscr{L}-\tup{p}\cdot\tup{v}
    \,=\, -\noscr{H}
    \,,
\end{align}
\end{smol}%
with \eq{\noscr{H}:=
\tup{p}\cdot\tup{v}-\noscr{L}} the usual nonextended Hamiltonian, with \eq{\tup{v}} expressed in terms of \eq{(\tup{q},\tup{p},t)} by inversion of \cref{extLag_p_def} (assumed possible). 

Next, in analogy to the usual relation  \eq{\noscr{H}=
\tup{p}\cdot\tup{v}-\noscr{L}}, consider the analogous extended expression \eq{\whtup{p}\cdot\whtup{u}-\whnoscr{L}}:
\begin{smol}
\begin{align} \label{extLag_legendre}
    \whtup{p}\cdot\whtup{u}-\whnoscr{L}
\;=\;
    \tup{p}\cdot\tup{u}+p_tu_t
    -u_t\noscr{L}
\;=\;
    u_t\big(
       \tup{p}\cdot\tup{v}-\noscr{L}+p_t
    \big)
\;=\;
    u_t(\noscr{H}+p_t)
    \,.
\end{align}
\end{smol}%
This would recover the definition of the extended Hamiltonian, \eq{\whnoscr{H}:=\cf (\noscr{H} + p_t)} from \cref{sec:ext_coord_EOMs}, if an identification can be made between the velocity coordinate \eq{u_t} and the conformal scaling function \eq{\cf(\tup{q},\tup{p},t)}. 
Such an identification can indeed be made, though it does not arise directly from the extended Legendre transformation, which  is nonregular (even if the original nonextended Legendre transformation is regular):
\footnote{The extended Lagrangian in \cref{extLag_def} is homogeneous of degree one in \eq{\whtup{u}}. 
That is, scaling \eq{\whtup{u} \mapsto k \whtup{u}= (k \tup{u},k u_t)}, for a scalar value \eq{k\neq 0}, leaves \eq{\tup{v}=\tup{u}/u_t} unchanged and thus \eq{\noscr{L}} unchanged, and therefore gives \eq{\whnoscr{L}=u_t \noscr{L} \mapsto k u_t \noscr{L} = k \whnoscr{L}}.} 
the velocity coordinate \eq{u_t} cannot be solved for in terms of the momenta.
 However, along any extended configuration curve it holds by definition that \eq{u_t=\rng{t}=\diff{t}{\varep}}, as per \cref{extVel_meaning}. An evolution parameter transformation prescribed by \eq{\mrm{d}t=\cf \mrm{d}\varep} (as considered previously) therefore gives the clear choice of this otherwise undetermined relation to be \eq{u_t=\cf}. Substitution of this choice into \cref{extLag_legendre} defines the corresponding extended Hamiltonian as:
\begin{smol}
\begin{align} \label{extHam_from_Legendre}
 \whnoscr{H}(\whtup{q},\whtup{p})
    :=
    \cf (\noscr{H} + p_t )
    \,.
\end{align}
\end{smol}%
Thus, the extended Lagrangian and its Legendre transformation independently recover the relation \eq{p_t=-\noscr{H}} associated with the zero level set and, upon imposing the prescribed evolution-parameter transformation, also yield the extended Hamiltonian \eq{\whnoscr{H}=\cf (\noscr{H} + p_t)}.

\end{appendices}

\let\oldbibitem\bibitem
\renewcommand{\bibitem}{\setlength{\itemsep}{0pt}\oldbibitem}

\phantomsection
\addcontentsline{toc}{section}{References}
\switchToTenPoint
\begin{small}
\bibliographystyle{ieeetr}
\bibliography{JOEreference}

@article{peterson2026projective,
  title={Projective transformations for linearized and regularized central-force dynamics: Hamiltonian formulation},
  author={Peterson, Joseph TA and Majji, Manoranjan and Junkins, John L},
  journal={Celestial Mechanics and Dynamical Astronomy},
  volume={138},
  number={5},
  pages={54},
  year={2026},
  publisher={Springer},
  doi = {10.1007/s10569-026-10304-3}
}

@article{struckmeier2005hamiltonian,
  title={Hamiltonian Dynamics on the Symplectic Extended Phase Space for Autonomous and Non-Autonomous Systems},
  author={Struckmeier, J{\"u}rgen},
  journal={Journal of Physics A: Mathematical and General},
  volume={38},
  number={6},
  pages={1257},
  year={2005},
  publisher={IOP Publishing}
}

@article{struckmeier2009extended,
  title={Extended Hamilton--Lagrange formalism and its application to Feynman's path integral for relativistic quantum physics},
  author={Struckmeier, J{\"u}rgen},
  journal={International Journal of Modern Physics E},
  volume={18},
  number={01},
  pages={79--108},
  year={2009},
  publisher={World Scientific}
}

@article{carinena1988timeScaling,
  author  = {Cari{\~n}ena, Jos{\'e} F. and
             Ibort, Luis A. and
             Lacomba, Ernesto A.},
  title   = {Time Scaling as an Infinitesimal Canonical Transformation},
  journal = {Celestial Mechanics},
  volume  = {42},
  pages   = {201--213},
  year    = {1988},
  doi     = {10.1007/BF01232957}
}

@article{CARINENA1991,
title = {Hamiltonian formalism and time transformations in the extended phase space {$T^*(Q \times R)$}},
author = {José F. Cariñena and Manuel F. Rañada},
journal = {Reports on Mathematical Physics},
volume = {30},
number = {1},
pages = {89-106},
year = {1991},
issn = {0034-4877},
doi = {10.1016/0034-4877(91)90043-M},
url = {https://www.sciencedirect.com/science/article/pii/003448779190043M}
}

@article{carinena2022infinitesimal,
  title={Infinitesimal Time Reparametrisation and its Applications},
  author={Cari{\~n}ena, Jos{\'e} F and Mart{\'\i}nez, Eduardo and Mu{\~n}oz-Lecanda, Miguel C},
  journal={Journal of Nonlinear Mathematical Physics},
  volume={29},
  number={3},
  pages={523--555},
  year={2022},
  publisher={Springer}
}

@article{deLeonSardon2017Cosymplectic,
author  = {de Le{\'o}n, Manuel and Sard{\'o}n, Cristina},
title   = {Cosymplectic and Contact Structures to Resolve Time-Dependent and Dissipative Hamiltonian Systems},
journal = {Journal of Physics A: Mathematical and Theoretical},
volume  = {50},
number  = {25},
pages   = {255205},
year    = {2017},
doi     = {10.1088/1751-8121/aa711d}
}

@book{MunozRoy2025geometrymechanics,
  author        = {Mu{\~n}oz-Lecanda, Miguel C. and Rom{\'a}n-Roy, Narciso},
  title         = {Geometry of Mechanics},
  series        = {Advanced Textbooks in Mathematics},
  publisher     = {World Scientific Publishing Company},
  year          = {2025},
  eprint        = {2401.12650},
  archivePrefix = {arXiv},
  primaryClass  = {math-ph},
  doi           = {10.1142/q0490},
  url           = {https://arxiv.org/abs/2401.12650}
}

@article{Munoz1991TimeDependent,
author  = {Echeverr{\'i}a-Enr{\'i}quez, Arturo
         and Mu{\~n}oz-Lecanda, Miguel C.
         and Rom{\'a}n-Roy, Narciso},
title   = {Geometrical Setting of Time-Dependent Regular Systems: Alternative Models},
journal = {Reviews in Mathematical Physics},
volume  = {3},
number  = {3},
pages   = {301--330},
year    = {1991},
doi     = {10.1142/S0129055X91000114}
}

@book{lanczos1986variational,
  title     = {The Variational Principles of Mechanics},
  author    = {Lanczos, Cornelius},
  edition   = {4th},
  year      = {1986},
  publisher = {Dover Publications},
  address   = {New York},
  isbn      = {978-0486650678}
}

@book{whittaker1937treatise,
  title     = {A Treatise on the Analytical Dynamics of Particles and Rigid Bodies: With an Introduction to the Problem of Three Bodies},
  author    = {Whittaker, Edmund Taylor},
  edition   = {4},
  publisher = {Cambridge University Press},
  year      = {1937},
  address   = {Cambridge},
  isbn      = {9780521358835}
}

@book{greenwood1997classical,
  author    = {Greenwood, Donald T.},
  title     = {Classical Dynamics},
  publisher = {Dover Publications},
  address   = {Mineola, NY},
  year      = {1997},
  pages     = {337},
  isbn      = {978-0486696904},
  note      = {Unabridged and corrected republication of the work first published by Prentice-Hall, Inc., Englewood Cliffs, N.J., in 1977}
}

@book{greenwood2003advanced,
  title={Advanced Dynamics},
  author={Greenwood, Donald T.},
  isbn={9780521826129},
  lccn={2007273288},
  url={https://books.google.com/books?id=kR2PeXP9wZoC},
  year={2003},
  publisher={Cambridge University Press}
}

@book{papastavridis2014analytical,
  author    = {Papastavridis, John G.},
  title     = {Analytical Mechanics: A Comprehensive Treatise on the Dynamics of Constrained Systems},
  publisher = {World Scientific Publishing Co.},
  year      = {2014},
  address   = {Singapore},
  edition   = {Reprint},
  isbn      = {978-981-4338-71-4},
  note      = {Reprint of the 2002 edition published by Oxford University Press}
}

@book{schaub2026analytical,
  Address   = {Reston, VA},
  Author    = {Hanspeter Schaub and John L. Junkins},
  Doi       = {10.2514/4.107597},
  Edition   = {5th},
  Publisher = {{AIAA} Education Series},
  Title     = {Analytical Mechanics of Space Systems},
  Year      = {2026}
}

@article{tveter1994hamilton,
  title={Hamilton's equations of motion for non-conservative systems},
  author={Tveter, Frank Thomas},
  journal={Celestial Mechanics and Dynamical Astronomy},
  volume={60},
  number={4},
  pages={409--419},
  year={1994},
  doi={10.1007/BF00692025},
  publisher={Springer}
}

@article{Tveter1998HamNoncon2,
    title = {Deriving the Hamilton equations of motion for a nonconservative system using a variational principle},
    author = {Tveter, Frank Thomas},
    journal = {Journal of Mathematical Physics},
    volume = {39},
    number = {3},
    pages = {1495-1500},
    year = {1998},
    month = {03},
    issn = {0022-2488},
    doi = {10.1063/1.532392},
    url = {https://doi.org/10.1063/1.532392},
    eprint = {https://pubs.aip.org/aip/jmp/article-pdf/39/3/1495/19166576/1495_1_online.pdf},
}

@article{bond1981canonical,
  title={Canonical orbital elements in terms of an arbitrary independent variable},
  author={Bond, Victor R and Janin, Guy},
  journal={Celestial mechanics},
  volume={23},
  number={2},
  pages={159--172},
  year={1981},
  publisher={Springer},
  doi = {10.1007/BF01229551},
  url = {https://link.springer.com/article/10.1007/BF01229551}
}

@book{stiefel1971linear,
  author    = {Stiefel, Eduard L. and Scheifele, Gerhard},
  title     = {{Linear and Regular Celestial Mechanics: Perturbed Two-Body Motion, Numerical Methods, Canonical Theory}},
  publisher = {Springer-Verlag},
  address   = {Berlin and New York},
  year      = {1971},
  doi       = {10.1007/978-3-642-65027-7}
}

@article{SHARMA2018energyInt,
title = {Energy-preserving variational integrators for forced Lagrangian systems},
journal = {Communications in Nonlinear Science and Numerical Simulation},
volume = {64},
pages = {159-177},
year = {2018},
issn = {1007-5704},
doi = {https://doi.org/10.1016/j.cnsns.2018.04.015},
url = {https://www.sciencedirect.com/science/article/pii/S1007570418301266},
author = {Harsh Sharma and Mayuresh Patil and Craig Woolsey}
}

@book{johns2011analytical,
  author    = {Johns, Oliver Davis},
  title     = {Analytical Mechanics for Relativity and Quantum Mechanics},
  edition   = {2nd},
  publisher = {Oxford University Press},
  year      = {2011},
  address   = {Oxford; New York},
  isbn      = {9780191001628},
  series    = {Oxford Graduate Texts}
}
\end{small}

\end{document}